\documentclass[american,aps,pra,reprint,tightenlines,superscriptaddress,twocolumn,twoside,longbibliography]{revtex4-1}
\pdfoutput=1

\usepackage[unicode=true,pdfusetitle, bookmarks=true,bookmarksnumbered=false,bookmarksopen=false, breaklinks=false,pdfborder={0 0 0},backref=false,colorlinks=false] {hyperref}
\hypersetup{ colorlinks,linkcolor=myurlcolor,citecolor=myurlcolor,urlcolor=myurlcolor}
\usepackage{braket,colortbl,amsthm,amsmath,amssymb,dsfont}
\definecolor{myurlcolor}{rgb}{0,0,0.7}
\usepackage{graphics,graphicx,relsize}
\usepackage{xcolor}
\usepackage{XCharter}
\usepackage[T1]{fontenc}
\usepackage{mathptmx}
\usepackage[left=1.6cm,right=1.6cm,top=2.45cm,bottom=2.45cm]{geometry}
\usepackage[caption=false]{subfig}
\usepackage{bigints}
\usepackage{accents}
 
\theoremstyle{plain}

\def\bea{\begin{eqnarray}}
\def\eea{\end{eqnarray}}
\def\ba{\begin{array}}
\def\ea{\end{array}}

\def\beq{\begin{equation}}
\def\eeq{\end{equation}}
\def\pc{{\hspace{-6px}c\hspace{2px}}}

\usepackage[normalem]{ulem}
 
\begin{document}

\title{Emergent non-Markovianity and dynamical quantification of the quantum switch}

\author{Vishal Anand}
\email{vishal.anand@research.iiit.ac.in}
\affiliation{Center for Security Theory and Algorithmic Research, \\
International Institute of Information Technology, Gachibowli, Hyderabad 500 032, India}

\author{Ananda G. Maity}
\email{anandamaity289@gmail.com}
\affiliation{Networked Quantum Devices Unit, 
Okinawa Institute of Science and Technology Graduate University, Onna-son, Okinawa 904-0495, Japan}
\author{Subhadip Mitra}
\email{subhadip.mitra@iiit.ac.in}
\affiliation{Center for Computational Natural Sciences and Bioinformatics,\\
International Institute of Information Technology, Gachibowli, Hyderabad 500 032, India}  
\affiliation{Center for Quantum Science and Technology,\\
International Institute of Information Technology, Gachibowli, Hyderabad 500 032, India} 

\author{Samyadeb Bhattacharya} 
\email{samyadeb.b@iiit.ac.in}
\affiliation{Center for Quantum Science and Technology,\\
International Institute of Information Technology, Gachibowli, Hyderabad 500 032, India} 

\affiliation{Center for Security Theory and Algorithmic Research, \\
International Institute of Information Technology, Gachibowli, Hyderabad 500 032, India}

\begin{abstract}
{\noindent We investigate the dynamical aspects of the quantum switch and find a particular form of quantum memory emerging out of the switch action. We first analyse the loss of information in a general quantum evolution subjected to a quantum switch and propose a measure to quantify the switch-induced memory. We then derive an uncertainty relation between information loss and switch-induced memory. We explicitly consider the example of depolarizing dynamics and show how it is affected by the action of a quantum switch. For a more detailed analysis, we consider both the control qubit and the final measurement on the control qubit as noisy and investigate the said uncertainty relation. Further, while deriving the Lindblad-type dynamics for the reduced operation of the switch action, we identify that the switch-induced memory actually leads to the emergence of non-Markovianity. Interestingly, we demonstrate that the emergent non-Markovianity can be explicitly attributed to the switch operation by comparing it with other standard measures of non-Markovianity. Our investigation thus paves the way forward to understanding the quantum switch as an emerging non-Markovian quantum memory.
}
\end{abstract}
\maketitle

\section{Introduction} 
\noindent

The superposition principle allows for multiple simultaneous evolutions, creating potential advantages in several quantum communication and quantum key distribution protocols by refining the effect of noise \cite{Gisin05}. Usually, even in quantum scenarios, the information carriers or channels are arranged in well-defined classical configurations. However, with an external control system called the quantum switch (QS) \cite{Chiribella13}, the causal order of multiple quantum evolutions or quantum channels can be put in superposition to create an \emph{indefinite} causal order. For an illustration, let us consider two quantum channels, $\Phi_1$ and $\Phi_2$, and a quantum state $\rho$. A control qubit determines the order of action of the two channels on $\rho$. When the control qubit is in the state $\ket{0}$, first $\Phi_1$ acts on $\rho$ followed by $\Phi_2$: $\left(\Phi_2 \circ \Phi_1\right)(\rho)$. The order is reversed when the control qubit is in the state $\ket{1}$, i.e., $\left(\Phi_1 \circ \Phi_2\right)(\rho)$. Hence, if the control qubit is prepared in the superposition state, $\ket{+}$ or $\ket{-}$ (where $\ket{\pm}\equiv (\ket{0}\pm \ket{1})/\sqrt{2}$), we get a superposition of the causal orders of the actions of the two channels. 

Indefiniteness in the order of quantum operations is beneficial over the standard quantum Shannon theory in several aspects. For example, it is better in testing the properties of quantum channels \cite{Chiribella12}, winning non-local games \cite{Oreshkov12}, achieving quantum computational advantages \cite{Araujo14}, minimizing quantum communication complexity \cite{Guerin16}, improving quantum communication \cite{Ebler18,Chiribella21,Banik21}, enhancing the precision of quantum metrology \cite{Zhao20}, or providing thermodynamic advantages \cite{Tamal20,Vedral20,Maity22,Liu23,Mukherjee24}, etc. (also see~\cite{Mukhopadhyay19,Ghosal22}). Recently, a second-quantized Shannon theory \cite{Kristjnsson2020} and a resource theory of causal connection \cite{Milz2022} have also been proposed. Several of these advantages have already been ascertained in experiments \cite{Procopio15,Rubino17,Goswami18(1)}. Because of the enormous application potential, in the literature, more attention has been paid to the applicability of the QS in quantum information-theoretic and communication protocols and its advantages than its dynamical aspects. Here, however, we mainly focus on the dynamical aspects of the QS and find a form of quantum memory that emerges from the QS dynamics. This is important since quantum memory is crucial for future developments of quantum technologies in long-distance quantum communications \cite{Chang19,Duan2001}, enhancing the capacity of long quantum channels \cite{Kretschmann05,DArrigo2007,Datta2009,task2}, or improving the efficiency of thermodynamic machines \cite{Bylicka2016,Taranto20,Taranto23}, etc. We also establish that the QS-induced memory (QSM) is equivalent to the non-Markovianity emerging from the resultant dynamics.

Since the past decade, quantum Non-Markovian processes are interpreted as a form of quantum memory in dynamical processes \cite{breuer1,blp1,rivas,RHPreview,vasile,lu,luo,fanchini,titas,haseli,sam2,sam4,sam5,sam6,sam7,sam8,sam9,Giarmatzi2021witnessingquantum,mallick2023}. In the theory of open quantum systems \cite{alicki}, the system is generally assumed to couple weakly to a static environment without any memory of the past, leading to memoryless Markovian processes. This gives rise to a one-way information flow from the system to the environment. However, in realistic scenarios, such an ideal assumption does not hold, and, almost always, there is some non-Markovian backflow of information from the environment to the system. The backflow behaves like a memory providing advantages in information processing, communication and computational tasks over the memoryless or Markovian operations \cite{task1,task2,task3,task4,task5}. There could be multiple reasons behind the occurrence of non-Markovianity, e.g., the strength of the system-environment interaction, the nature of the bath state, etc. Therefore, the system-environment interaction is one of the key resources for triggering such non-Markovian dynamics \cite{breuer,RHPreview,Vegareview,BLPreview}. Naturally, the QS also allows us to create and regulate the effect of non-Markovianity by manipulating the interaction between the environment and the system degrees of freedom via the control qubit. Hence, considering the control qubit to be a part of the environment, we can interpret that the emergent memory is stemming from a carefully controlled system-environment coupling. The ability to control memory effects by manipulating the system-environment interaction is beneficial in several information-processing tasks. For example, it creates potential advantages over classical technology \cite{Chiribella08(1),Chiribella09,White2020}, enhances the efficiencies of thermodynamic machines \cite{Bylicka2016,Taranto20,Taranto23}, preserves coherence and quantum correlation \cite{Barreiro2010}, allows the implementation of randomized benchmarking and error correction \cite{Ball16,Figueroa-Romero21,FigueroaRomero2022} or performing optimal dynamical decoupling \cite{Addis2015,Biercuk2009}, etc.

In this paper, we investigate the dynamics of QS and characterize the non-Markovian memory emerging from it. We look at the \emph{information loss} \cite{SamyaErg} in an \emph{ergodic} Markovian-quantum evolution under the QS. The notion of ergodic quantum operations stems from the well-known \emph{ergodic hypothesis}, according to which a system evolving under the influence of a static environment over a sufficiently long time will always evolve to a steady state, whose properties depend on the environment and the nature of the interaction. For example, in the case of thermal environments, the system will deterministically evolve to a thermal state with the same temperature as the environment. In quantum scenarios, it implies that ergodic quantum operations have a singular fixed point or, in other words, a particular steady state.
For our purpose, we choose a generic Markovian operation satisfying the ergodicity property for the evolution operation on which a QS is applied.

The information loss at a time $t$ can be measured by the difference of the geometric distances between two states at $t$ and initially. Even though the difference never decreases with time for memoryless dynamics~\cite{SamyaErg}, it does so under a QS. This indicates a non-trivial relation between the information loss for a given dynamics under a QS and the QSM. We show that the sum of the information loss and the QSM is actually lower-bounded by the distance between the initial states. We consider a special case of the generalized Pauli channel (depolarizing channel) and investigate the dynamical evolution. We find that under the action of QS, it allows reverse information transmission from the environment to the system.
We prove that the information-carrying capability of the completely depolarizing channel under QS is due to the induced non-Markovianity from the switch operation alone, and not due to the convex combination of two separate evolutions. For completeness, we also quantify the amount of noise a QS can tolerate. We introduce noise at both the control qubit and the final measurement on the control bit and investigate how the system evolutions get affected by the noises.
 
Further, we analyse non-Markovian quantum operations from their divisibility perspective \cite{rivas}. Divisible quantum operations are those which can be perceived as a sequence of an arbitrary number of completely positive trace-preserving (CPTP) dynamical maps acting on a quantum system. Whereas, non-divisible operations elucidate the information backflow \cite{blp1,BLP2}, which can be perceived as a sufficient benchmark of the presence of non-Markovian memory in the concerning quantum operation.

The paper is organized as follows. In the next section, we briefly introduce the QS. In Section \ref{s3}, we evaluate the information loss of general quantum evolution with or without the QS, define the QSM measures and derive their relation. Further, we define the ergodic quantum channel formally and prove some relevant statements as a consolidated backdrop for our investigation. In section \ref{s4}, we investigate the dynamical evolution under QS. We then explore how the QS behaves in the presence of various possible noises and investigate its tolerance against them. In section \ref{s5}, we derive a Lindbald-type master equation for qubit depolarizing dynamics subjected to QS and investigate how non-Markovianity emerges out of QS dynamics from the perspective of a qubit evolution. We then find interesting connections between the QS with some well-known measures of non-Markovianity. Finally, in Section \ref{s6}, we conclude and remark about possible future directions.

\begin{figure}[!t]
\centering
\includegraphics[height=5cm,width=9cm]{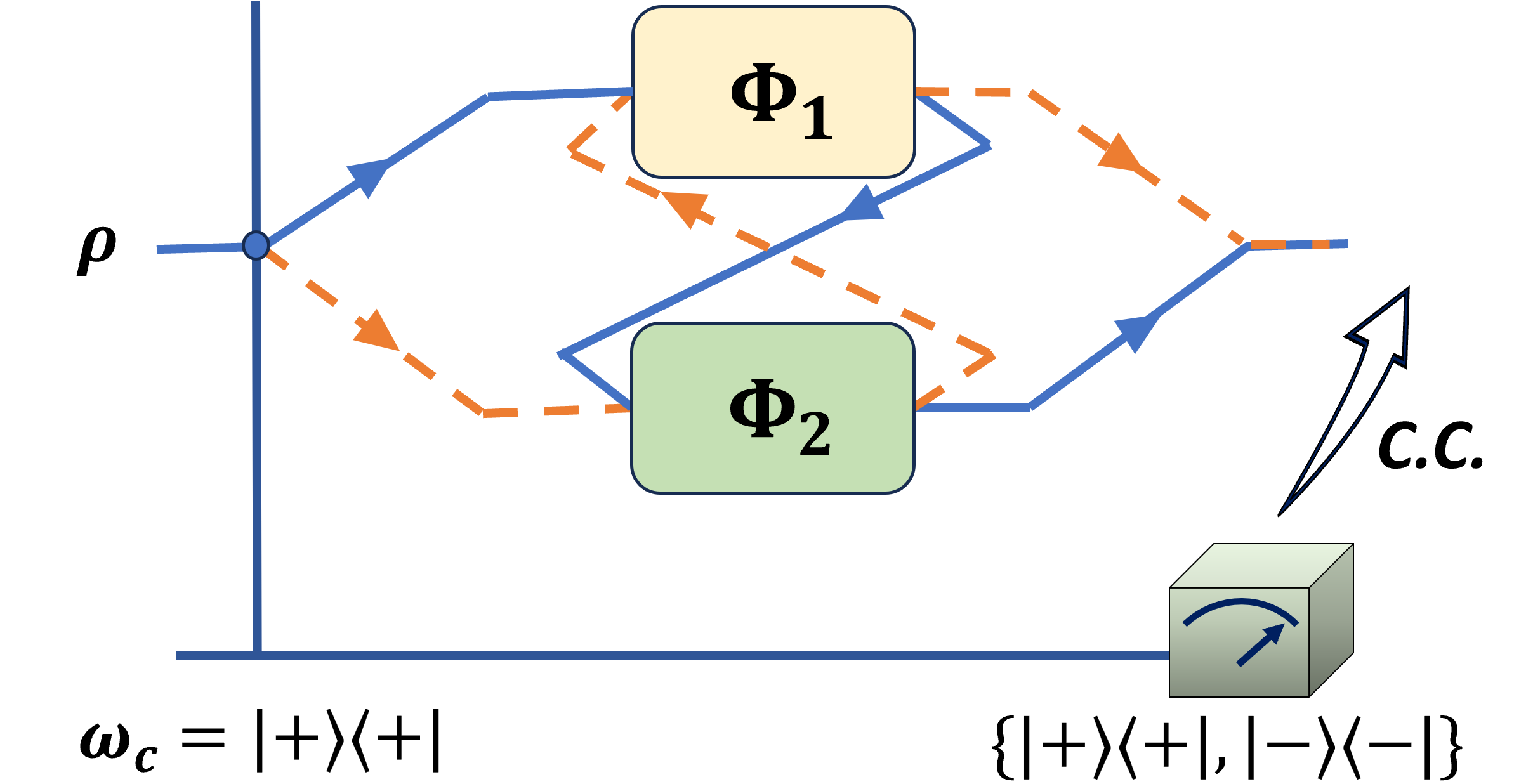}
\caption{The quantum switch. When the control qubit $\omega_c$ is prepared in $\ket{0}\bra{0}$ state, the state $\rho$ can evolve first through channel $\Phi_1$ and then through $\Phi_2$ (the blue solid line) and when the control qubit is prepared in $\ket{1}\bra{1}$ state, the state $\rho$ can evolve first through channel $\Phi_2$ and then through $\Phi_1$ (the orange dotted line). However, when the control qubit is prepared in $\ket{+}\bra{+}$ state, the order of the action of two channels is in superposition. Finally, the control qubit is measured in $\lbrace \ket{\pm}\bra{\pm} \rbrace$ basis.}\label{SWITCH_pic}
\end{figure}

\section{Quantum SWITCH}\label{s2}
A quantum channel $\Phi$ between systems $A$ and $B$ is a completely positive trace-preserving map from $L(\mathcal{H}_A)$ to $L(\mathcal{H}_B)$, where $\mathcal{H}_A$ and $\mathcal{H}_B$ are the Hilbert spaces corresponding to the input system $A$ and the output system $B$, respectively, and $L(\mathcal{H}_i)$ are the set of bounded linear operators on $\mathcal{H}_i$. The action of $\Phi$ on an input state $\rho \in L(\mathcal{H}_A)$ can be expressed in the Kraus representation (operator-sum representation) form as $\Phi (\rho) = \sum_i K_i \rho K_i^{\dagger}$, where $\lbrace K_i \rbrace$ are the Kraus operators for $\Phi$. If there are two channels, $\Phi_1$ and $\Phi_2$, they can act either in parallel or in series. The parallel action of the channels can be realized as $\Phi_1 \otimes \Phi_2$. The series actions can be realized in more than one way: $\Phi_1$ followed by $\Phi_2$ (denoted by $\Phi_2 \circ \Phi_1$) or $\Phi_2$ followed by $\Phi_1$ (denoted by $\Phi_1 \circ \Phi_2$). If the causal order is definite, then among these two possibilities, any one of either $\Phi_2 \circ \Phi_1$ or $\Phi_1 \circ \Phi_2$ is allowed. However, as mentioned in the Introduction, the order of the action of two channels can be made indefinite with an additional ancillary system called the control qubit ($\omega_c$) \cite{Chiribella13,Ebler18,Chiribella21,Banik21}. When $\omega_c$ is initialized in the $\ket{0}\bra{0}$ state, the $\Phi_2 \circ \Phi_1$ configuration acts on the state $\rho$, whereas the $\Phi_1 \circ \Phi_2$ configuration acts when $\omega_c$ is initialized in the $\ket{1}\bra{1}$ state. If ${K^{(1)}_i}$ represents the Kraus operator for $\Phi_1$ and ${K^{(2)}_j}$ for $\Phi_2$, the general Kraus operator can be represented as
\begin{equation}
    S_{i j}=K_{j}^{(2)} \circ K_{i}^{(1)} \otimes|0\rangle_c\left\langle_{\pc} 0\right|+K_{i}^{(1)} \circ K_{j}^{(2)} \otimes|1\rangle_c\left\langle_{\pc} 1\right|.
\end{equation}
The overall evolution of the combined system can be written as
\begin{equation}
S\left(\Phi_1, \Phi_2 \right)\left(\rho \otimes \omega_{c}\right)=\sum_{i, j} S_{i j}\left(\rho \otimes \omega_{c}\right) S_{i j}^{\dagger},
\end{equation}
where $S\left(\Phi_1, \Phi_2 \right) (.)$ is the quantum superchannel that maps the input quantum channels into a global quantum channel \cite{Chiribella2008,Gour2019,Quintino2019}.
In the end, the control qubit is measured on the coherent basis $\lbrace \ket{+}\bra{+}, \ket{-}\bra{-}\rbrace$. Then, for each outcome, the reduced state corresponding to the target qubit becomes,
  \begin{equation}
     {}_c\bra{\pm}S(\Phi_1,\Phi_2)(\rho \otimes \omega_c) \ket{\pm}_c.
 \end{equation}
This is essentially how the QS works, as we show schematically in Fig.~\ref{SWITCH_pic}.

We note that since the QS functions as a superchannel, one can also describe it in the process matrix formalism~\cite{Oreshkov12} (a general framework in quantum mechanics that does not require a fixed causal order between events) as a superposition of direct pure processes \cite{Araujo2017,Guerin18} and define it uniquely by its action on unitary operations \cite{Dong2023}. In this formalism, a causal inequality characterizes indefinite causal orders in a device-independent framework~\cite{Oreshkov12}. Even though the QS does not violate the causal inequality~\cite{Araujo2015,Branciard2016}, nevertheless, it is causally non-separable (i.e., it shows a weaker form of indefinite causal order). This is similar to the relationship between Bell inequality violations and quantum entanglement: just as some entangled states do not violate Bell inequalities, some causally non-separable processes, such as the QS, do not violate the causal inequality. The causal non-separability of the QS can be certified using a causal witness, which verifies whether a process is causally non-separable~\cite{Araujo2015,Branciard2016}. Unlike causal inequalities, however, a causal witness is not device-independent and requires the experimenter to have accurate knowledge of the quantum system's description. Recently, under specific conditions and assumptions, semi-~\cite{Bavaresco2019,Dourdent22} and fully device-independent~\cite{vanderLugt2023} approaches for certifying causal non-separability have been proposed, offering a new direction for studying causal indefiniteness in a more robust manner. For a detailed discussion of QS in terms of process, we refer the interested reader to Refs. \cite{Araujo2015,Araujo2017,Dong2023}. However, in this paper, we do not discuss the process matrix formalism further and restrict ourselves only to the Kraus operator description of the QS.

\section{Information Loss and the Measure of quantum SWITCH}\label{s3}
Before proceeding further, we need to introduce some basic concepts, like the notion of time-averaged quantum states and ergodic evolution, to understand the relationship between information loss and the QSM. The ergodic hypothesis of statistical mechanics states that if a physical system (be it classical or quantum) evolves over a long period, the time-averaged state of the system will be equal to its thermal state with a temperature equal to that of the bath, with which the system was interacting. In other words, for an observable $f$, if the (long-)time average $\langle f \rangle_{t}$ is equal to the ensemble average $\langle f \rangle_{\text{ens}}$ over the equilibrium state, the system is ergodic. If a quantum operation $\Phi_t(\cdot)$ acts on a state $\rho$ from $t=0$ to $t= T$, we can express the long-time-averaged state as 
\begin{equation}
\overline{\Phi_\infty(\rho)} = \lim_{T\to\infty}\frac{1}T\int_{0}^{T} \Phi_t(\rho)\; dt.
\end{equation}
Long-time averaging of any observable can thus be defined in the Schr\"{o}dinger picture as 
\begin{align}
    \langle f \rangle_{\infty} &= \lim_{T\to \infty}\frac{1}{T}\int_0^T\text{Tr}\left[f\Phi_t(\rho)\right] dt \nonumber\\
    &= \text{Tr}\left[ f \left(\lim_{T\to\infty}\frac{\int_0^T\Phi_t(\rho)dt}{T}\right)\right] \nonumber \\
    &=\text{Tr}[f\;\overline{\Phi_\infty(\rho)}].
\end{align}
We now define the ergodic evolution as follows~\citep{Burgarth_2013,SamyaErg}.\\

\noindent{$\blacksquare$~\bf Definition 1:}\label{def1} A quantum evolution is ergodic if it has a singular fixed point and the time-averaged state equals that fixed point.\\ 

A dynamics or quantum evolution is said to possess a singular fixed point if there exists exactly one state that remains unaffected by the quantum evolution---for example, a depolarizing channel with the maximally mixed state, a thermal channel with the thermal state, an amplitude damping channel with the ground state, etc. Thus, if $\tau$ is the singular fixed point of a given quantum channel $\Phi_t(\cdot)$, i.e., if $\Phi_t(\tau)=\tau$, the channel is called ergodic if $\overline{\Phi_\infty(\rho)}=\tau$ for an arbitrary initial state $\rho$. We denote such channels as $\Phi^\epsilon_t(\cdot)$. In Appendix~\ref{sec:appergo}, we present an example of a qubit ergodic Pauli channel, which is relevant to our study. Note that, by definition, ergodic quantum channels are the true {\it memoryless} channels since the time-averaged states do not depend on the initial state at all.

For some channels, the singular fixed point $\tau$ remains unaltered even under the action of a QS. If we denote the action of such channels as $\Phi^S_t(.)$, then $\Phi^S_t(\tau) = \tau$. Below, we prove that such a condition is true even for generalized Pauli channels in arbitrary dimensions.\\

\noindent{$\blacksquare$~\bf Statement 1:}\label{stmt1} For generalized Pauli channels in arbitrary dimensions, their singular fixed point (maximally mixed state) remains unchanged under the action of the quantum switch. 

\proof The generalized Pauli channel for a $d$-dimensional quantum system can be represented by the following map \citep{genPauli},
\begin{equation}\label{depol}
    \Lambda_{W} (\rho) = \sum_{k,l=0}^{d-1} p_{kl} W_{kl}\rho W_{kl}^{\dagger}
\end{equation}
where $W_{kl}=\sum_{m=0}^{d-1}\omega^{mk}\ket{m}\bra{m+l}$ with $\omega =e^{2\pi i/d}$ are the unitary Weyl operators. The Weyl operators satisfy the following well-known properties:
\begin{equation}
   W_{kl}W_{rs}= \omega^{ks} W_{k+r,l+s}~~\text{and}~~W_{kl}^{\dagger}=\omega^{kl}W_{-k,-l} 
\end{equation}
where the indices are modulo-$d$ integers. For such channels, the maximally mixed state $\frac{\mathbb{I}}{d}$ is the fixed point. We show that the maximally mixed state remains the fixed point after the action of the QS. The local effect of the action of the QS on the
system qubit undergoing a Pauli depolarising dynamics can be written as
\begin{align}
\Lambda_{W}^{S}(\rho) =& \frac{\sum_{k,l,r,s} \widetilde{W}_{klrs}\rho  \widetilde{W}_{klrs}^{\dagger}}{\text{Tr}\left[\rho \sum_{k,l,r,s}  \widetilde{W}_{klrs}^{\dagger} \widetilde{W}_{klrs}\right]},\\
{\text{ with}}\ \ \ \ \ \widetilde{W}_{klrs} =& \frac{\sqrt{p_{kl}p_{rs}}}2(W_{kl}W_{rs}+W_{rs}W_{kl})~\mbox{(see Appendix~\ref{sec:appkraus})}.\nonumber
\end{align}
Using these relations and the properties of dummy indices $k,l,r,s$, we get 
\begin{align}
\sum_{k,l,r,s} \widetilde{W}_{klrs}\frac{\mathbb{I}}{d}  \widetilde{W}_{klrs}^{\dagger} &= \frac{1}{d}\sum_{k,l,r,s} \widetilde{W}_{klrs}  \widetilde{W}_{klrs}^{\dagger}\nonumber\\
&=\frac{1}{2d}\sum_{k,l,r,s}p_{kl}p_{rs}(1+\omega^{ks-rl})\mathbb{I}.
\end{align}
Therefore, 
\begin{equation}
\Lambda_{W}^{S}\left(\frac{\mathbb{I}}{d}\right)=\frac{\frac{1}{d}\sum_{k,l,r,s}p_{kl}p_{rs}(1+\omega^{ks-rl})\mathbb{I}}{\sum_{k,l,r,s}p_{kl}p_{rs}(1+\omega^{ks-rl})}=\frac{\mathbb{I}}{d},    
\end{equation}
which proves the statement. \qed\\

\noindent
We use the ergodic generalized Pauli channel in our further analysis.

\subsection{Uncertainty relation between QS-induced memory and information loss}
Let $D(\rho_1, \rho_2)$ be the geometric distance between two states $\rho_1$ and $\rho_2$ satisfying the necessary properties of a distance measure, i.e., it is symmetric, non-negative and obeys the triangular inequality. In particular, if $D(\rho_1, \rho_2)$ is the trace distance measure then $D(\rho_1, \rho_2) \equiv \frac{1}{2}|| \rho_1 - \rho_2||_1$, where $||A||_1 = \text{Tr} [\sqrt{A^{\dagger}A}]$. Under a noisy Markovian evolution, $\Phi_t$, the trace distance between two states becomes a monotonically decreasing function \cite{blp1,BLP2}, implying that the information about the states is susceptible to the influence of noise. With this in mind, we define the following quantity.\\

\noindent{$\blacksquare$~\bf Definition 2:}\label{def2} If $\rho_1 (0)$ and $\rho_2 (0)$ are two initial states evolving under $\Phi_t(\cdot)$, the quantity $\Delta \mathcal{I}(\rho_1(t),\rho_2(t))$, defined as 
\begin{align}
\Delta \mathcal{I}(\rho_1(t),\rho_2(t)) \equiv&\  D(\rho_1(0), \rho_2(0))\nonumber\\ &\ - D(\Phi_t(\rho_1(0)), \Phi_t(\rho_2(0))),
\end{align} 
is a measure of the {\it information loss} across the channel.\\

\noindent
The quantity $ \Delta \mathcal{I}(\rho_1(t),\rho_2(t))$ quantifies the difference between the initial distinguishability of the two states and their distinguishability after the action of the noisy evolution $\Phi_t$. If we consider an ergodic operation  $\Phi_t^{\epsilon}(\cdot)$ with a singular fixed point $\tau$ and take $\rho_1(0) = \rho$ and $\rho_2(0) = \tau$ then the information loss becomes
\[
   \Delta \mathcal{I}_\epsilon(\rho(t)) = D(\rho, \tau) - D(\Phi_t^{\epsilon}(\rho), \tau),
\]
since $\tau$ remains unaffected by the evolution. To make the measure state independent, we can optimize over the initial state $\rho$ and define the measure as 
\begin{equation}
     \Delta \mathcal{I}_\epsilon^M(\rho(t)) = \max_{\rho}\left [ D(\rho, \tau) - D(\Phi_t^{\epsilon}(\rho), \tau)\right] .
\end{equation}
Now, if $\tau$ is also a fixed point under QS, then the information loss under the QS dynamics will be
\[
   \Delta \mathcal{I}_S(\rho(t)) = D(\rho, \tau) - D(\Phi_t^{S}(\rho), \tau), 
\]
and the state-independent maximized measure will be 
\begin{equation}
     \Delta \mathcal{I}_S^M(\rho(t)) = \max_{\rho}\left [ D(\rho, \tau) - D(\Phi_t^{S}(\rho), \tau)\right] .
\end{equation}

We can now define the QS-induced memory as follows.\\

\noindent{$\blacksquare$~\bf Definition 3:}\label{def3} For any bilinear CPTP supermap \cite{Oreshkov12,Araujo2017} acting on two copies of a noisy ergodic quantum operation $\Phi_t^\epsilon(\cdot)$ to produce a linear quantum process $\mathcal{W}\left(\Phi_t^\epsilon(\cdot),\Phi_t^\epsilon(\cdot)\right)$ keeping the singular fixed point of $\Phi_t^\epsilon(\cdot)$ unchanged, we define the {\it induced memory} $\mathcal{Q}_I$ as
\begin{equation*}
    \mathcal{Q}_I(t) \equiv D(\Phi^I_t(\rho), \Phi_t^{\epsilon}(\rho)),\label{eq:QSM}\\
\end{equation*}
where $D(\Phi^I_t(\rho)=\mathcal{W}\left(\Phi_t^\epsilon(\rho),\Phi_t^\epsilon(\rho)\right)$ is the induced linear process over $\rho$. Therefore, for the specific case of the QS, we can express the induced memory (QSM) as
\begin{equation*}
    \mathcal{Q}_S(t) \equiv D(\Phi^S_t(\rho), \Phi_t^{\epsilon}(\rho)).\label{eq:QSM}\\
\end{equation*}
and the corresponding state-independent optimized QSM as 
\begin{equation}
     \mathcal{Q}_S^M(t) \equiv \max_{\rho}\left [D(\Phi^S_t(\rho), \Phi_t^{\epsilon}(\rho))\right ].\label{eq:QSM1}\\ 
\end{equation}

\noindent
Clearly, when the evolution $\Phi(\cdot)$ is unaffected by the switch, $\Phi^S_t =\Phi_t^{\epsilon}$ and hence $ \mathcal{Q}_S$ will be zero. 

We can relate $\Delta \mathcal{I}_S(\rho(t))$ with $\Delta \mathcal{I}_\epsilon(\rho(t))$ as
\begin{align}
\Delta \mathcal{I}_S(\rho(t)) =&\ D(\rho, \tau) - D(\Phi^S_t(\rho), \tau)\nonumber\\
=&\ D(\rho, \tau) - D(\Phi^S_t(\rho) - \Phi_t^\epsilon(\rho), \tau - \Phi_t^\epsilon(\rho))\nonumber\\
\geq&\ D(\rho, \tau) - D(\Phi^S_t(\rho), \Phi_t^\epsilon(\rho)) - D(\Phi_t^\epsilon(\rho), \tau)\nonumber\\
\geq&\ D(\rho, \tau) - \mathcal{Q}_S(t) - D(\Phi_t^\epsilon(\rho), \tau), \nonumber   
\end{align}
which implies,
\begin{equation}\label{uncer1}
\Delta \mathcal{I}_S(\rho(t)) + \mathcal{Q}_S(t) \geq \Delta\mathcal{I}_\epsilon(\rho(t)).
\end{equation}
Here, we have used the triangle inequality: $D(A, B) + D(B, C) \geq D(C, A)$ and the symmetric property of the distance measure: $D(A, B) = D(B, A)$. The uncertainty relation between the information loss and the QSM in Eq.~\eqref{uncer1} is one of the main results of this paper. We will refer to it as the quantum switch-induced information inequality (QSI) later.

If we evaluate these measures for the time-averaged states under the QS, the information loss and the QSM measure reduce to 
\begin{align}
\Delta\overline{\mathcal{I}}_S=D(\rho, \tau)-D(\overline{\Phi^S_\infty(\rho)},\tau)  
\end{align}
and
\begin{align}
\overline{\mathcal{Q}}_S=D(\overline{\Phi^S_\infty(\rho)},\overline{\Phi_\infty^\epsilon(\rho)})=D(\overline{\Phi^S_\infty(\rho)},\tau),
\end{align}
respectively, and the QSI in Eq.~\eqref{uncer1} reduces to an equality:
\begin{equation}\label{uncer2}
    \Delta\overline{\mathcal{I}}_S+ \overline{\mathcal{Q}}_S = D(\rho, \tau) = \Delta\overline{\mathcal{I}}_\epsilon.
\end{equation}
The above equation indicates that as the QSM increases, the information loss decreases and hence, the QS is a resource for information storage undergoing noisy quantum operations. \\

\begin{figure}[!t]
\centering
\includegraphics[width=0.95\columnwidth]{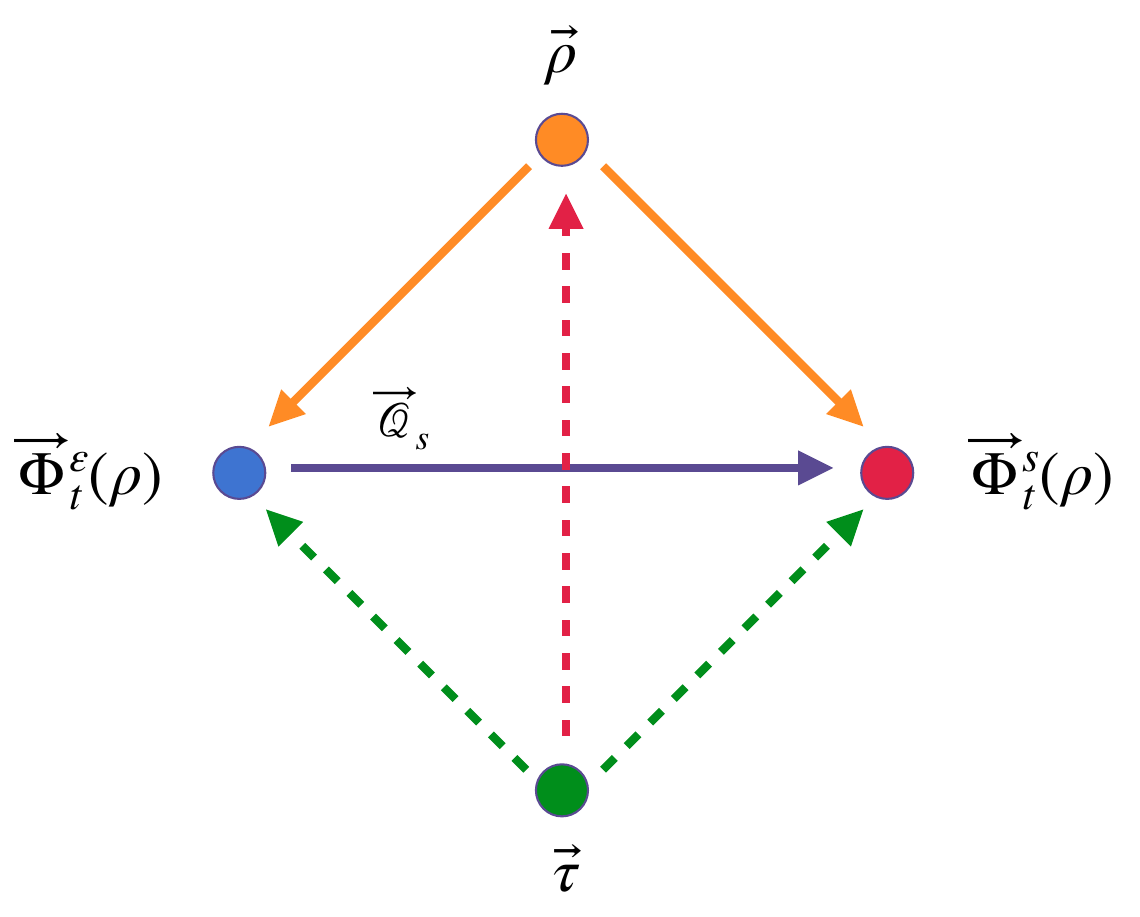}
\caption{In this figure, we consider the quantum states as elements of a vector space. The states $\overrightarrow{\rho}$, $\overrightarrow{\tau}$, $\overrightarrow{\Phi}^{\epsilon}_t(\rho)$, and $\overrightarrow{\Phi}^{S}_t(\rho)$ are described in the text.}
\label{fig:Vector}
\end{figure}

The relation in Eq.~\eqref{uncer1} can also be understood from the pictorial representation in Fig.~\ref{fig:Vector}. From the triangle law, we get
\begin{align}
    & \left[\overrightarrow{\tau}-\overrightarrow{\Phi}^{\epsilon}_t(\rho)\right] + \left[\overrightarrow{\Phi}^{S}_t(\rho)-\overrightarrow{\Phi}^{\epsilon}_t(\rho)\right] = \left[\overrightarrow{\tau}-\overrightarrow{\Phi}^{S}_t(\rho)\right]\nonumber\\
 \implies &\ D(\Phi^{\epsilon}_t(\rho)-\tau)\ +\  D(\Phi^{S}_t(\rho)-\Phi^{\epsilon}_t(\rho))\ \geq\ D(\Phi^{S}_t(\rho)-\tau)\nonumber\\
 \implies &\ D(\rho ,\tau)\ -\  D(\Phi^{S}_t(\rho) ,\tau)\ +\ D(\Phi^{S}_t(\rho),\Phi^{\epsilon}_t(\rho))\nonumber \\
 &\hspace{1.15cm}\ \geq D(\rho,\tau)\ -\  D(\Phi^{\epsilon}_t(\rho),\tau),
\end{align}
which implies Eq.~\eqref{uncer1}. Now, let $\rho^*$ be the initial state that maximizes the RHS, i.e., $\Delta\mathcal{I}_\epsilon^M(t)=D(\rho^*,\tau) -  D(\Phi^{\epsilon}_t(\rho^*),\tau)$. Since Eq.~\eqref{uncer1} holds for the evolution from $\rho^*$ as well, we have 
\begin{align}
  \left[\Delta\mathcal{I}_S(\rho(t))\right. &+ \left.\mathcal{Q}_S(t)\right]_{\rho(0)=\rho*} \nonumber\\
  &=\ D(\rho^* ,\tau) -  D(\Phi^{S}_t(\rho^*) ,\tau) +D(\Phi^{S}_t(\rho^*),\Phi^{\epsilon}_t(\rho^*)) \nonumber\\
  &\geq\ \Delta\mathcal{I}_\epsilon^M(t). 
\end{align}
Now, since $\Delta\mathcal{I}_S^M (t)\geq \Delta\mathcal{I}_S(t)$ and $\mathcal{Q}_S^M(t)\geq \mathcal{Q}_S(t)$, 
the uncertainty relation in Eq.~\eqref{uncer1} is naturally extended for the maximized state-independent measures:
 \begin{equation}
     \Delta\mathcal{I}_S^M (t) + \mathcal{Q}_S^M(t) \geq \Delta\mathcal{I}_\epsilon^M(t).  
 \end{equation}

The above uncertainty relation is not exclusive to the QS; these relations also hold for other linear control dynamics or external drives that induce memory into the system, keeping the fixed point unchanged. The uncertainty applies to any bilinear supermap~\cite{Oreshkov12,Araujo2017,Guerin18}, $\mathcal{W}\left(\Phi,\Phi\right)$, acting on two copies of a noisy operation to produce a linear quantum process keeping the singular fixed point of $\Phi(\cdot)$ unchanged.

\section{Dynamical evolution under the action of quantum SWITCH and emergence of non-Markovianity}\label{s4}
\noindent
In this section, we study an explicit example of dynamical evolution and investigate how the evolution changes under the action of the QS. We analyse the QSM (1) from the perspective of non-Markovianity by deriving the Lindblad-type generator for the switch operation and (2) from the point of view of non-ergodicity as proposed in the previous section. For this purpose, we consider qubit depolarizing operation as described below.

\subsection{Evolution under completely depolarizing dynamics in a definite causal structure}

We choose the Markovian qubit-depolarizing channel for the case study. We call a channel Markovian if it is completely positive (CP) divisible. A dynamical map $\Phi_{(t,t_0)}$ is said to be CP divisible if it can be represented as $\Phi_{(t, t_0)} = \Phi_{(t,t_s)} \circ \Phi_{(t_s,t_0)}$, $\forall t_s$ with $t_0 \leq t_s \leq t$ and $\Phi_{(t,t_s)}$ is completely positive. The notation $\Phi_{(t,t_s)}$ means the quantum channel $\Phi(\cdot)$ acts for the time period $t_s$ to $t$. When $t_0=0$, we abbreviate the notation as $\Phi_t$, as in the previous section. Note that there are various other criteria for identifying quantum non-Markovianity in the literature, of which information backflow \cite{blp1,BLP2} is relevant for our discussion. Information backflow is the reverse information flow from the environment to the system, mathematically quantified by the anomalous non-monotonic behaviours of known monotones (like the trace distance between two states) under dynamical evolution \cite{blp1}. It is well-known that breaking CP-divisibility is a necessary condition for information backflow, but not vice versa.

We consider a CP-divisible map $\Phi_t$ possessing a Lindblad-type generator $\mathcal{L}_t$, i.e.,  $\Phi_t\equiv \exp{\left(\int_0^t\mathcal{L}_sds\right)}$. The generator of the dynamics can be expressed as,
\begin{align}
\frac{d}{d t} \rho(t)=\mathcal{L}_{t}(\rho(t)),    
\end{align} 
where 
\begin{align*}\mathcal{L}_{t}(X)=\sum_{i} \Gamma_{i}(t)\left(A_{i} X A_{i}^{\dagger}-\frac{1}{2}\left\{A_{i}^{\dagger} A_{i}, X\right\}\right)
\end{align*} 
with $\Gamma_{i}(t)$ being the Lindblad coefficients and $A_{i}$, the Lindblad operators.  The necessary and sufficient condition for an operation being CP divisible is that all the Lindblad coefficients $\Gamma_{i}(t)$ are positive $\forall(i, t)$ \cite{lindblad,gorini}. For our purpose, the evolution of a system under definite causal order is considered to be Markovian.

Starting with a definite causal order, we consider the following master equation,
\begin{align}\label{depolGen}
\frac{d}{d t} \rho(t)=\sum_{i=1}^{3} \gamma_{i}(t)\left[\sigma_{i} \rho(t) \sigma_{i}-\rho(t)\right]
\end{align}
where $\gamma_{i}(t)$ are the Lindblad coefficients and $\sigma_{i}$ are the Pauli matrices. The qubit is represented by
\begin{align}
\Phi_t(\rho)\equiv\rho(t)=\left(\begin{array}{ll}
\rho_{11}(t) & \rho_{12}(t) \\
\rho_{21}(t) & \rho_{22}(t)
\end{array}\right).
\end{align}
For the completely depolarizing channel, the corresponding dynamics are represented by a CP trace-preserving dynamical map,
\begin{align}
\rho_{11}(t)=&\ \rho_{11}(0)\left(\frac{1+e^{-2 \xi_{1}(t)}}{2}\right)+\rho_{22}(0)\left(\frac{1-e^{-2 \xi_{1}(t)}}{2}\right),\nonumber\\
\rho_{22}(t)=&\ 1-\rho_{11}(t),\quad
\rho_{12/21}(t)=\ \rho_{12/21}(0) e^{-2\xi_{2}(t)},
\end{align}
with 
\begin{align*}
\xi_{1}(t)=&\ \int_{0}^{t} \left[\gamma_{1}(s) + \gamma_{2}(s)\right] d s\quad\text{and}\\
\xi_{2}(t)=&\ \int_{0}^{t}\left[\gamma_{2}(s) + \gamma_{3}(s)\right] d s.
\end{align*}
The corresponding Kraus operators are given by,
\begin{align}
& K_{1}=\sqrt{A_{2}(t)}\left(\begin{array}{ll}0 & 1 \\0 & 0\end{array}\right), \quad K_{2}=\sqrt{A_{2}(t)}\left(\begin{array}{ll}0 & 0 \\1 & 0\end{array}\right),\nonumber \\
& K_{3}=\sqrt{\frac{A_{1}(t)+A_{3}(t)}{2}}\left(\begin{array}{cc}e^{i \theta(t)} & 0 \\0 & 1\end{array}\right),\nonumber \\
& K_{4}=\sqrt{\frac{A_{1}(t)-A_{3}(t)}{2}}\left(\begin{array}{cc}-e^{i \theta(t)} & 0 \\0 & 1\end{array}\right),
\end{align}
where, 
\begin{align*}
A_{1}(t)=&\ \frac{1+e^{-2 \tilde{\xi}_{1}(t)}}{2},\quad A_{2}(t)=\ \frac{1-e^{-2 \tilde{\xi}_{1}(t)}}{2},\\
A_{3}(t)=&\ e^{-2 \tilde{\xi}_{2}(t)},\quad 
\theta(t)=\ \tan^{-1} \left(\frac{\operatorname{Im}\ A_{3}(t)}{\operatorname{Re}\ A_{3}(t)}\right)=0.
\end{align*}
Notice that the evolution is Markovian for the simple choice of the Lindblad coefficients, $\gamma_{1}(t)=\gamma_{2}(t)=\gamma_3(t)=\gamma >0$. Unless stated otherwise, we choose this set of Lindblad coefficients throughout the paper.

\subsection{Evolution under quantum switch}
\begin{figure}[!t]\label{FigUncer1}
\centering
\includegraphics[width=0.95\columnwidth]{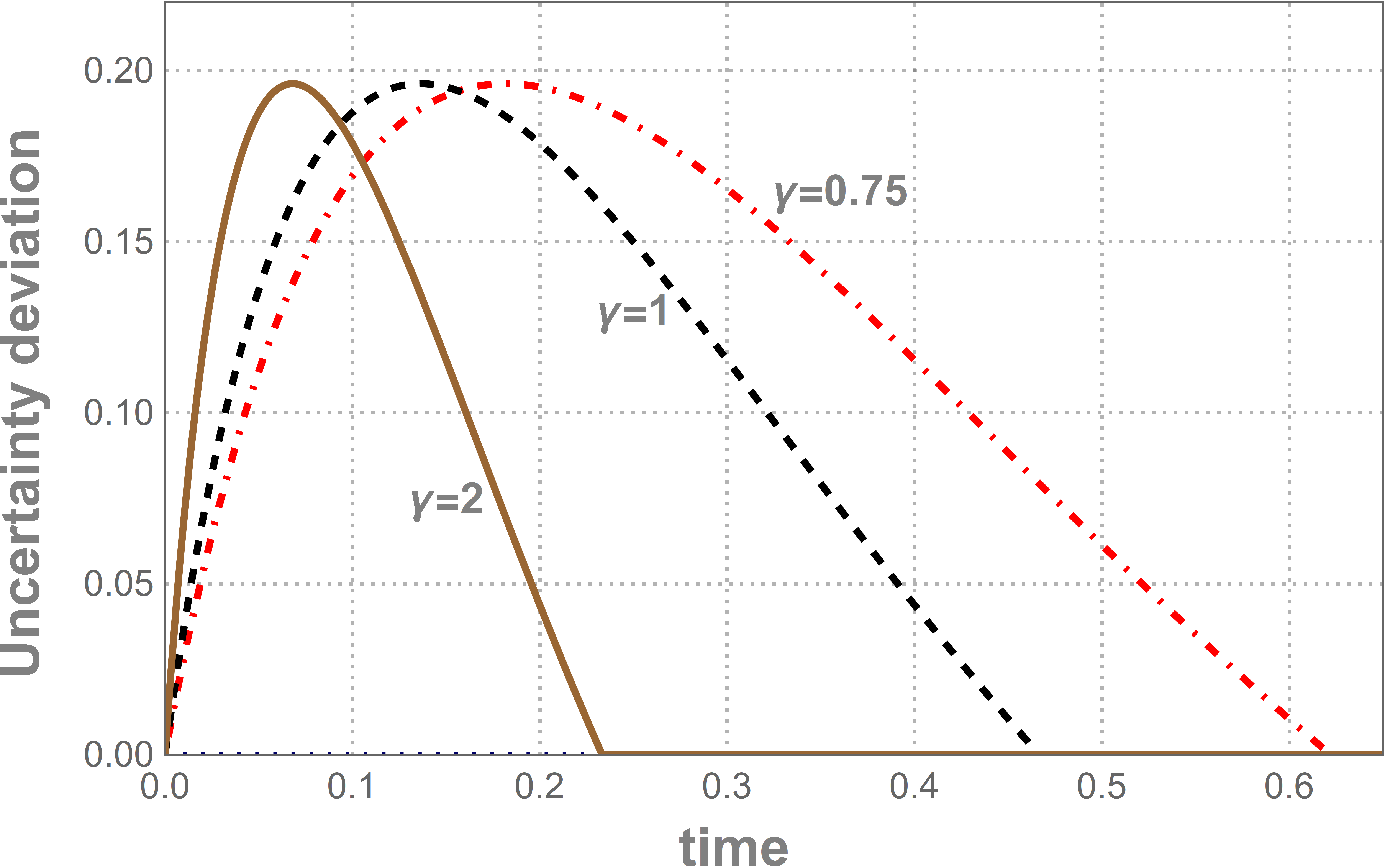}
\caption{In this plot we verify the QSI in Eq.~\eqref{uncer1} for the stated qubit dynamics. The uncertainty deviation i.e, the difference between the left-hand side and the right-hand side of Eq.~\eqref{uncer1} for $\rho=\ket{1}\bra{1}$ for different values of $\gamma$ are plotted. The difference [$=\Delta \mathcal{I}_S(\rho(t)) + \mathcal{Q}_S(t) - \Delta\mathcal{I}_\epsilon(\rho(t))$] increases initially but comes down to zero to settle there. As expected from the inequality relation, it is never negative.}
\label{fig:lhsmrhs}
\end{figure}

To see how the states evolve under the QS, we prepare two dynamical maps and put them in a superposition of causal orders with an additional control qubit, initially prepared in the $|+\rangle_{c}\left\langle_{\pc} +\right|$ state. We consider both channels to be the same depolarizing channel, i.e., $\Phi_{1} = \Phi_{2} = \mathcal{E}$. In general, the control qubit can be measured in any coherent basis but, for our purpose, we measure it in the $\left\{|\pm\rangle_{c}\left\langle_{\pc} \pm\right|\right\}$ basis. After measuring, the state $\rho$ corresponding to the `$+$' outcome can be expressed as
\begin{align}\label{switchmap1}
&\hspace{-0.2cm}\Phi^S_t(\rho)= \frac{1}{\mathcal{A}_1}\left\langle_{\pc} +\left|S\left(\rho \otimes|+\rangle_{c}\left\langle_{\pc} +\right|\right)\right|+\right\rangle_{c} \nonumber\\
&=\ \left(\begin{array}{cc}
A(t) \rho_{11}(0)+B(t) \rho_{22}(0) & C(t) \rho_{12}(0) \\
C(t) \rho_{21}(0) & B(t) \rho_{11}(0)+A(t) \rho_{22}(0)
\end{array}\right),
\end{align}
such that $A(t)=\left(1+C(t)\right)/2$, $B(t)=\left(1-C(t)\right)/2$, and  
\begin{align}
C(t)=&\ \frac{ \mathcal{G}^2 - 2 \mathcal{G} +9}{ 5 \mathcal{G}^2 + 6 \mathcal{G}-3},\nonumber
\end{align}
with $\mathcal{G}=e^{4\gamma t}$. Here,  $\mathcal{A}_1$ is the normalization factor and the suffix `$c$' represents the control qubit. In Fig.~\ref{fig:lhsmrhs}, we demonstrate the validity of the QSI in Eq.~\eqref{uncer1} for different values of $\gamma$. It proves the concerned QS dynamics obeys QSI perfectly. Moreover, the figure shows that initially the expression is driven away from the equality, but as time passes and the dynamics approaches the steady state, equality is reached. This, in turn, validates the equality of Eq.~\eqref{uncer2} for the long-time-averaged states. Below, we investigate the validity of this QSI for a more general situation of QS with arbitrary qubit control and measurement.
\subsection{QSI for noisy quantum switches}
We consider two different ways noise can be introduced to the QS. In the first case, the noise is quantum while in the second case, the noise is classical. By quantum noise, we mean that for the control qubit, an arbitrary pure state is used instead of the $\ket{+}$ state and for the measurement, another arbitrary pure state is used in the place of $\ket{+}\bra{+}$. For the case of the classical noise, we consider a mixed state in the $\ket{\pm}$ basis as the control and an arbitrary positive operator-valued measure (POVM) on the same basis. Note that here our goal is to evaluate the noise tolerance of the quantum switch itself. Hence, we introduce the noise directly to the switching process (which essentially involves preparing the control qubit and conducting the final measurement on the control qubit), irrespective of the nature of the channels upon which it operates.

However, before that, it is necessary to understand whether the QSI is valid for these general situations. For that, we need to verify \textbf{Statement 1} for such a generalized QS evolution. In Appendix~\ref{sec:appendixLD}, we prove it for such a general consideration with an arbitrary control qubit and final measurement. We show that for the generalized Pauli channel, the fixed point remains unchanged under the QS operation. This is the only prerequisite for the validity of Eqs.~\eqref{uncer1} and \eqref{uncer2} and hence it asserts that QSI is perfectly legible for the following considerations.

\subsubsection{Quantum noise}\label{sec:subsecVA}

The control qubit is prepared in the state, $\omega_c = \sqrt{p}|0\rangle + \sqrt{(1-p)}|1\rangle$ and the measurement on the control system is performed in the $\lbrace \ket{\mathcal{M}_q}\bra{\mathcal{M}_q}, \mathbb{I} - \ket{\mathcal{M}_q}\bra{\mathcal{M}_q}\rbrace$ basis with $\ket{\mathcal{M}_q} = \sqrt{q}|0\rangle + \sqrt{(1-q)}|1\rangle$, with $p,~q$ being arbitrary probabilities. After the measurement, the state of the generic target system corresponding to the $'+'$ outcome becomes
\begin{align}\label{switchmap2}
&\Phi^S_t(\rho)=\frac{1}{\mathcal{A}_2}{ }_{c}\left\langle \mathcal{M}_q\left|S(\rho \otimes\omega_c) \right| \mathcal{M}_q\right\rangle_{c} \nonumber \\
&=\left(\begin{array}{cc}
A_{pq}(t) \rho_{11}(0)+B_{pq}(t) \rho_{22}(0) & C_{pq}(t) \rho_{12}(0) \\
C_{pq}(t) \rho_{21} & \hspace{-0.6cm}B_{pq}(t) \rho_{11}(0)+A_{pq}(t) \rho_{22}(0)
\end{array}\right)
\end{align}
with $A_{pq}(t) = (1+C_{pq}(t))/2$, $B_{pq}(t) = (1-C_{pq}(t))/2$, and 
\begin{align*}
C_{pq}(t) = \frac{f_p f_q ( \mathcal G^2 -2\mathcal G+5)+p (4 q-2)+2(1-q)}{f_p f_q ( \mathcal G^2 +6\mathcal G-3)+p (4 q-2)\mathcal G^2}    
\end{align*}
where $f_x = \sqrt{x(1-x)}$ and $\mathcal{A}_2$ is the normalization factor. 

\begin{figure}[!t]
\centering
\includegraphics[width=0.95\columnwidth]{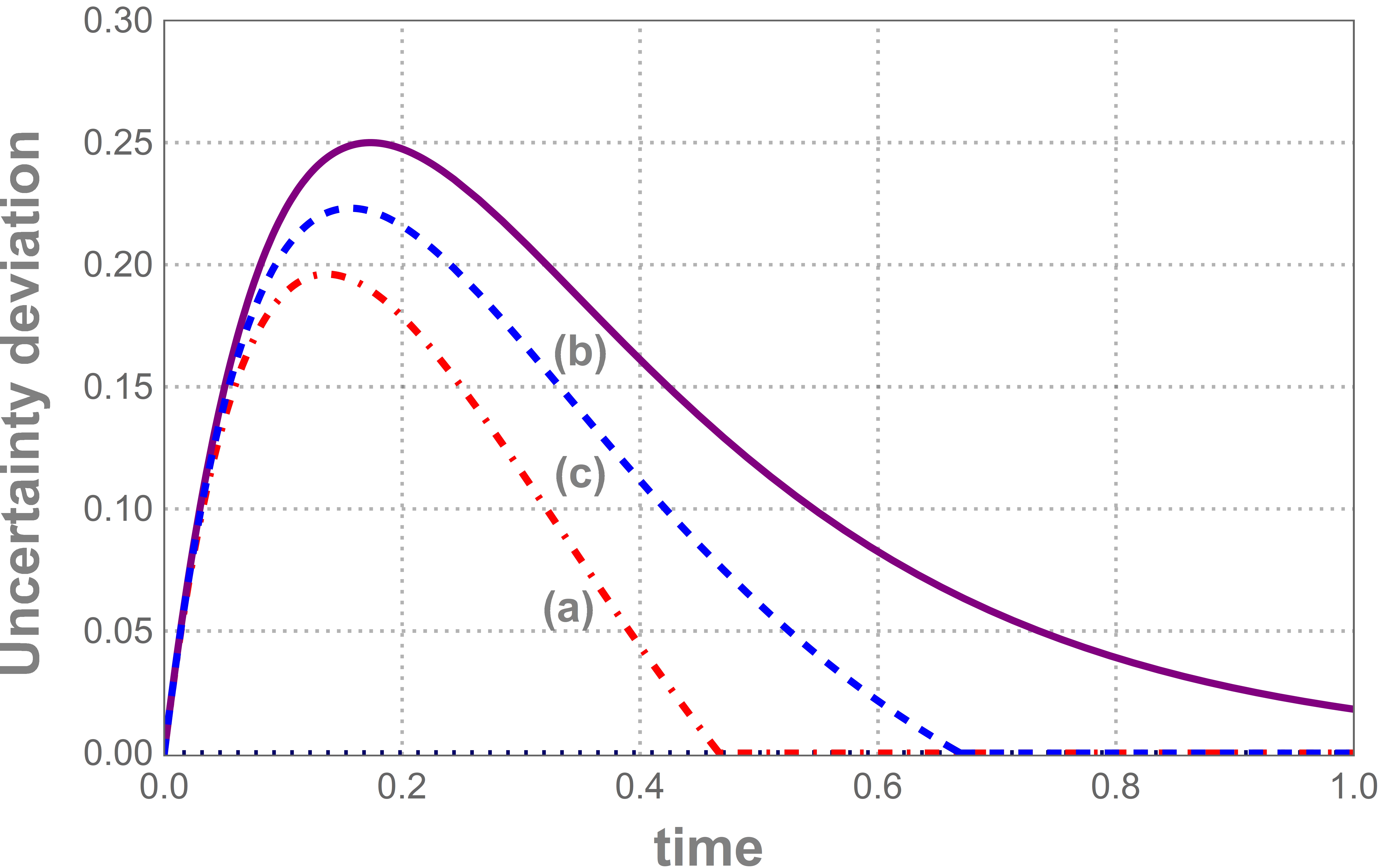}
\caption{Here we depict the QSI for a noisy QS described in Section~\ref{sec:subsecVA}. The difference between the left and right-hand sides of Eq.~\eqref{uncer1} for $\rho=\ket{1}\bra{1}$ and $\gamma=1$. For $\{p,\ q\}\ =$ (a) $\{1/2,\ 1/2\}$, (b)  $\{2/5,\ 1\}$, and (c) $\{4/5,\ 9/10\}$}.
\label{figUncerN}
\end{figure}

Fig.~\ref{figUncerN} confirms that the QSI is valid in this case, however, it takes more time to reach the equality with the varying noise parameter.

\subsubsection{Classical noise}\label{sec:subsecVB}
We now consider the case where the control qubit is prepared in the Fourier basis, i.e., $\omega_c = p|+\rangle_{c}\langle+| + (1-p)|-\rangle_{c}\langle-|$, and the measurement on the control qubit is performed in the POVM set, $\lbrace \mathcal{M}_{q_1,q_2}, \mathbb{I} - \mathcal{M}_{q_1,q_2}\rbrace$, where $\mathcal{M}_{q_1,q_2} = q_1|+\rangle_{c}\langle+| + q_2|-\rangle_{c}\langle-|$. The final state of the target system after the measurement is performed on the above basis would be given by:
\begin{align}\label{switchmap3}
&\Phi^S_t(\rho)=\frac{1}{\mathcal{A}_3} {\rm Tr}_c \left(\mathcal{M}_{q_1,q_2} S(\rho \otimes\omega_c)\right) =\nonumber\\
&\small \left(\begin{array}{cc}
A_{pq_1q_2}(t) \rho_{11}(0)+B_{pq_1q_2}(t) \rho_{22}(0) & C_{pq_1q_2}(t) \rho_{12}(0) \\
C_{pq_1q_2}(t) \rho_{21} & \hspace{-0.65cm}B_{pq_1q_2}(t) \rho_{11}(0)+A_{pq_1q_2}(t) \rho_{22}(0)
\end{array}\right)
\end{align}
with $A_{p q_1q_2}(t)=\left(1+ C_{p q_1q_2}(t)\right)/2$, $B_{p q_1q_2}(t)=\left(1- C_{p q_1q_2}(t)\right)/2$, and
\begin{widetext}
 \begin{equation*}
C_{p q_1q_2}(t) =
       \frac{(-1+10p+\mathcal{G}^2(-1+2p)+\mathcal{G}(2-4p))q_1 - (-9+10p+\mathcal{G}^2(-1+2p)+\mathcal{G}(2-4p))q_2}{(3-6p+\mathcal{G}^2(3+2p)+6\mathcal{G}(-1+2p))q_1 - (3-6p+\mathcal{G}^2(-5+2p)+\mathcal{G}(-1+2p))q_2},
\end{equation*}   
\end{widetext}
where $\mathcal{A}_3$ is the normalization factor. 

\begin{figure}[!b]\label{fifUncerN2}
\centering
\includegraphics[width=0.95\columnwidth]{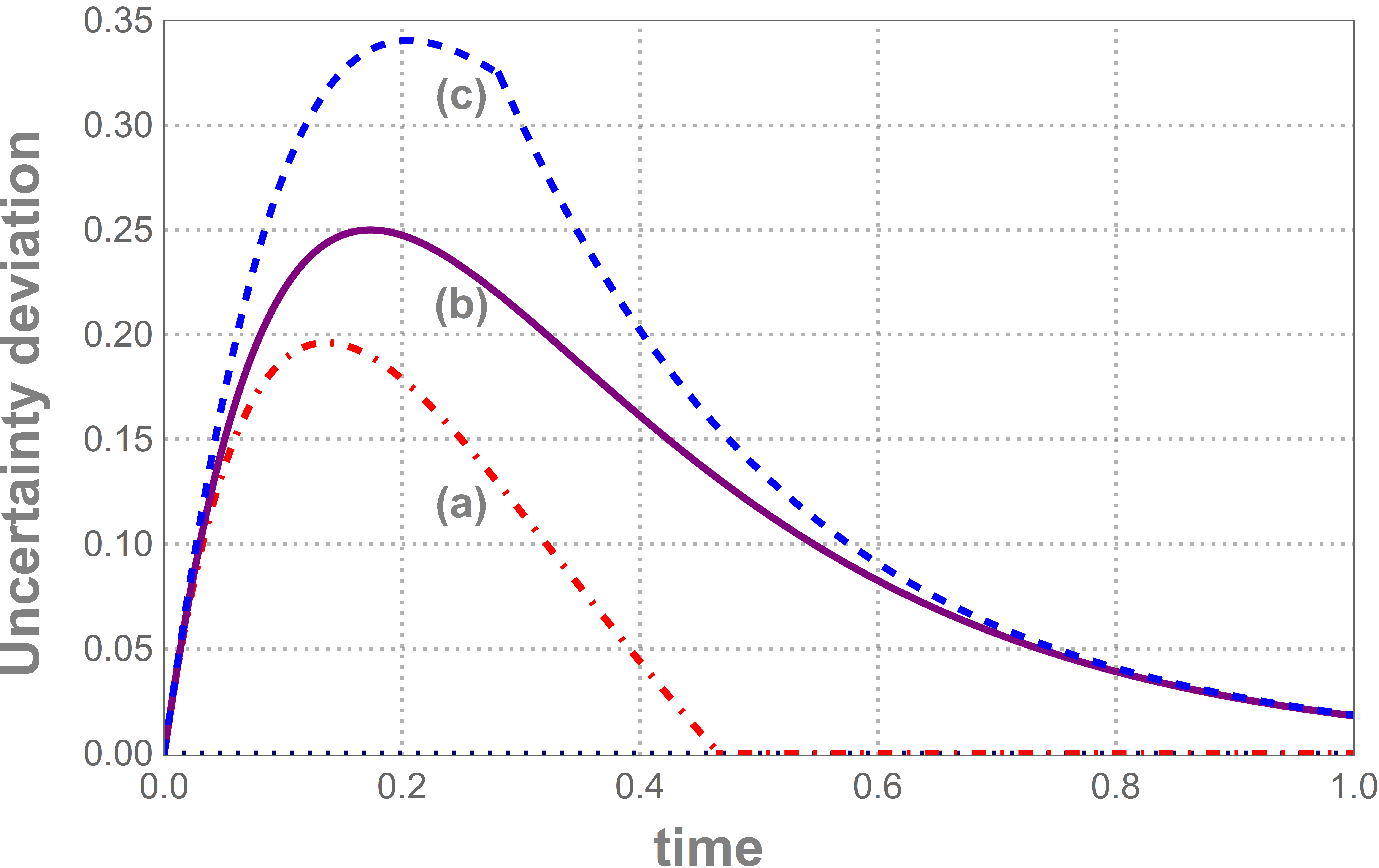}
\caption{Here we depict the uncertainty deviation for the ``classical noise case" for $\rho=\ket{1}\bra{1}$, with $\{p,\ q_1,q_2\}\ =$ (a) $\{1,\ 1,\ 0\}$, (b)  $\{1/2,\ 1/2,\ 1/2\}$, and (c) $\{4/5,\ 1/10,\ 9/10\}$}.
\label{fig:lhsmrhs2}
\end{figure}
Fig.~\ref{fig:lhsmrhs2} confirms the validity of QSI for this noisy QS, though, similar to the case of quantum noise, it takes more time to reach the equilibrium with the addition of noise.

\section{Lindblad dynamics of the switched channel}\label{s5}
We are now in a position to derive the canonical Lindblad-type master equation for the dynamical map generated under the action of the QS. It is important to note that after the switching action, when the control bit is finally measured, we consider the dynamics of only one outcome (`$+$') even though both outcomes (`$+$',`$-$') are possible. Therefore, it may seem that after normalizing, the operation may not be linear anymore, and hence deriving the Lindblad equation is not possible. However, in our case, the trace of the final state density matrix [Eq.~\eqref{switchmap1}] is independent of the initial state (since $A(t)+B(t)=1$) and thus, the linearity is left intact. In particular, we prove that for any generalized Pauli channel represented by Eq.~\eqref{depol}, the final map after post-selection is independent of the initial state.\\

\noindent{$\blacksquare$~\textbf{Statement 2:}}\label{stmt2} For any identical pair of generalized Pauli channels, the final map obtained after the switch action and post-selection  presented in Eq.~\eqref{switchmap1} is linear.

\proof Using the properties of the Weyl operators, we can show that 
\begin{align}
\text{Tr}\left[\rho\sum_{k,l,r,s}\widetilde{W}_{klrs}^\dagger\widetilde{W}_{klrs}\right]
=&\ \text{Tr}\left[\rho\sum_{k,l,r,s}\frac{p_{kl}p_{rs}}{2}(1+\omega^{ks - rl})\mathbb{I}\right]\nonumber\\
=&\ \sum_{k,l,r,s}\frac{p_{kl}p_{rs}}{2}(1+\omega^{ks - rl}).
\end{align}
This shows that the trace is independent of the input density matrix and hence the linearity is preserved. \qed
\\

In Appendix~\ref{sec:appendixLB}, we briefly sketch the method to obtain the Lindblad operator for a dynamical map. Applying it on Eq.~\eqref{switchmap1}, we get the corresponding Lindblad-type master equation of the form
\begin{equation}\label{lindswitch}
    \frac{d}{dt}\rho(t)= \Gamma_S(t)\sum_{i=1}^3 \left(\sigma_i.\rho(t).\sigma_i-\rho(t)\right),
\end{equation}
with 
\begin{align}\label{gamma}
\Gamma_S(t)=\ \ \ 8 \gamma&\ \times\ \frac{\cosh (4 \gamma  t)-2 \sinh (4 \gamma  t)+3}{5 \cosh (4 \gamma  t)-4 \sinh (4 \gamma  t)-1 }\nonumber\\
&\ \times\ \frac{1}{4 \sinh (4 \gamma  t)+\cosh (4 \gamma  t)+3}\nonumber\\
=\ 16 \gamma&\ \times\ \frac{(-\mathcal G)(\mathcal G^2-6\mathcal G-3)}{(\mathcal G^2-2\mathcal G +9)(5\mathcal G^2+6\mathcal G - 3)}
\end{align}
The above coefficient becomes negative with time as shown in Fig.~\ref{fig:sixgamma}. This implies the QS has converted the initial Markovian operation (i.e., the completely depolarizing channel) into a non-Markovian one.
\label{fig:sixgamma}

\subsection{Emergent non-Markovianity from the quantum switch}
We further explore various aspects of non-Markovianity \citep{breuer1,blp1,rivas,RHPreview,vasile,lu,luo,fanchini,titas,haseli,sam2,sam4,sam5,sam6,sam7,sam8,sam9} emerging from the QS dynamics. Quantum non-Markovianity can be measured in several ways. One such measure is based on the divisibility of a dynamical map which was first proposed by Rivas, Huelga and Plenio (RHP) \cite{rivas}. \\

\noindent{\bf RHP measure:} The dynamical map $\Phi_{\Delta t} \equiv \Phi (t+\Delta t, t)$ is CP if and only if $(\mathbb{I}\otimes\Phi_{\Delta t})\ket{\psi}\bra{\psi} \geq 0$ for all $\Delta t$ \cite{rivas}, where $\mathbb{I}$ stands for the identity map. Utilizing the trace-preserving condition, we can use the identity $||(\mathbb{I}\otimes\Phi_{\Delta t})\ket{\psi}\bra{\psi}||_1 = 1$ if and only if $\Phi_{\Delta t}$ is CP; otherwise $||(\mathbb{I}\otimes\Phi_{\Delta t})\ket{\psi}\bra{\psi}||_1 > 1$. Using this property, the  RHP measure is defined as
\begin{equation}
    g(t)=\lim_{\Delta t\rightarrow 0}\frac{||\mathbb{I}\otimes\Phi_{\Delta t}\ket{\psi}\bra{\psi}||_1-1}{\Delta t},
\end{equation}
where $\Phi_{\Delta t}=\mathbb{I}+\Delta t {\mathcal L}_t$. The integral $\int_0^{\infty} g(t) dt$ can be considered as a measure of non-Markovianity. For the switch operation, it is straightforward to calculate that $g(t)=6|\Gamma_S(t)|$. 
Therefore, from the perspective of the divisibility of dynamical maps, we can measure the 
QS-induced non-Markovianity with
\begin{align}
    N_{S} = \int_{T_-}^{\infty}g(t)dt=\int_{T_-}^{\infty}6|\Gamma_S(t)|dt,
\end{align} 
or the normalized measure,  
\begin{align}
\mathcal{N}_S=\frac{N_S}{1+N_S}.
\end{align} 

\begin{figure*}\label{nm1}
\centering
\captionsetup[subfigure]{labelformat=empty}
\subfloat[(a)]{\includegraphics[width=0.49\textwidth]{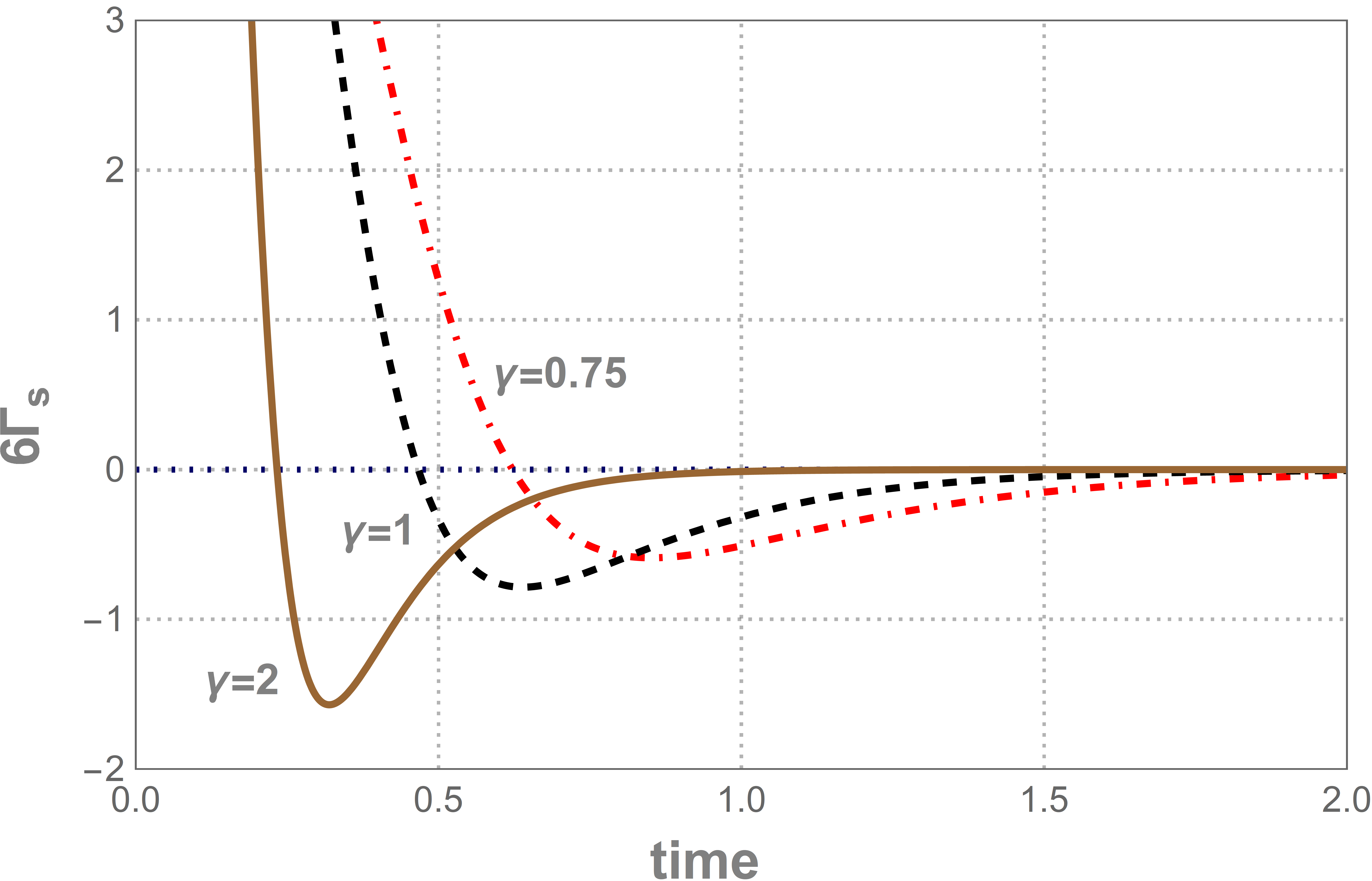}\label{fig:SixGammaA}}\hfill
\subfloat[\quad\quad\quad(b)]{\includegraphics[width=0.49\textwidth]{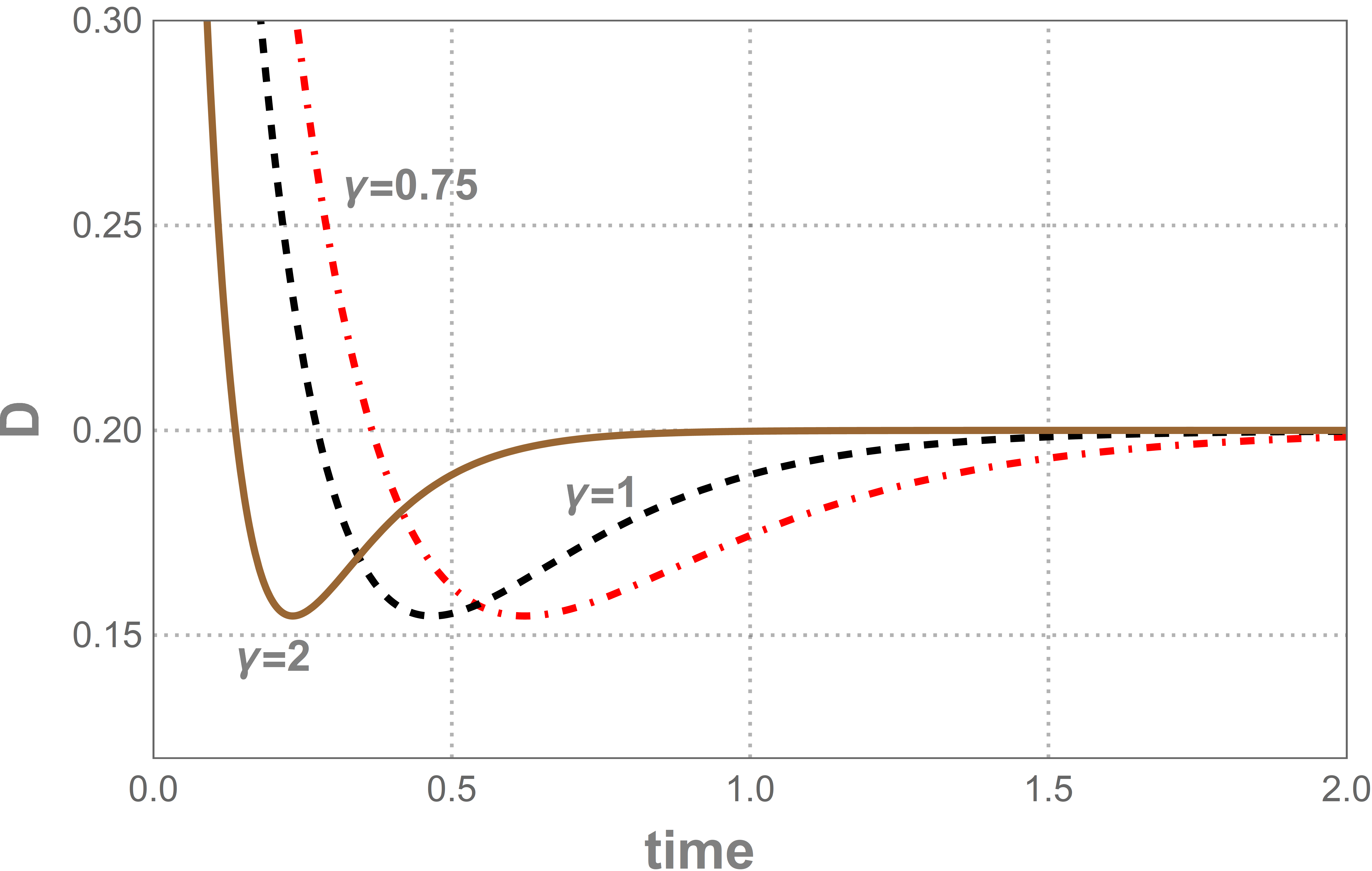}\label{fig:SixGammaB}}
\caption{Here, we represent the evolution of RHP and BLP measure for the given qubit dynamics, for three different values of $\gamma$ as mentioned in the plot. It depicts the emergent non-Markovianity due to the action of QS. It also shows that the information backflow (plot (b)) emerges at the same time when divisibility of the channel breaks down (plot (a)) }
\label{fig:sixgamma}
\end{figure*}

In Fig.~\ref{fig:SixGammaA}, the RHP measure of non-Markovianity is depicted for different values of $\gamma$. It clearly shows the time dependence of the emergent non-Markovianity due to the QS.\\

\noindent{\bf BLP measure:} Another well-known measure of non-Markovianity was proposed by Breuer, Laine, and Piilo (BLP) \cite{BLP2,blp1}. They characterized non-Markovian dynamics by the information backflow from the environment to the system. Usually, the distinguishability between two states decreases under Markovian dynamics as information moves from the system to the environment. Thus, an increase in the distinguishability of a pair of evolving states indicates information backflow under the dynamics. According to the BLP measure, a dynamics is non-Markovian if there exists a pair of states whose distinguishability increases for some time $t$. The time derivative of the distance between two states $\rho$ and $\tau$ evolving under the QS is given as
\begin{align}
\mathcal{B} \equiv \frac{d}{dt}D(\Phi^S_t(\rho),\tau).    
\end{align}
The above expression implies that the second state, $\tau$ is actually the fixed point of the dynamics. This choice reduces the complexity of calculation while retaining the physical meaning of the measure with its generality. The BLP measure of non-Markovianity is calculated by integrating $\mathcal{B}$ over the time where $\mathcal{B} > 0$ and then optimizing over all possible input states $\rho$,
\begin{align}
N_{\rm inf}=&\ \max_{\rho}\int_{T_-}^\infty\mathcal{B}dt\nonumber\\
=&\ \max_{\rho}\left[D(\Phi^S_\infty(\rho),\tau)-D(\Phi^S_{T_-}(\rho),\tau)\right].
\end{align}
For the qubit-depolarizing dynamics, this quantity can be readily derived as follows. Let us consider an arbitrary initial state $\rho=\frac{1}{2}\left(\mathbb{I}+\vec{n}\cdot\vec{\sigma}\right)$, where $\vec{n}=\{n_x,n_y,n_z\}$ with $n_x^2+n_y^2+n_z^2 \leq 1$ and the other symbols have there usual meanings. For the qubit depolarizing channel, the fixed point is $\tau=\mathbb{I}/2$. For the switch dynamics given in Eq.~\eqref{switchmap1}, the dynamic trace distance between  $\Phi_t^S(\rho)$ and $\mathbb{I}/2$ can be calculated as  
\begin{align}
    D(\Phi^S_\infty(\rho),\tau)- &D(\Phi^S_{T_-}(\rho),\tau)\nonumber\\
    =&\ \left\{C(\infty)-C(T_-)\right\}\sqrt{n_x^2+n_y^2+n_z^2},
\end{align}
where $C(t)$ is given in Eq.~\eqref{switchmap1}. Since the (state-dependent) quantity under the square root can be at most $1$, we get
\begin{align}
N_{\rm inf} = C(\infty)-C(T_-).
\end{align}
Similarly, we can calculate the switch-induced memory as 
\begin{align}
\mathcal{Q}_S(\infty)=D({\Phi}^S_\infty(\rho),\tau) = C(\infty).
\end{align}
This establishes the direct connection between the emergent non-Marvonianity and the QSM. Evidently, the switch-induced information backflow is just the QSM with a negative offset of $C(T_-)$, a constant. Moreover, we can consider the normalized BLP measure, 
\begin{equation}
\mathcal{N}_{\rm BLP}=\frac{N_{\rm inf}}{1+N_{\rm inf}},
\end{equation}
to get 
\begin{equation}
\mathcal{Q}_S(\infty)=\frac{\mathcal{N}_{\rm BLP}}{1-\mathcal{N}_{\rm BLP}}+C(T_-).
\end{equation}
We can also consider a normalized QSM measure:
\begin{equation}
\mathcal{N}_S^\infty=\frac{\mathcal{Q}_S(\infty)}{1+\mathcal{Q}_S(\infty)},
\end{equation}
which gives us the following relation:
\begin{equation}
\mathcal{N}_S^\infty=\frac{\mathcal{N}_{\rm BLP}+C(T_-)(1-\mathcal{N}_{\rm BLP})}{1+C(T_-)(1-\mathcal{N}_{\rm BLP})}.
\end{equation}
In highly non-Markovian cases where $\mathcal{N}_{\rm BLP}\rightarrow 1$, it gives $\mathcal{N}_S^\infty\rightarrow \mathcal{N}_{\rm BLP}$.

In Fig.~\ref{fig:SixGammaB}, the BLP measure of non-Markovianity is shown for different values of $\gamma$, which shows the evolution of information backflow of the emergent non-Markovian dynamics with time. From Fig.~\ref{fig:sixgamma}, it is evident that, at a certain characteristic time,  there is a clear transition from Markovian to non-Markovian behaviour for the concerned dynamics. Below, we analyse this particular issue.

\subsection{Characteristic time}
Let $T_-$ be the characteristic time--- the earliest time when information backflow starts. This is the time when the Lindblad coefficient in Fig.~\ref{fig:sixgamma} turns negative; below, we show this explicitly. Let us consider two arbitrary states,
\begin{equation}
    \chi_i (t) = \frac{1}{2}(\mathbb{I} + \vec{\sigma}\cdot\vec{n}_i(t))
\end{equation}
with $i=1,2$, and $\vec{n}_i(t) = \{n_i^1(t), n_i^2(t), n_i^3(t)\}$. The trace distance between these two states can be written as 
\begin{equation}
    D\left(\chi_1 (t),\chi_2 (t)\right) = \sqrt{(a_1(t))^2+ (a_2(t))^2+ (a_3(t))^2},
\end{equation}
where $a_k (t) = n_1^k(t) - n_2^k(t)$. Since, at $t=T_-$, the trace distance attains an extremum value, its time derivative vanishes, i.e., \begin{align}
    \frac{d}{dt}  D\left(\chi_1 (T_-),\chi_2(T_-)\right) = 0.
\end{align} 
After simplification, this condition reduces to 
 \begin{align}
    \lim_{\epsilon\to 0}\;\sum_{i=1}^{3}\left[ (a_i(T_-+\epsilon))^2 -(a_i(T_-))^2\right] = 0.
\end{align}
Expanding, we get $a_i(T_-+\epsilon)= a_i(T_-)(1 + \epsilon\Gamma_s(T_-))$.
This implies, $((a_1 (T_-))^2+ (a_2(T_-))^2 + (a_3(T_-))^2)\Gamma_s(T_-) = 0$. Clearly, the time derivative of the trace distance becomes zero and then positive, exactly when $\Gamma_s(t)$ becomes zero and then negative. Therefore, the characteristic time $T_-$ can be evaluated from the equation $\Gamma_s(T_-)=0$. We thus get 
 \begin{align}
&\cosh (4 \gamma  T_-)-2 \sinh (4 \gamma  T_-)+3=0\nonumber\\
{\rm or,}\quad& \mathcal G_-^2-6\mathcal G_--3=0,~~\mbox{with}~~\mathcal G_-=e^{4\gamma T_-}.
\end{align}
Solving this, we obtain the expression of characteristic time as,
\begin{equation}\label{charac}
    T_-=\frac{1}{4\gamma}\ln \left(2\sqrt{3}+3\right).
\end{equation}

In the literature of quantum non-Markovianity, the system-environment correlation is considered one of the primary reasons behind the generation of non-Markovian dynamics. In the case of the QS, the control qubit can be interpreted as a part of the environment and the QSM can be understood as the emergent non-Markovianity. Here it is important to remember that CP-divisible operations are not convex. Therefore for two such operations, $\Phi_1$ and $\Phi_2$, the operation $p\Phi_1\circ\Phi_2+(1-p)\Phi_2\circ\Phi_1$ may not be CP divisible. One may question the usefulness of the QS for the emergence of non-Markovianity, as discussed previously. Indeed, if one uses two different CP-divisible operations, the QS is not unique in generating non-Markovian dynamics that can be called a ``self-switching process''. However, if one uses the same operations as $\Phi_1=\Phi_2=\Phi$, neither convex combination nor any series or parallel combinations of those operations can produce non-Markovian dynamics, and hence, in the present scenario, it is induced solely due to the switching process. Note that the phenomenon of inducing memory is not exclusive to QS, and therefore, the formulation can also be extended to other CPTP bilinear supermaps.

\begin{figure*}
\centering
\captionsetup[subfigure]{labelformat=empty}
\subfloat[(a)]{\includegraphics[height=6.25cm]{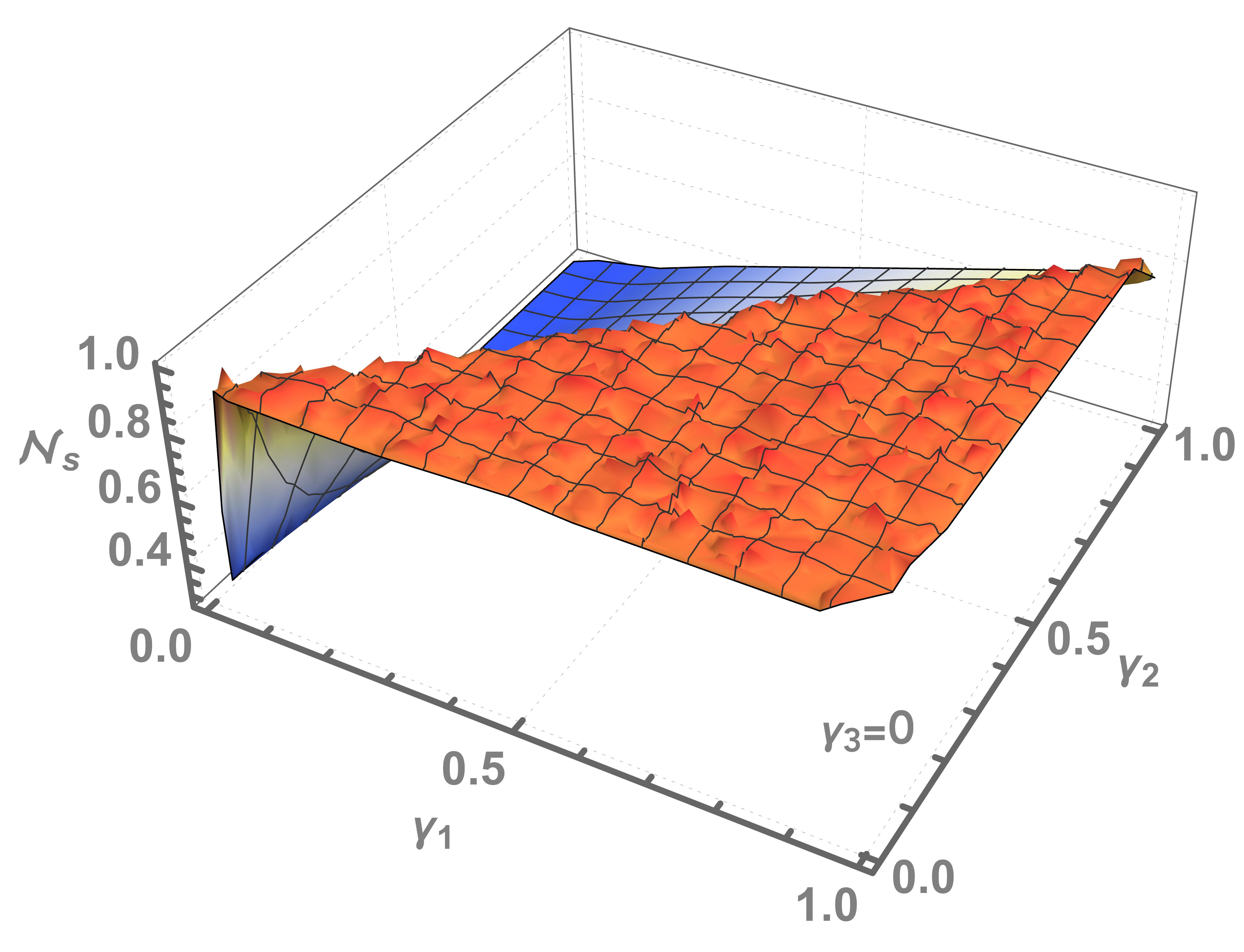}\label{fig:g3zero}}\hfill
\subfloat[]{\includegraphics[height=6.25cm]{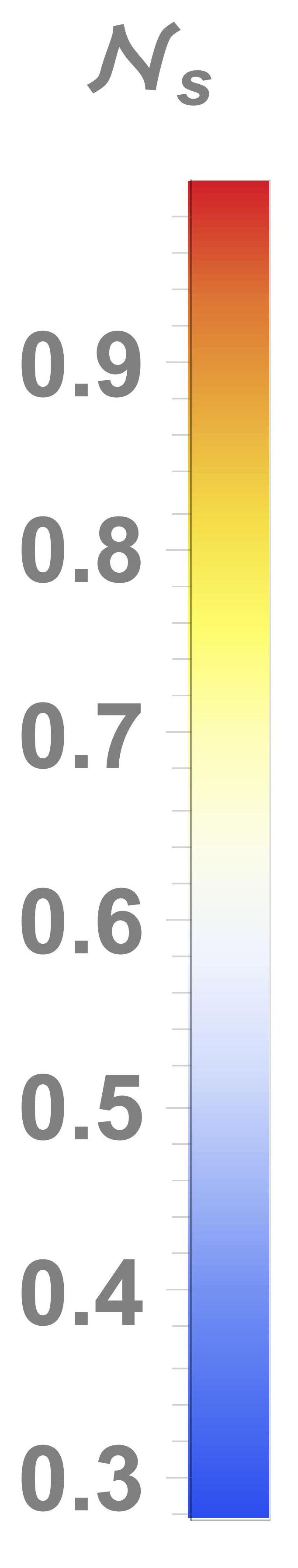}\label{fig:g3zero}}\hfill
\subfloat[\quad\quad\quad(b)]{\includegraphics[height=6.25cm]{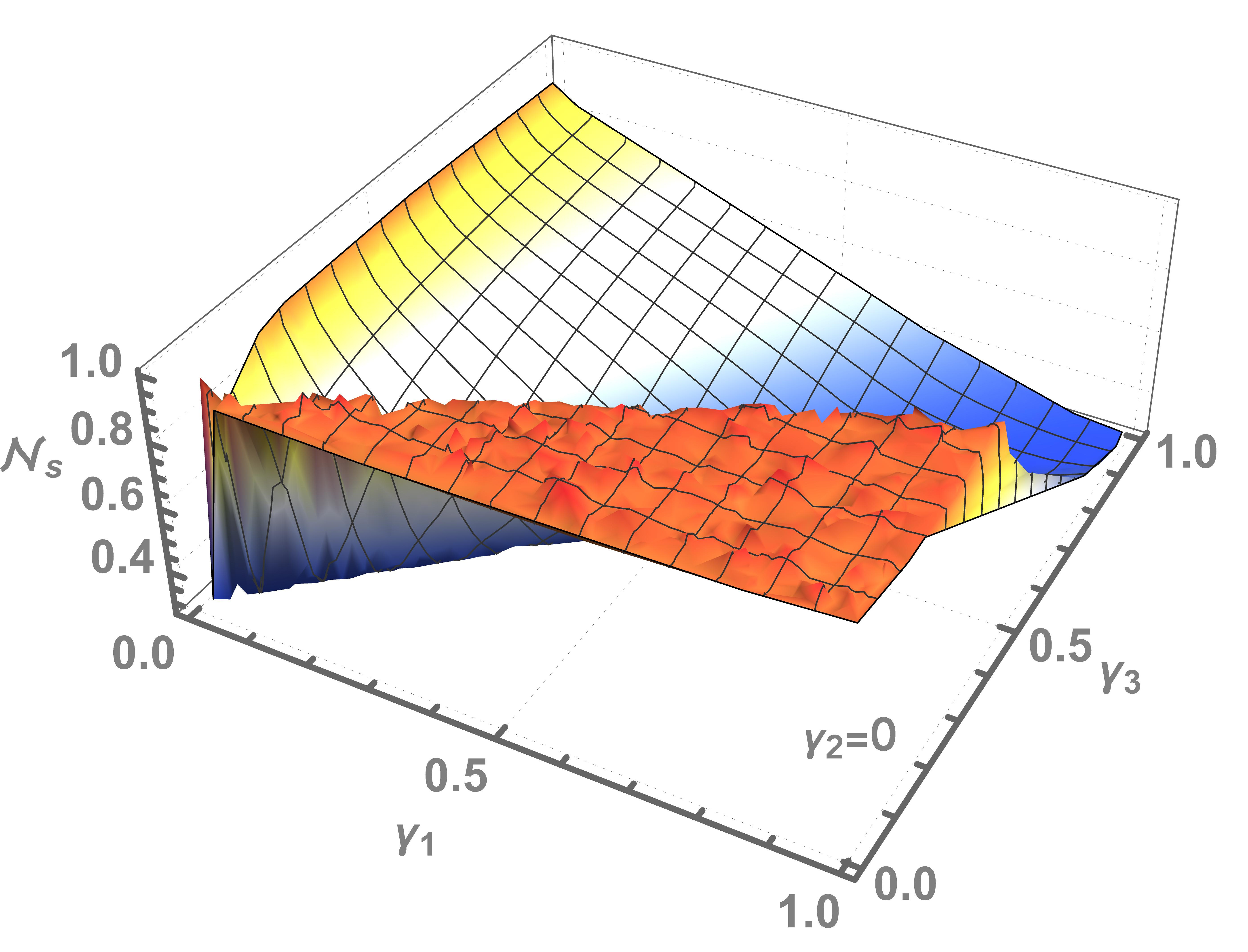}\label{fig:g2zero}}
\caption{In this plot, we show the emergent non-Markovianity in terms of $\mathcal{N}_S$ for QS action over a  general Pauli channel given in Eq.~\eqref{depolGen} with different Lindblad coefficients. In case (a) we consider $\gamma_3=0$ and vary $\gamma_1$ and $\gamma_2$ from 0 to 1. We see that for high values of them, the QS channel is highly non-Markovian. In case (b), we see similar features for $\gamma_2=0$. }
\label{fig:plots}
\end{figure*}

Finally, we study the emergent non-Markovianity due to the QS for a general qubit Pauli channel given in Eq.~\eqref{depolGen} with constant but different Lindblad coefficients, in terms of the normalized RHP measure $\mathcal{N}_S$. The result is shown in Fig.~\ref{fig:plots} for multiple Lindblad coefficients. The figure shows that by manipulating the coefficients, we can produce highly non-Markovian dynamics, starting from a purely Markovian depolarizing evolution.

\section{Discussion}\label{s6}
A series of recent papers have established the usefulness of the superposition of alternative causal orders in information-theoretic tasks~\cite{Chiribella12,Oreshkov12,Araujo14,Guerin16,Ebler18,Chiribella21,Banik21,Zhao20,Tamal20,Vedral20,Maity22,Liu23,Mukherjee24,Mukhopadhyay19,Ghosal22}, indicating the existence of a resource theoretic structure~\cite{Milz2022}. However, further research is needed to establish a proper framework for better characterization and quantification. For this purpose, we intend to consider this issue from the perspective of resource interconversion, considering the memory induced by the QS as emerging non-Markovianity, for which there is already a resource-theoretic framework~\cite{sam4}. With this motivation, in this paper, we explore the dynamical behaviour of the QS and characterize the non-Markovian memory that emerges from it. We first investigate how the loss of information of a general quantum evolution changes when it is subjected to the QS and, on this backdrop, we propose a quantification of the switch-induced memory, QSM. We then derive an uncertainty relation between the information loss and the QSM, which captures the interplay between information storage capacity and the QSM. Furthermore, we show that after a sufficiently long period of time, as the dynamics approaches the steady state, the uncertainty inequality reduces to an equality, implying a complementarity between the long-time averaged loss of information and the QSM. We then consider an example of completely depolarizing dynamics for qubit and explore its behaviour under the action of QS. We make both the control qubit and the final measurement on the control qubit noisy and look at the amount of noise that a quantum switch can tolerate. Further, we derive the reduced operation of the switch action on the qubit both in terms of the Kraus representation and Lindblad  evolution and verify the uncertainty relation in this particular case.
We then attribute this effect of the QS for activating a completely depolarizing channel to the non-Markovian memory that emerges from the switch-induced memory. Comparing it with other standard measures of non-Markovianity, we show that the long-term memory of a QS is equivalent to the emergent non-Markovianity.

Our work is particularly relevant from several backdrops of quantum information theory. While investigating the dynamical behaviour of a quantum switch, we find that the quantum switch actually carries some non-Markovian memory. The emergence of such non-Markovian memory induced by the quantum switch is quite important for further developments of quantum technology and near-term quantum devices. Furthermore, this investigation also allows us to quantify the amount of memory a quantum switch can possess. In other words, our study opens up a new avenue for quantifying the amount of noise a quantum switch can tolerate. To the best of our knowledge, this is the first study for adequate quantification of the quantum switch, a well-established quantum resource for future quantum technologies. However, it is important to note that similar effects can also be achieved by other quantum circuits with quantum control \cite{Wechs21}. For instance, Ref.~\cite{sup-switch1,Abbott20} showed earlier that the classical communication capacity of a channel could be enhanced both by the QS or the superposition of quantum channels. The authors of Ref.~\cite{sup-switch1} also showed that the superposition of channels always enhances communication capacity, unlike the quantum switch, which does not always achieve this. Similar questions were investigated in Ref.~\cite{sup-switch2} as well. These differences raise questions about the nature of the induced memories for both processes. However, further investigation is required to analyze such control processes within the framework of induced non-Markovian memory, as constructed in Definition 2, indicating a promising direction for future research.

\appendix

\section{Calculation of the time-averaged state of the dynamical map}\label{sec:appergo}
As mentioned earlier, for a completely depolarizing channel, the dynamical map is given by the following equations:
\begin{align*}
\rho_{11}(t)=&\ \rho_{11}(0)\left(\frac{1+e^{-2 \xi_{1}(t)}}{2}\right)+\rho_{22}(0)\left(\frac{1-e^{-2 \xi_{1}(t)}}{2}\right),\nonumber\\
\rho_{22}(t)=&\ 1-\rho_{11}(t),\nonumber\\
\rho_{12}(t)=&\ \rho_{12}(0) e^{-\xi_{2}(t)},
\end{align*}
with 
\begin{align*}
\xi_{1}(t)=&\ \int_{0}^{t} \left[\gamma_{1}(s) + \gamma_{2}(s)\right] d s\quad\text{and}\\
\xi_{2}(t)=&\ \int_{0}^{t}\left[\gamma_{2}(s) + \gamma_{3}(s)\right] d s.
\end{align*}
Now the time-averaged state for the above dynamics is given by
\begin{align*}
\overline{\rho} &= \lim_{T\to\infty}\frac{\int_{0}^{T} \Phi_t(\rho)\,dt}{\int_{0}^{T} dt}\\
&=\lim_{T\to\infty}\frac{\bigints_{0}^{T}\left(\begin{array}{ll}
\rho_{11}(t) & \rho_{12}(t) \\
\rho_{21}(t) & \rho_{22}(t)
\end{array}\right)\,dt}{\int_{0}^{T} dt}
&=\left(\begin{array}{ll}
\frac{1}{2} & 0 \\
0 & \frac{1}{2}
\end{array}\right)
&= \frac{\mathbb{I}}{2}.
\end{align*}
If $\tau$ is a fixed point of the dynamical map $\Phi$, then
\begin{align*}
\mathcal{L}_t(\tau) & = \sum_{i=0}^{3} \gamma_{i}\left[\sigma_{i} \tau \sigma_{i}-\tau \right] = 0.  
\end{align*}
Solving the above equations, we get 
\begin{align*}
\tau =\left(\begin{array}{ll}
\frac{1}{2} & 0 \\
0 & \frac{1}{2}
\end{array}\right)
= \frac{\mathbb{I}}{2}.
\end{align*}
This shows that the ergodicity condition is satisfied.

\section{Kraus representation of the quantum switch}\label{sec:appkraus}
The evolution of the combined state of the target system and the control qubit under the switch is given by 
$$
S\left(\mathcal{E}, \mathcal{E}\right)\left(\rho \otimes \omega_{c}\right)=\sum_{i, j} S_{i j}\left(\rho \otimes \omega_{c}\right) S_{i j}^{\dagger}
$$

The state of the target system after measuring the control qubit in the $\left\{|\pm\rangle_{c}\langle\pm|\right\}$ basis is given by ${ }_{c}\left\langle+\left|S\left(\mathcal{E}, \mathcal{E}\right)\left(\rho \otimes|+\rangle_{c}\langle+|\right)\right|+\right\rangle_{c}
$. This can also be written as
$$\sum_{i, j} M_{i j}\rho  M_{i j}^{\dagger}$$
with
$$M_{i j} = \frac{K_i K_j + K_j K_i}{2}$$
and $\sum_{i, j}  M_{i j}^{\dagger} M_{i j} \leq \mathbb{I}$. The equality holds if $[K_i, K_j] = 0$ or $[K_i, K_j^{\dagger}] = 0$ or both. Therefore, the switch action
$$\Phi^S_t(\rho) = \frac{\sum_{i, j} M_{i j}\rho  M_{i j}^{\dagger}}{\text{Tr}\left[\rho \sum_{i, j}  M_{i j}^{\dagger} M_{i j}\right]}$$

Now, this operation is CPTP, but the linearity is only confirmed if $\text{Tr}\left[\rho \sum_{i, j}  M_{i j}^{\dagger} M_{i j}\right]$ is independent of $\rho$. For our case it is independent of $\rho$, so linearity is confirmed.

\section{Proof of Statements 1 and 2 for the control qubit in the superposition state and measurement with a parameter}\label{sec:appendixLD}

Here we will prove \textbf{Statement 1} and \textbf{Statement 2} for a general switch operation which includes all the cases discussed in Sections~\ref{s3}, \ref{s4}, and \ref{s5}. A general quantum switch operation can be represented by 
\[ \Phi_t^S(\rho)= \frac{1}{\mathcal{P}}Tr_c\left[M_\alpha\left(\sum_{ij}W_{ij}\rho\otimes\omega_cW_{ij}^\dagger\right)\right],\]
with $\omega_c=\sum_\kappa q_\kappa\ket{\psi_\kappa}\bra{\psi_\kappa}$ being a general qubit control state, $M_\alpha =A_\alpha^\dagger A_\alpha$ being a general Positive operator valued measure, $\mathcal{P}$ being the normalization factor, $W_{ij}=K_iK_j\otimes\ket{0}\bra{0}+K_jK_i\otimes\ket{1}\bra{1}$ where $K_i$s are the Kraus operators for the original quantum operation and $Tr_c[~~]$ representing partial trace over the control basis. For this setting, the reduced Kraus representation can be expressed as 

\[\tilde{K}_{ij}=\chi_1 K_iK_j+\chi_2K_jK_i,\] 
with $\chi_1=\sum_{l,\kappa}\sqrt{q_\kappa}\bra{\psi_l}A_\alpha\ket{0}\langle 0|\psi_\kappa\rangle$ and $\chi_2=\sum_{l,\kappa}\sqrt{q_\kappa}\bra{\psi_l}A_\alpha\ket{1}\langle 1|\psi_\kappa\rangle$. Therefore the switch operation can be represented as 
\[\Phi_t^S(\rho) = \frac{1}{\mathcal{P}}\sum_{ij}\tilde{K}_{ij}\rho\tilde{K}^\dagger_{ij},\] with $\mathcal{P}=Tr\left[\sum_{ij}\tilde{K}_{ij}\rho\tilde{K}^\dagger_{ij}\right]$. Using this general representation for arbitrary dimensional depolarizing operations represented in Eq.~\eqref{depol} and following the same method, we can prove \textbf{Statement 1} and \textbf{Statement 2} for such general switch actions. It is to be noted that the dynamical maps constructed in Eqs.~\eqref{switchmap1}, \eqref{switchmap2}, and \eqref{switchmap3} are all special cases of the general form discussed in this appendix.

\section{Construction of the master equation}\label{sec:appendixLB}
Let us now consider a dynamical map of the form
\begin{equation}\label{dynamicalmap}
    \rho (t) = \Omega [\rho (0)].
\end{equation}
Further, consider the equation of motion corresponding to the previous dynamical equation to be 
\begin{equation}\label{eqom}
    \Dot{\rho} (t) = \Tilde{\Lambda} [\rho (t)]
\end{equation}
where $\Tilde{\Lambda} [.]$ is the generator of the dynamics. Now following Ref. \cite{Hall14,Bhattacharya17}, we can find the master equation and generator of the dynamics. 

Let $\lbrace \mathcal{G}_i\rbrace$ denotes the orthonormal basis set with the properties $\mathcal{G}_0 = \mathbb{I}/\sqrt{2}$, $\mathcal{G}_i^{\dagger} = \mathcal{G}_i$, $\mathcal{G}_i$ are traceless except $\mathcal{G}_0$ and $\text{Tr}[\mathcal{G}_i\mathcal{G}_j] = \delta_{ij}$. The map in Eq.~\eqref{dynamicalmap} can be represented as
\begin{equation*}
    \Omega [\rho (0)] = \sum_{m,n} \text{Tr}[\mathcal{G}_m \Omega [\mathcal{G}_n]] \text{Tr}[\mathcal{G}_n \rho (0)] \mathcal{G}_m = [F(t)r(0)] \mathcal{G}^T
\end{equation*}
where $F_{mn}=\text{Tr}[\mathcal{G}_m \Omega [\mathcal{G}_n]]$ and $r_n(s) = \text{Tr}[\mathcal{G}_n \rho(s)]$. Taking time-derivative of the above equation we shall get
\begin{equation*}
    \dot{\rho}(t) = [\dot{F}(t)r(0)] \mathcal{G}^T.
\end{equation*}
Let us now consider a matrix $L$ with elements $L_{mn} = \text{Tr}[\mathcal{G}_m \Tilde{\Lambda} [\mathcal{G}_n]]$. We can therefore represent Eq.~\eqref{eqom} as
\begin{equation*}\label{Eq13}
    \dot{\rho}(t) = \sum_{m,n} \text{Tr}[\mathcal{G}_m]\Tilde{\Lambda} [\mathcal{G}_n] \text{Tr}[\mathcal{G}_n \rho (t)] \mathcal{G}_m = [L(t)r(t)] \mathcal{G}^T.
\end{equation*}
Comparing the above two equations, we find
\begin{equation*}
    \dot{F}(t) = L(t) F(t) \implies L(t)= \dot{F}(t) F(t)^{-1}.
\end{equation*}
One may note that $L(t)$ can be obtained if $F(t)^{-1}$ exists and $F(0) = \mathbb{I}$. From this $L(t)$ matrix, we can derive the corresponding master equation following the methods given in Refs.~\cite{Hall14,Bhattacharya17}. It is to be noted that of the dynamical maps constructed in Eqs.~\eqref{switchmap1}, \eqref{switchmap2}, and \eqref{switchmap3}, all have the following form 
\begin{equation}
\begin{array}{ll}
    \rho_{11}(t)= \left(\frac{1+\mathcal{C}(t)}{2}\right)\rho_{11}(0)+\left(\frac{1-\mathcal{C}(t)}{2}\right)\rho_{22}(0),\\
    \\
    \rho_{22}(t)= \left(\frac{1-\mathcal{C}(t)}{2}\right)\rho_{11}(0)+\left(\frac{1+\mathcal{C}(t)}{2}\right)\rho_{22}(0),\\
    \\
    \rho_{12}(t)=\mathcal{C}(t)\rho_{12}(0),~~ \rho_{21}(t)=\rho_{12}^*(t),
\end{array}
\end{equation}
where $\mathcal{C}(t)$ is a real function of time. For these evolutions, the $L(t)$ matrices will be of the form 
\begin{equation}
L(t)= \left(\begin{matrix}
    0 & 0 & 0 & 0 \\
    0 & \frac{d}{dt}\ln\mathcal{C}(t) & 0 & 0\\
    0 & 0 & \frac{d}{dt}\ln\mathcal{C}(t) & 0\\
    0 & 0 & 0 & \frac{d}{dt}\ln\mathcal{C}(t)
      \end{matrix}\right).
\end{equation}
The corresponding master equation will be of the form 
\begin{equation}
    \frac{d}{dt}\rho(t)= \Gamma_{\mathcal{C}}(t) \sum_{i}\left[\sigma_i\rho(t)\sigma_i-\rho(t) \right],
\end{equation}
with \[\Gamma_{\mathcal{C}}(t)= -\frac{1}{4}\frac{d}{dt}\ln\mathcal{C}(t).\]

\bibliography{QSWITCH}

\begin{thebibliography}{90}%
\makeatletter
\providecommand \@ifxundefined [1]{%
 \@ifx{#1\undefined}
}%
\providecommand \@ifnum [1]{%
 \ifnum #1\expandafter \@firstoftwo
 \else \expandafter \@secondoftwo
 \fi
}%
\providecommand \@ifx [1]{%
 \ifx #1\expandafter \@firstoftwo
 \else \expandafter \@secondoftwo
 \fi
}%
\providecommand \natexlab [1]{#1}%
\providecommand \enquote  [1]{``#1''}%
\providecommand \bibnamefont  [1]{#1}%
\providecommand \bibfnamefont [1]{#1}%
\providecommand \citenamefont [1]{#1}%
\providecommand \href@noop [0]{\@secondoftwo}%
\providecommand \href [0]{\begingroup \@sanitize@url \@href}%
\providecommand \@href[1]{\@@startlink{#1}\@@href}%
\providecommand \@@href[1]{\endgroup#1\@@endlink}%
\providecommand \@sanitize@url [0]{\catcode `\\12\catcode `\$12\catcode
  `\&12\catcode `\#12\catcode `\^12\catcode `\_12\catcode `\%12\relax}%
\providecommand \@@startlink[1]{}%
\providecommand \@@endlink[0]{}%
\providecommand \url  [0]{\begingroup\@sanitize@url \@url }%
\providecommand \@url [1]{\endgroup\@href {#1}{\urlprefix }}%
\providecommand \urlprefix  [0]{URL }%
\providecommand \Eprint [0]{\href }%
\providecommand \doibase [0]{http://dx.doi.org/}%
\providecommand \selectlanguage [0]{\@gobble}%
\providecommand \bibinfo  [0]{\@secondoftwo}%
\providecommand \bibfield  [0]{\@secondoftwo}%
\providecommand \translation [1]{[#1]}%
\providecommand \BibitemOpen [0]{}%
\providecommand \bibitemStop [0]{}%
\providecommand \bibitemNoStop [0]{.\EOS\space}%
\providecommand \EOS [0]{\spacefactor3000\relax}%
\providecommand \BibitemShut  [1]{\csname bibitem#1\endcsname}%
\let\auto@bib@innerbib\@empty
\bibitem [{\citenamefont {Gisin}\ \emph {et~al.}(2005)\citenamefont {Gisin},
  \citenamefont {Linden}, \citenamefont {Massar},\ and\ \citenamefont
  {Popescu}}]{Gisin05}%
  \BibitemOpen
  \bibfield  {author} {\bibinfo {author} {\bibfnamefont {N.}~\bibnamefont
  {Gisin}}, \bibinfo {author} {\bibfnamefont {N.}~\bibnamefont {Linden}},
  \bibinfo {author} {\bibfnamefont {S.}~\bibnamefont {Massar}}, \ and\ \bibinfo
  {author} {\bibfnamefont {S.}~\bibnamefont {Popescu}},\ }\bibfield  {title}
  {\enquote {\bibinfo {title} {Error filtration and entanglement purification
  for quantum communication},}\ }\href {\doibase 10.1103/PhysRevA.72.012338}
  {\bibfield  {journal} {\bibinfo  {journal} {Phys. Rev. A}\ }\textbf {\bibinfo
  {volume} {72}},\ \bibinfo {pages} {012338} (\bibinfo {year}
  {2005})}\BibitemShut {NoStop}%
\bibitem [{\citenamefont {Chiribella}\ \emph {et~al.}(2013)\citenamefont
  {Chiribella}, \citenamefont {D'Ariano}, \citenamefont {Perinotti},\ and\
  \citenamefont {Valiron}}]{Chiribella13}%
  \BibitemOpen
  \bibfield  {author} {\bibinfo {author} {\bibfnamefont {G.}~\bibnamefont
  {Chiribella}}, \bibinfo {author} {\bibfnamefont {G.~M.}\ \bibnamefont
  {D'Ariano}}, \bibinfo {author} {\bibfnamefont {P.}~\bibnamefont {Perinotti}},
  \ and\ \bibinfo {author} {\bibfnamefont {B.}~\bibnamefont {Valiron}},\
  }\bibfield  {title} {\enquote {\bibinfo {title} {Quantum computations without
  definite causal structure},}\ }\href
  {https://doi.org/10.1103/PhysRevA.88.022318} {\bibfield  {journal} {\bibinfo
  {journal} {Phys. Rev. A}\ }\textbf {\bibinfo {volume} {88}},\ \bibinfo
  {pages} {022318} (\bibinfo {year} {2013})}\BibitemShut {NoStop}%
\bibitem [{\citenamefont {Chiribella}(2012)}]{Chiribella12}%
  \BibitemOpen
  \bibfield  {author} {\bibinfo {author} {\bibfnamefont {G.}~\bibnamefont
  {Chiribella}},\ }\bibfield  {title} {\enquote {\bibinfo {title} {Perfect
  discrimination of no-signalling channels via quantum superposition of causal
  structures},}\ }\href {https://doi.org/10.1103/PhysRevA.86.040301} {\bibfield
   {journal} {\bibinfo  {journal} {Phys. Rev. A}\ }\textbf {\bibinfo {volume}
  {86}},\ \bibinfo {pages} {040301(R)} (\bibinfo {year} {2012})}\BibitemShut
  {NoStop}%
\bibitem [{\citenamefont {Oreshkov}\ \emph {et~al.}(2012)\citenamefont
  {Oreshkov}, \citenamefont {Costa},\ and\ \citenamefont {\v{C}.
  Brukner}}]{Oreshkov12}%
  \BibitemOpen
  \bibfield  {author} {\bibinfo {author} {\bibfnamefont {O.}~\bibnamefont
  {Oreshkov}}, \bibinfo {author} {\bibfnamefont {F.}~\bibnamefont {Costa}}, \
  and\ \bibinfo {author} {\bibnamefont {\v{C}. Brukner}},\ }\bibfield  {title}
  {\enquote {\bibinfo {title} {Quantum correlations with no causal order},}\
  }\href {https://www.nature.com/articles/ncomms2076} {\bibfield  {journal}
  {\bibinfo  {journal} {Nat. Commun.}\ }\textbf {\bibinfo {volume} {3}}
  (\bibinfo {year} {2012})}\BibitemShut {NoStop}%
\bibitem [{\citenamefont {Ara\'{u}jo}\ \emph {et~al.}(2014)\citenamefont
  {Ara\'{u}jo}, \citenamefont {Costa},\ and\ \citenamefont {\v{C}.
  Brukner}}]{Araujo14}%
  \BibitemOpen
  \bibfield  {author} {\bibinfo {author} {\bibfnamefont {M.}~\bibnamefont
  {Ara\'{u}jo}}, \bibinfo {author} {\bibfnamefont {F.}~\bibnamefont {Costa}}, \
  and\ \bibinfo {author} {\bibnamefont {\v{C}. Brukner}},\ }\bibfield  {title}
  {\enquote {\bibinfo {title} {Computational advantage from quantum-controlled
  ordering of gates},}\ }\href
  {https://link.aps.org/doi/10.1103/PhysRevLett.113.250402} {\bibfield
  {journal} {\bibinfo  {journal} {Phys. Rev. Lett.}\ }\textbf {\bibinfo
  {volume} {113}},\ \bibinfo {pages} {250402} (\bibinfo {year}
  {2014})}\BibitemShut {NoStop}%
\bibitem [{\citenamefont {Gu\'{e}rin}\ \emph {et~al.}(2016)\citenamefont
  {Gu\'{e}rin}, \citenamefont {Feix}, \citenamefont {Ara\'{u}jo},\ and\
  \citenamefont {\v{C}. Brukner}}]{Guerin16}%
  \BibitemOpen
  \bibfield  {author} {\bibinfo {author} {\bibfnamefont {P.~A.}\ \bibnamefont
  {Gu\'{e}rin}}, \bibinfo {author} {\bibfnamefont {A.}~\bibnamefont {Feix}},
  \bibinfo {author} {\bibfnamefont {M.}~\bibnamefont {Ara\'{u}jo}}, \ and\
  \bibinfo {author} {\bibnamefont {\v{C}. Brukner}},\ }\bibfield  {title}
  {\enquote {\bibinfo {title} {Exponential communication complexity advantage
  from quantum superposition of the direction of communication},}\ }\href
  {https://doi.org/10.1103/PhysRevLett.117.100502} {\bibfield  {journal}
  {\bibinfo  {journal} {Phys. Rev. Lett.}\ }\textbf {\bibinfo {volume} {117}},\
  \bibinfo {pages} {100502} (\bibinfo {year} {2016})}\BibitemShut {NoStop}%
\bibitem [{\citenamefont {Ebler}\ \emph {et~al.}(2018)\citenamefont {Ebler},
  \citenamefont {Salek},\ and\ \citenamefont {Chiribella}}]{Ebler18}%
  \BibitemOpen
  \bibfield  {author} {\bibinfo {author} {\bibfnamefont {D.}~\bibnamefont
  {Ebler}}, \bibinfo {author} {\bibfnamefont {S.}~\bibnamefont {Salek}}, \ and\
  \bibinfo {author} {\bibfnamefont {G.}~\bibnamefont {Chiribella}},\ }\bibfield
   {title} {\enquote {\bibinfo {title} {Enhanced communication with the
  assistance of indefinite causal order},}\ }\href
  {https://doi.org/10.1103/PhysRevLett.120.120502} {\bibfield  {journal}
  {\bibinfo  {journal} {Phys. Rev. Lett.}\ }\textbf {\bibinfo {volume} {120}},\
  \bibinfo {pages} {120502} (\bibinfo {year} {2018})}\BibitemShut {NoStop}%
\bibitem [{\citenamefont {Chiribella}\ \emph {et~al.}(2021)\citenamefont
  {Chiribella}, \citenamefont {Banik}, \citenamefont {Bhattacharya},
  \citenamefont {Guha}, \citenamefont {Alimuddin}, \citenamefont {Roy},
  \citenamefont {Saha}, \citenamefont {Agrawal},\ and\ \citenamefont
  {Kar}}]{Chiribella21}%
  \BibitemOpen
  \bibfield  {author} {\bibinfo {author} {\bibfnamefont {Giulio}\ \bibnamefont
  {Chiribella}}, \bibinfo {author} {\bibfnamefont {Manik}\ \bibnamefont
  {Banik}}, \bibinfo {author} {\bibfnamefont {Some~Sankar}\ \bibnamefont
  {Bhattacharya}}, \bibinfo {author} {\bibfnamefont {Tamal}\ \bibnamefont
  {Guha}}, \bibinfo {author} {\bibfnamefont {Mir}\ \bibnamefont {Alimuddin}},
  \bibinfo {author} {\bibfnamefont {Arup}\ \bibnamefont {Roy}}, \bibinfo
  {author} {\bibfnamefont {Sutapa}\ \bibnamefont {Saha}}, \bibinfo {author}
  {\bibfnamefont {Sristy}\ \bibnamefont {Agrawal}}, \ and\ \bibinfo {author}
  {\bibfnamefont {Guruprasad}\ \bibnamefont {Kar}},\ }\bibfield  {title}
  {\enquote {\bibinfo {title} {Indefinite causal order enables perfect quantum
  communication with zero capacity channels},}\ }\href {\doibase
  10.1088/1367-2630/abe7a0} {\bibfield  {journal} {\bibinfo  {journal} {New
  Journal of Physics}\ }\textbf {\bibinfo {volume} {23}},\ \bibinfo {pages}
  {033039} (\bibinfo {year} {2021})}\BibitemShut {NoStop}%
\bibitem [{\citenamefont {Bhattacharya}\ \emph {et~al.}(2021)\citenamefont
  {Bhattacharya}, \citenamefont {Maity}, \citenamefont {Guha}, \citenamefont
  {Chiribella},\ and\ \citenamefont {Banik}}]{Banik21}%
  \BibitemOpen
  \bibfield  {author} {\bibinfo {author} {\bibfnamefont {Some~Sankar}\
  \bibnamefont {Bhattacharya}}, \bibinfo {author} {\bibfnamefont {Ananda~G.}\
  \bibnamefont {Maity}}, \bibinfo {author} {\bibfnamefont {Tamal}\ \bibnamefont
  {Guha}}, \bibinfo {author} {\bibfnamefont {Giulio}\ \bibnamefont
  {Chiribella}}, \ and\ \bibinfo {author} {\bibfnamefont {Manik}\ \bibnamefont
  {Banik}},\ }\bibfield  {title} {\enquote {\bibinfo {title} {Random-receiver
  quantum communication},}\ }\href {\doibase 10.1103/PRXQuantum.2.020350}
  {\bibfield  {journal} {\bibinfo  {journal} {PRX Quantum}\ }\textbf {\bibinfo
  {volume} {2}},\ \bibinfo {pages} {020350} (\bibinfo {year}
  {2021})}\BibitemShut {NoStop}%
\bibitem [{\citenamefont {Zhao}\ \emph {et~al.}(2020)\citenamefont {Zhao},
  \citenamefont {Yang},\ and\ \citenamefont {Chiribella}}]{Zhao20}%
  \BibitemOpen
  \bibfield  {author} {\bibinfo {author} {\bibfnamefont {Xiaobin}\ \bibnamefont
  {Zhao}}, \bibinfo {author} {\bibfnamefont {Yuxiang}\ \bibnamefont {Yang}}, \
  and\ \bibinfo {author} {\bibfnamefont {Giulio}\ \bibnamefont {Chiribella}},\
  }\bibfield  {title} {\enquote {\bibinfo {title} {Quantum metrology with
  indefinite causal order},}\ }\href {\doibase 10.1103/PhysRevLett.124.190503}
  {\bibfield  {journal} {\bibinfo  {journal} {Phys. Rev. Lett.}\ }\textbf
  {\bibinfo {volume} {124}},\ \bibinfo {pages} {190503} (\bibinfo {year}
  {2020})}\BibitemShut {NoStop}%
\bibitem [{\citenamefont {Guha}\ \emph {et~al.}(2020)\citenamefont {Guha},
  \citenamefont {Alimuddin},\ and\ \citenamefont {Parashar}}]{Tamal20}%
  \BibitemOpen
  \bibfield  {author} {\bibinfo {author} {\bibfnamefont {Tamal}\ \bibnamefont
  {Guha}}, \bibinfo {author} {\bibfnamefont {Mir}\ \bibnamefont {Alimuddin}}, \
  and\ \bibinfo {author} {\bibfnamefont {Preeti}\ \bibnamefont {Parashar}},\
  }\bibfield  {title} {\enquote {\bibinfo {title} {Thermodynamic advancement in
  the causally inseparable occurrence of thermal maps},}\ }\href {\doibase
  10.1103/PhysRevA.102.032215} {\bibfield  {journal} {\bibinfo  {journal}
  {Phys. Rev. A}\ }\textbf {\bibinfo {volume} {102}},\ \bibinfo {pages}
  {032215} (\bibinfo {year} {2020})}\BibitemShut {NoStop}%
\bibitem [{\citenamefont {Felce}\ and\ \citenamefont
  {Vedral}(2020)}]{Vedral20}%
  \BibitemOpen
  \bibfield  {author} {\bibinfo {author} {\bibfnamefont {David}\ \bibnamefont
  {Felce}}\ and\ \bibinfo {author} {\bibfnamefont {Vlatko}\ \bibnamefont
  {Vedral}},\ }\bibfield  {title} {\enquote {\bibinfo {title} {Quantum
  refrigeration with indefinite causal order},}\ }\href {\doibase
  10.1103/PhysRevLett.125.070603} {\bibfield  {journal} {\bibinfo  {journal}
  {Phys. Rev. Lett.}\ }\textbf {\bibinfo {volume} {125}},\ \bibinfo {pages}
  {070603} (\bibinfo {year} {2020})}\BibitemShut {NoStop}%
\bibitem [{\citenamefont {Maity}\ and\ \citenamefont
  {Bhattacharya}(2024)}]{Maity22}%
  \BibitemOpen
  \bibfield  {author} {\bibinfo {author} {\bibfnamefont {Ananda~G}\
  \bibnamefont {Maity}}\ and\ \bibinfo {author} {\bibfnamefont {Samyadeb}\
  \bibnamefont {Bhattacharya}},\ }\bibfield  {title} {\enquote {\bibinfo
  {title} {Activating information backflow with the assistance of quantum
  switch},}\ }\href {\doibase 10.1088/1751-8121/ad41a7} {\bibfield  {journal}
  {\bibinfo  {journal} {Journal of Physics A: Mathematical and Theoretical}\
  }\textbf {\bibinfo {volume} {57}},\ \bibinfo {pages} {215302} (\bibinfo
  {year} {2024})}\BibitemShut {NoStop}%
\bibitem [{\citenamefont {Liu}\ \emph {et~al.}(2022)\citenamefont {Liu},
  \citenamefont {Ebler},\ and\ \citenamefont {Dahlsten}}]{Liu23}%
  \BibitemOpen
  \bibfield  {author} {\bibinfo {author} {\bibfnamefont {Xiangjing}\
  \bibnamefont {Liu}}, \bibinfo {author} {\bibfnamefont {Daniel}\ \bibnamefont
  {Ebler}}, \ and\ \bibinfo {author} {\bibfnamefont {Oscar}\ \bibnamefont
  {Dahlsten}},\ }\bibfield  {title} {\enquote {\bibinfo {title} {Thermodynamics
  of quantum switch information capacity activation},}\ }\href {\doibase
  10.1103/PhysRevLett.129.230604} {\bibfield  {journal} {\bibinfo  {journal}
  {Phys. Rev. Lett.}\ }\textbf {\bibinfo {volume} {129}},\ \bibinfo {pages}
  {230604} (\bibinfo {year} {2022})}\BibitemShut {NoStop}%
\bibitem [{\citenamefont {Mukherjee}\ \emph {et~al.}(2024)\citenamefont
  {Mukherjee}, \citenamefont {Mallick}, \citenamefont {Yanamandra},
  \citenamefont {Bhattacharya},\ and\ \citenamefont {Maity}}]{Mukherjee24}%
  \BibitemOpen
  \bibfield  {author} {\bibinfo {author} {\bibfnamefont {Saheli}\ \bibnamefont
  {Mukherjee}}, \bibinfo {author} {\bibfnamefont {Bivas}\ \bibnamefont
  {Mallick}}, \bibinfo {author} {\bibfnamefont {Sravani}\ \bibnamefont
  {Yanamandra}}, \bibinfo {author} {\bibfnamefont {Samyadeb}\ \bibnamefont
  {Bhattacharya}}, \ and\ \bibinfo {author} {\bibfnamefont {Ananda~G.}\
  \bibnamefont {Maity}},\ }\bibfield  {title} {\enquote {\bibinfo {title}
  {Interplay between the hilbert-space dimension of a control system and the
  memory induced by a quantum switch},}\ }\href {\doibase
  10.1103/PhysRevA.110.042624} {\bibfield  {journal} {\bibinfo  {journal}
  {Phys. Rev. A}\ }\textbf {\bibinfo {volume} {110}},\ \bibinfo {pages}
  {042624} (\bibinfo {year} {2024})}\BibitemShut {NoStop}%
\bibitem [{\citenamefont {Mukhopadhyay}\ and\ \citenamefont
  {Pati}(2020)}]{Mukhopadhyay19}%
  \BibitemOpen
  \bibfield  {author} {\bibinfo {author} {\bibfnamefont {Chiranjib}\
  \bibnamefont {Mukhopadhyay}}\ and\ \bibinfo {author} {\bibfnamefont
  {Arun~Kumar}\ \bibnamefont {Pati}},\ }\bibfield  {title} {\enquote {\bibinfo
  {title} {Superposition of causal order enables quantum advantage in
  teleportation under very noisy channels},}\ }\href {\doibase
  10.1088/2399-6528/abbd77} {\bibfield  {journal} {\bibinfo  {journal} {Journal
  of Physics Communications}\ }\textbf {\bibinfo {volume} {4}},\ \bibinfo
  {pages} {105003} (\bibinfo {year} {2020})}\BibitemShut {NoStop}%
\bibitem [{\citenamefont {Ghosal}\ \emph {et~al.}(2023)\citenamefont {Ghosal},
  \citenamefont {Ghosal}, \citenamefont {Das},\ and\ \citenamefont
  {Maity}}]{Ghosal22}%
  \BibitemOpen
  \bibfield  {author} {\bibinfo {author} {\bibfnamefont {Pratik}\ \bibnamefont
  {Ghosal}}, \bibinfo {author} {\bibfnamefont {Arkaprabha}\ \bibnamefont
  {Ghosal}}, \bibinfo {author} {\bibfnamefont {Debarshi}\ \bibnamefont {Das}},
  \ and\ \bibinfo {author} {\bibfnamefont {Ananda~G.}\ \bibnamefont {Maity}},\
  }\bibfield  {title} {\enquote {\bibinfo {title} {Quantum superposition of
  causal structures as a universal resource for local implementation of
  nonlocal quantum operations},}\ }\href {\doibase 10.1103/PhysRevA.107.022613}
  {\bibfield  {journal} {\bibinfo  {journal} {Phys. Rev. A}\ }\textbf {\bibinfo
  {volume} {107}},\ \bibinfo {pages} {022613} (\bibinfo {year}
  {2023})}\BibitemShut {NoStop}%
\bibitem [{\citenamefont {Kristj{\'{a}}nsson}\ \emph
  {et~al.}(2020)\citenamefont {Kristj{\'{a}}nsson}, \citenamefont {Chiribella},
  \citenamefont {Salek}, \citenamefont {Ebler},\ and\ \citenamefont
  {Wilson}}]{Kristjnsson2020}%
  \BibitemOpen
  \bibfield  {author} {\bibinfo {author} {\bibfnamefont {Hl{\'{e}}r}\
  \bibnamefont {Kristj{\'{a}}nsson}}, \bibinfo {author} {\bibfnamefont
  {Giulio}\ \bibnamefont {Chiribella}}, \bibinfo {author} {\bibfnamefont
  {Sina}\ \bibnamefont {Salek}}, \bibinfo {author} {\bibfnamefont {Daniel}\
  \bibnamefont {Ebler}}, \ and\ \bibinfo {author} {\bibfnamefont {Matthew}\
  \bibnamefont {Wilson}},\ }\bibfield  {title} {\enquote {\bibinfo {title}
  {Resource theories of communication},}\ }\href {\doibase
  10.1088/1367-2630/ab8ef7} {\bibfield  {journal} {\bibinfo  {journal} {New
  Journal of Physics}\ }\textbf {\bibinfo {volume} {22}},\ \bibinfo {pages}
  {073014} (\bibinfo {year} {2020})}\BibitemShut {NoStop}%
\bibitem [{\citenamefont {Milz}\ \emph {et~al.}(2022)\citenamefont {Milz},
  \citenamefont {Bavaresco},\ and\ \citenamefont {Chiribella}}]{Milz2022}%
  \BibitemOpen
  \bibfield  {author} {\bibinfo {author} {\bibfnamefont {Simon}\ \bibnamefont
  {Milz}}, \bibinfo {author} {\bibfnamefont {Jessica}\ \bibnamefont
  {Bavaresco}}, \ and\ \bibinfo {author} {\bibfnamefont {Giulio}\ \bibnamefont
  {Chiribella}},\ }\bibfield  {title} {\enquote {\bibinfo {title} {Resource
  theory of causal connection},}\ }\href {\doibase 10.22331/q-2022-08-25-788}
  {\bibfield  {journal} {\bibinfo  {journal} {{Quantum}}\ }\textbf {\bibinfo
  {volume} {6}},\ \bibinfo {pages} {788} (\bibinfo {year} {2022})}\BibitemShut
  {NoStop}%
\bibitem [{\citenamefont {et~al.}(2015)}]{Procopio15}%
  \BibitemOpen
  \bibfield  {author} {\bibinfo {author} {\bibfnamefont {L.~M.~Procopio}\
  \bibnamefont {et~al.}},\ }\bibfield  {title} {\enquote {\bibinfo {title}
  {Experimental superposition of orders of quantum gates},}\ }\href
  {https://www.nature.com/articles/ncomms8913} {\bibfield  {journal} {\bibinfo
  {journal} {Nat. Commun.}\ }\textbf {\bibinfo {volume} {6}} (\bibinfo {year}
  {2015})}\BibitemShut {NoStop}%
\bibitem [{\citenamefont {al}(2017)}]{Rubino17}%
  \BibitemOpen
  \bibfield  {author} {\bibinfo {author} {\bibfnamefont {G.~Rubino}\
  \bibnamefont {al}},\ }\bibfield  {title} {\enquote {\bibinfo {title}
  {Experimental verification of an indefinite causal order},}\ }\href
  {http://advances.sciencemag.org/content/3/3/e1602589} {\bibfield  {journal}
  {\bibinfo  {journal} {Science Advances}\ }\textbf {\bibinfo {volume} {3}},\
  \bibinfo {pages} {e1602589} (\bibinfo {year} {2017})}\BibitemShut {NoStop}%
\bibitem [{\citenamefont {Goswami}\ \emph {et~al.}(2018)\citenamefont
  {Goswami}, \citenamefont {Giarmatzi}, \citenamefont {Kewming}, \citenamefont
  {Costa}, \citenamefont {Branciard}, \citenamefont {Romero},\ and\
  \citenamefont {White}}]{Goswami18(1)}%
  \BibitemOpen
  \bibfield  {author} {\bibinfo {author} {\bibfnamefont {K.}~\bibnamefont
  {Goswami}}, \bibinfo {author} {\bibfnamefont {C.}~\bibnamefont {Giarmatzi}},
  \bibinfo {author} {\bibfnamefont {M.}~\bibnamefont {Kewming}}, \bibinfo
  {author} {\bibfnamefont {F.}~\bibnamefont {Costa}}, \bibinfo {author}
  {\bibfnamefont {C.}~\bibnamefont {Branciard}}, \bibinfo {author}
  {\bibfnamefont {J.}~\bibnamefont {Romero}}, \ and\ \bibinfo {author}
  {\bibfnamefont {A.~G.}\ \bibnamefont {White}},\ }\bibfield  {title} {\enquote
  {\bibinfo {title} {Indefinite causal order in a quantum switch},}\ }\href
  {https://doi.org/10.1103/PhysRevLett.121.090503} {\bibfield  {journal}
  {\bibinfo  {journal} {Phys. Rev. Lett.}\ }\textbf {\bibinfo {volume} {121}},\
  \bibinfo {pages} {090503} (\bibinfo {year} {2018})}\BibitemShut {NoStop}%
\bibitem [{\citenamefont {Chang}\ \emph {et~al.}(2019)\citenamefont {Chang},
  \citenamefont {Li}, \citenamefont {Wu}, \citenamefont {Jiang}, \citenamefont
  {Zhang}, \citenamefont {Pu}, \citenamefont {Chang},\ and\ \citenamefont
  {Duan}}]{Chang19}%
  \BibitemOpen
  \bibfield  {author} {\bibinfo {author} {\bibfnamefont {W.}~\bibnamefont
  {Chang}}, \bibinfo {author} {\bibfnamefont {C.}~\bibnamefont {Li}}, \bibinfo
  {author} {\bibfnamefont {Y.-K.}\ \bibnamefont {Wu}}, \bibinfo {author}
  {\bibfnamefont {N.}~\bibnamefont {Jiang}}, \bibinfo {author} {\bibfnamefont
  {S.}~\bibnamefont {Zhang}}, \bibinfo {author} {\bibfnamefont {Y.-F.}\
  \bibnamefont {Pu}}, \bibinfo {author} {\bibfnamefont {X.-Y.}\ \bibnamefont
  {Chang}}, \ and\ \bibinfo {author} {\bibfnamefont {L.-M.}\ \bibnamefont
  {Duan}},\ }\bibfield  {title} {\enquote {\bibinfo {title} {Long-distance
  entanglement between a multiplexed quantum memory and a telecom photon},}\
  }\href {\doibase 10.1103/PhysRevX.9.041033} {\bibfield  {journal} {\bibinfo
  {journal} {Phys. Rev. X}\ }\textbf {\bibinfo {volume} {9}},\ \bibinfo {pages}
  {041033} (\bibinfo {year} {2019})}\BibitemShut {NoStop}%
\bibitem [{\citenamefont {Duan}\ \emph {et~al.}(2001)\citenamefont {Duan},
  \citenamefont {Lukin}, \citenamefont {Cirac},\ and\ \citenamefont
  {Zoller}}]{Duan2001}%
  \BibitemOpen
  \bibfield  {author} {\bibinfo {author} {\bibfnamefont {L.-M.}\ \bibnamefont
  {Duan}}, \bibinfo {author} {\bibfnamefont {M.~D.}\ \bibnamefont {Lukin}},
  \bibinfo {author} {\bibfnamefont {J.~I.}\ \bibnamefont {Cirac}}, \ and\
  \bibinfo {author} {\bibfnamefont {P.}~\bibnamefont {Zoller}},\ }\bibfield
  {title} {\enquote {\bibinfo {title} {Long-distance quantum communication with
  atomic ensembles and linear optics},}\ }\href {\doibase 10.1038/35106500}
  {\bibfield  {journal} {\bibinfo  {journal} {Nature}\ }\textbf {\bibinfo
  {volume} {414}},\ \bibinfo {pages} {413--418} (\bibinfo {year}
  {2001})}\BibitemShut {NoStop}%
\bibitem [{\citenamefont {Kretschmann}\ and\ \citenamefont
  {Werner}(2005)}]{Kretschmann05}%
  \BibitemOpen
  \bibfield  {author} {\bibinfo {author} {\bibfnamefont {Dennis}\ \bibnamefont
  {Kretschmann}}\ and\ \bibinfo {author} {\bibfnamefont {Reinhard~F.}\
  \bibnamefont {Werner}},\ }\bibfield  {title} {\enquote {\bibinfo {title}
  {Quantum channels with memory},}\ }\href {\doibase
  10.1103/PhysRevA.72.062323} {\bibfield  {journal} {\bibinfo  {journal} {Phys.
  Rev. A}\ }\textbf {\bibinfo {volume} {72}},\ \bibinfo {pages} {062323}
  (\bibinfo {year} {2005})}\BibitemShut {NoStop}%
\bibitem [{\citenamefont {D{\textquotesingle}Arrigo}\ \emph
  {et~al.}(2007)\citenamefont {D{\textquotesingle}Arrigo}, \citenamefont
  {Benenti},\ and\ \citenamefont {Falci}}]{DArrigo2007}%
  \BibitemOpen
  \bibfield  {author} {\bibinfo {author} {\bibfnamefont {A}~\bibnamefont
  {D{\textquotesingle}Arrigo}}, \bibinfo {author} {\bibfnamefont
  {G}~\bibnamefont {Benenti}}, \ and\ \bibinfo {author} {\bibfnamefont
  {G}~\bibnamefont {Falci}},\ }\bibfield  {title} {\enquote {\bibinfo {title}
  {Quantum capacity of dephasing channels with memory},}\ }\href {\doibase
  10.1088/1367-2630/9/9/310} {\bibfield  {journal} {\bibinfo  {journal} {New
  Journal of Physics}\ }\textbf {\bibinfo {volume} {9}},\ \bibinfo {pages}
  {310--310} (\bibinfo {year} {2007})}\BibitemShut {NoStop}%
\bibitem [{\citenamefont {Datta}\ and\ \citenamefont
  {Dorlas}(2009)}]{Datta2009}%
  \BibitemOpen
  \bibfield  {author} {\bibinfo {author} {\bibfnamefont {Nilanjana}\
  \bibnamefont {Datta}}\ and\ \bibinfo {author} {\bibfnamefont {Tony}\
  \bibnamefont {Dorlas}},\ }\bibfield  {title} {\enquote {\bibinfo {title}
  {Classical capacity of quantum channels with general markovian correlated
  noise},}\ }\href {\doibase 10.1007/s10955-009-9726-0} {\bibfield  {journal}
  {\bibinfo  {journal} {Journal of Statistical Physics}\ }\textbf {\bibinfo
  {volume} {134}},\ \bibinfo {pages} {1173--1195} (\bibinfo {year}
  {2009})}\BibitemShut {NoStop}%
\bibitem [{\citenamefont {Bylicka}\ \emph {et~al.}(2014)\citenamefont
  {Bylicka}, \citenamefont {Chruściński},\ and\ \citenamefont
  {Maniscalco}}]{task2}%
  \BibitemOpen
  \bibfield  {author} {\bibinfo {author} {\bibfnamefont {B}~\bibnamefont
  {Bylicka}}, \bibinfo {author} {\bibfnamefont {D}~\bibnamefont
  {Chruściński}}, \ and\ \bibinfo {author} {\bibfnamefont {S}~\bibnamefont
  {Maniscalco}},\ }\bibfield  {title} {\enquote {\bibinfo {title}
  {Non-markovianity and reservoir memory of quantum channels: a quantum
  information theory perspective},}\ }\href
  {https://www.nature.com/articles/srep05720} {\bibfield  {journal} {\bibinfo
  {journal} {Scientific Reports}\ }\textbf {\bibinfo {volume} {4}},\ \bibinfo
  {pages} {5720} (\bibinfo {year} {2014})}\BibitemShut {NoStop}%
\bibitem [{\citenamefont {Bylicka}\ \emph {et~al.}(2016)\citenamefont
  {Bylicka}, \citenamefont {Tukiainen}, \citenamefont
  {Chru{\'{s}}ci{\'{n}}ski}, \citenamefont {Piilo},\ and\ \citenamefont
  {Maniscalco}}]{Bylicka2016}%
  \BibitemOpen
  \bibfield  {author} {\bibinfo {author} {\bibfnamefont {Bogna}\ \bibnamefont
  {Bylicka}}, \bibinfo {author} {\bibfnamefont {Mikko}\ \bibnamefont
  {Tukiainen}}, \bibinfo {author} {\bibfnamefont {Dariusz}\ \bibnamefont
  {Chru{\'{s}}ci{\'{n}}ski}}, \bibinfo {author} {\bibfnamefont {Jyrki}\
  \bibnamefont {Piilo}}, \ and\ \bibinfo {author} {\bibfnamefont {Sabrina}\
  \bibnamefont {Maniscalco}},\ }\bibfield  {title} {\enquote {\bibinfo {title}
  {Thermodynamic power of non-markovianity},}\ }\href {\doibase
  10.1038/srep27989} {\bibfield  {journal} {\bibinfo  {journal} {Scientific
  Reports}\ }\textbf {\bibinfo {volume} {6}} (\bibinfo {year} {2016}),\
  10.1038/srep27989}\BibitemShut {NoStop}%
\bibitem [{\citenamefont {Taranto}\ \emph {et~al.}(2020)\citenamefont
  {Taranto}, \citenamefont {Bakhshinezhad}, \citenamefont {Sch\"uttelkopf},
  \citenamefont {Clivaz},\ and\ \citenamefont {Huber}}]{Taranto20}%
  \BibitemOpen
  \bibfield  {author} {\bibinfo {author} {\bibfnamefont {Philip}\ \bibnamefont
  {Taranto}}, \bibinfo {author} {\bibfnamefont {Faraj}\ \bibnamefont
  {Bakhshinezhad}}, \bibinfo {author} {\bibfnamefont {Philipp}\ \bibnamefont
  {Sch\"uttelkopf}}, \bibinfo {author} {\bibfnamefont {Fabien}\ \bibnamefont
  {Clivaz}}, \ and\ \bibinfo {author} {\bibfnamefont {Marcus}\ \bibnamefont
  {Huber}},\ }\bibfield  {title} {\enquote {\bibinfo {title} {Exponential
  improvement for quantum cooling through finite-memory effects},}\ }\href
  {\doibase 10.1103/PhysRevApplied.14.054005} {\bibfield  {journal} {\bibinfo
  {journal} {Phys. Rev. Appl.}\ }\textbf {\bibinfo {volume} {14}},\ \bibinfo
  {pages} {054005} (\bibinfo {year} {2020})}\BibitemShut {NoStop}%
\bibitem [{\citenamefont {Taranto}\ \emph {et~al.}(2023)\citenamefont
  {Taranto}, \citenamefont {Bakhshinezhad}, \citenamefont {Bluhm},
  \citenamefont {Silva}, \citenamefont {Friis}, \citenamefont {Lock},
  \citenamefont {Vitagliano}, \citenamefont {Binder}, \citenamefont {Debarba},
  \citenamefont {Schwarzhans}, \citenamefont {Clivaz},\ and\ \citenamefont
  {Huber}}]{Taranto23}%
  \BibitemOpen
  \bibfield  {author} {\bibinfo {author} {\bibfnamefont {Philip}\ \bibnamefont
  {Taranto}}, \bibinfo {author} {\bibfnamefont {Faraj}\ \bibnamefont
  {Bakhshinezhad}}, \bibinfo {author} {\bibfnamefont {Andreas}\ \bibnamefont
  {Bluhm}}, \bibinfo {author} {\bibfnamefont {Ralph}\ \bibnamefont {Silva}},
  \bibinfo {author} {\bibfnamefont {Nicolai}\ \bibnamefont {Friis}}, \bibinfo
  {author} {\bibfnamefont {Maximilian~P.E.}\ \bibnamefont {Lock}}, \bibinfo
  {author} {\bibfnamefont {Giuseppe}\ \bibnamefont {Vitagliano}}, \bibinfo
  {author} {\bibfnamefont {Felix~C.}\ \bibnamefont {Binder}}, \bibinfo {author}
  {\bibfnamefont {Tiago}\ \bibnamefont {Debarba}}, \bibinfo {author}
  {\bibfnamefont {Emanuel}\ \bibnamefont {Schwarzhans}}, \bibinfo {author}
  {\bibfnamefont {Fabien}\ \bibnamefont {Clivaz}}, \ and\ \bibinfo {author}
  {\bibfnamefont {Marcus}\ \bibnamefont {Huber}},\ }\bibfield  {title}
  {\enquote {\bibinfo {title} {Landauer versus nernst: What is the true cost of
  cooling a quantum system?}}\ }\href {\doibase 10.1103/PRXQuantum.4.010332}
  {\bibfield  {journal} {\bibinfo  {journal} {PRX Quantum}\ }\textbf {\bibinfo
  {volume} {4}},\ \bibinfo {pages} {010332} (\bibinfo {year}
  {2023})}\BibitemShut {NoStop}%
\bibitem [{\citenamefont {Breuer}\ \emph {et~al.}(2004)\citenamefont {Breuer},
  \citenamefont {Burgarth},\ and\ \citenamefont {Petruccione}}]{breuer1}%
  \BibitemOpen
  \bibfield  {author} {\bibinfo {author} {\bibfnamefont {Heinz-Peter}\
  \bibnamefont {Breuer}}, \bibinfo {author} {\bibfnamefont {Daniel}\
  \bibnamefont {Burgarth}}, \ and\ \bibinfo {author} {\bibfnamefont
  {Francesco}\ \bibnamefont {Petruccione}},\ }\bibfield  {title} {\enquote
  {\bibinfo {title} {Non-markovian dynamics in a spin star system: Exact
  solution and approximation techniques},}\ }\href {\doibase
  10.1103/PhysRevB.70.045323} {\bibfield  {journal} {\bibinfo  {journal} {Phys.
  Rev. B}\ }\textbf {\bibinfo {volume} {70}},\ \bibinfo {pages} {045323}
  (\bibinfo {year} {2004})}\BibitemShut {NoStop}%
\bibitem [{\citenamefont {Laine}\ \emph {et~al.}(2010)\citenamefont {Laine},
  \citenamefont {Piilo},\ and\ \citenamefont {Breuer}}]{blp1}%
  \BibitemOpen
  \bibfield  {author} {\bibinfo {author} {\bibfnamefont {Elsi-Mari}\
  \bibnamefont {Laine}}, \bibinfo {author} {\bibfnamefont {Jyrki}\ \bibnamefont
  {Piilo}}, \ and\ \bibinfo {author} {\bibfnamefont {Heinz-Peter}\ \bibnamefont
  {Breuer}},\ }\bibfield  {title} {\enquote {\bibinfo {title} {Measure for the
  non-markovianity of quantum processes},}\ }\href {\doibase
  10.1103/PhysRevA.81.062115} {\bibfield  {journal} {\bibinfo  {journal} {Phys.
  Rev. A}\ }\textbf {\bibinfo {volume} {81}},\ \bibinfo {pages} {062115}
  (\bibinfo {year} {2010})}\BibitemShut {NoStop}%
\bibitem [{\citenamefont {Rivas}\ \emph {et~al.}(2010)\citenamefont {Rivas},
  \citenamefont {Huelga},\ and\ \citenamefont {Plenio}}]{rivas}%
  \BibitemOpen
  \bibfield  {author} {\bibinfo {author} {\bibfnamefont {\'Angel}\ \bibnamefont
  {Rivas}}, \bibinfo {author} {\bibfnamefont {Susana~F.}\ \bibnamefont
  {Huelga}}, \ and\ \bibinfo {author} {\bibfnamefont {Martin~B.}\ \bibnamefont
  {Plenio}},\ }\bibfield  {title} {\enquote {\bibinfo {title} {Entanglement and
  non-markovianity of quantum evolutions},}\ }\href {\doibase
  10.1103/PhysRevLett.105.050403} {\bibfield  {journal} {\bibinfo  {journal}
  {Phys. Rev. Lett.}\ }\textbf {\bibinfo {volume} {105}},\ \bibinfo {pages}
  {050403} (\bibinfo {year} {2010})}\BibitemShut {NoStop}%
\bibitem [{\citenamefont {Rivas}\ \emph {et~al.}(2014)\citenamefont {Rivas},
  \citenamefont {Huelga},\ and\ \citenamefont {Plenio}}]{RHPreview}%
  \BibitemOpen
  \bibfield  {author} {\bibinfo {author} {\bibfnamefont {Angel}\ \bibnamefont
  {Rivas}}, \bibinfo {author} {\bibfnamefont {Susana~F}\ \bibnamefont
  {Huelga}}, \ and\ \bibinfo {author} {\bibfnamefont {Martin~B}\ \bibnamefont
  {Plenio}},\ }\bibfield  {title} {\enquote {\bibinfo {title} {Quantum
  non-markovianity: characterization, quantification and detection},}\ }\href
  {http://stacks.iop.org/0034-4885/77/i=9/a=094001} {\bibfield  {journal}
  {\bibinfo  {journal} {Reports on Progress in Physics}\ }\textbf {\bibinfo
  {volume} {77}},\ \bibinfo {pages} {094001} (\bibinfo {year}
  {2014})}\BibitemShut {NoStop}%
\bibitem [{\citenamefont {Vasile}\ \emph {et~al.}(2011)\citenamefont {Vasile},
  \citenamefont {Maniscalco}, \citenamefont {Paris}, \citenamefont {Breuer},\
  and\ \citenamefont {Piilo}}]{vasile}%
  \BibitemOpen
  \bibfield  {author} {\bibinfo {author} {\bibfnamefont {Ruggero}\ \bibnamefont
  {Vasile}}, \bibinfo {author} {\bibfnamefont {Sabrina}\ \bibnamefont
  {Maniscalco}}, \bibinfo {author} {\bibfnamefont {Matteo G.~A.}\ \bibnamefont
  {Paris}}, \bibinfo {author} {\bibfnamefont {Heinz-Peter}\ \bibnamefont
  {Breuer}}, \ and\ \bibinfo {author} {\bibfnamefont {Jyrki}\ \bibnamefont
  {Piilo}},\ }\bibfield  {title} {\enquote {\bibinfo {title} {Quantifying
  non-markovianity of continuous-variable gaussian dynamical maps},}\ }\href
  {\doibase 10.1103/PhysRevA.84.052118} {\bibfield  {journal} {\bibinfo
  {journal} {Phys. Rev. A}\ }\textbf {\bibinfo {volume} {84}},\ \bibinfo
  {pages} {052118} (\bibinfo {year} {2011})}\BibitemShut {NoStop}%
\bibitem [{\citenamefont {Lu}\ \emph {et~al.}(2010)\citenamefont {Lu},
  \citenamefont {Wang},\ and\ \citenamefont {Sun}}]{lu}%
  \BibitemOpen
  \bibfield  {author} {\bibinfo {author} {\bibfnamefont {Xiao-Ming}\
  \bibnamefont {Lu}}, \bibinfo {author} {\bibfnamefont {Xiaoguang}\
  \bibnamefont {Wang}}, \ and\ \bibinfo {author} {\bibfnamefont {C.~P.}\
  \bibnamefont {Sun}},\ }\bibfield  {title} {\enquote {\bibinfo {title}
  {Quantum fisher information flow and non-markovian processes of open
  systems},}\ }\href {\doibase 10.1103/PhysRevA.82.042103} {\bibfield
  {journal} {\bibinfo  {journal} {Phys. Rev. A}\ }\textbf {\bibinfo {volume}
  {82}},\ \bibinfo {pages} {042103} (\bibinfo {year} {2010})}\BibitemShut
  {NoStop}%
\bibitem [{\citenamefont {Luo}\ \emph {et~al.}(2012)\citenamefont {Luo},
  \citenamefont {Fu},\ and\ \citenamefont {Song}}]{luo}%
  \BibitemOpen
  \bibfield  {author} {\bibinfo {author} {\bibfnamefont {Shunlong}\
  \bibnamefont {Luo}}, \bibinfo {author} {\bibfnamefont {Shuangshuang}\
  \bibnamefont {Fu}}, \ and\ \bibinfo {author} {\bibfnamefont {Hongting}\
  \bibnamefont {Song}},\ }\bibfield  {title} {\enquote {\bibinfo {title}
  {Quantifying non-markovianity via correlations},}\ }\href {\doibase
  10.1103/PhysRevA.86.044101} {\bibfield  {journal} {\bibinfo  {journal} {Phys.
  Rev. A}\ }\textbf {\bibinfo {volume} {86}},\ \bibinfo {pages} {044101}
  (\bibinfo {year} {2012})}\BibitemShut {NoStop}%
\bibitem [{\citenamefont {Fanchini}(2014)}]{fanchini}%
  \BibitemOpen
  \bibfield  {author} {\bibinfo {author} {\bibfnamefont {F.~F.~et.al}\
  \bibnamefont {Fanchini}},\ }\bibfield  {title} {\enquote {\bibinfo {title}
  {Non-markovianity through accessible information},}\ }\href {\doibase
  10.1103/PhysRevLett.112.210402} {\bibfield  {journal} {\bibinfo  {journal}
  {Phys. Rev. Lett.}\ }\textbf {\bibinfo {volume} {112}},\ \bibinfo {pages}
  {210402} (\bibinfo {year} {2014})}\BibitemShut {NoStop}%
\bibitem [{\citenamefont {Chanda}\ and\ \citenamefont
  {Bhattacharya}(2016)}]{titas}%
  \BibitemOpen
  \bibfield  {author} {\bibinfo {author} {\bibfnamefont {Titas}\ \bibnamefont
  {Chanda}}\ and\ \bibinfo {author} {\bibfnamefont {Samyadeb}\ \bibnamefont
  {Bhattacharya}},\ }\bibfield  {title} {\enquote {\bibinfo {title}
  {Delineating incoherent non-markovian dynamics using quantum coherence},}\
  }\href {\doibase http://dx.doi.org/10.1016/j.aop.2016.01.004} {\bibfield
  {journal} {\bibinfo  {journal} {Annals of Physics}\ }\textbf {\bibinfo
  {volume} {366}},\ \bibinfo {pages} {1 -- 12} (\bibinfo {year}
  {2016})}\BibitemShut {NoStop}%
\bibitem [{\citenamefont {Haseli}(2014)}]{haseli}%
  \BibitemOpen
  \bibfield  {author} {\bibinfo {author} {\bibfnamefont {S.~et.al}\
  \bibnamefont {Haseli}},\ }\bibfield  {title} {\enquote {\bibinfo {title}
  {Non-markovianity through flow of information between a system and an
  environment},}\ }\href {\doibase 10.1103/PhysRevA.90.052118} {\bibfield
  {journal} {\bibinfo  {journal} {Phys. Rev. A}\ }\textbf {\bibinfo {volume}
  {90}},\ \bibinfo {pages} {052118} (\bibinfo {year} {2014})}\BibitemShut
  {NoStop}%
\bibitem [{\citenamefont {Mukhopadhyay}\ \emph {et~al.}(2017)\citenamefont
  {Mukhopadhyay}, \citenamefont {Bhattacharya}, \citenamefont {Misra},\ and\
  \citenamefont {Pati}}]{sam2}%
  \BibitemOpen
  \bibfield  {author} {\bibinfo {author} {\bibfnamefont {Chiranjib}\
  \bibnamefont {Mukhopadhyay}}, \bibinfo {author} {\bibfnamefont {Samyadeb}\
  \bibnamefont {Bhattacharya}}, \bibinfo {author} {\bibfnamefont {Avijit}\
  \bibnamefont {Misra}}, \ and\ \bibinfo {author} {\bibfnamefont {Arun~Kumar}\
  \bibnamefont {Pati}},\ }\bibfield  {title} {\enquote {\bibinfo {title}
  {Dynamics and thermodynamics of a central spin immersed in a spin bath},}\
  }\href {\doibase 10.1103/PhysRevA.96.052125} {\bibfield  {journal} {\bibinfo
  {journal} {Phys. Rev. A}\ }\textbf {\bibinfo {volume} {96}},\ \bibinfo
  {pages} {052125} (\bibinfo {year} {2017})}\BibitemShut {NoStop}%
\bibitem [{\citenamefont {Bhattacharya}\ \emph
  {et~al.}(2020{\natexlab{a}})\citenamefont {Bhattacharya}, \citenamefont
  {Bhattacharya},\ and\ \citenamefont {Majumdar}}]{sam4}%
  \BibitemOpen
  \bibfield  {author} {\bibinfo {author} {\bibfnamefont {Samyadeb}\
  \bibnamefont {Bhattacharya}}, \bibinfo {author} {\bibfnamefont {Bihalan}\
  \bibnamefont {Bhattacharya}}, \ and\ \bibinfo {author} {\bibfnamefont {A~S}\
  \bibnamefont {Majumdar}},\ }\bibfield  {title} {\enquote {\bibinfo {title}
  {Convex resource theory of non-markovianity},}\ }\href {\doibase
  10.1088/1751-8121/abd191} {\bibfield  {journal} {\bibinfo  {journal} {Journal
  of Physics A: Mathematical and Theoretical}\ }\textbf {\bibinfo {volume}
  {54}},\ \bibinfo {pages} {035302} (\bibinfo {year}
  {2020}{\natexlab{a}})}\BibitemShut {NoStop}%
\bibitem [{\citenamefont {Bhattacharya}\ \emph
  {et~al.}(2020{\natexlab{b}})\citenamefont {Bhattacharya}, \citenamefont
  {Bhattacharya},\ and\ \citenamefont {Majumdar}}]{sam5}%
  \BibitemOpen
  \bibfield  {author} {\bibinfo {author} {\bibfnamefont {Samyadeb}\
  \bibnamefont {Bhattacharya}}, \bibinfo {author} {\bibfnamefont {Bihalan}\
  \bibnamefont {Bhattacharya}}, \ and\ \bibinfo {author} {\bibfnamefont {A~S}\
  \bibnamefont {Majumdar}},\ }\bibfield  {title} {\enquote {\bibinfo {title}
  {Thermodynamic utility of non-markovianity from the perspective of resource
  interconversion},}\ }\href {\doibase 10.1088/1751-8121/aba0ee} {\bibfield
  {journal} {\bibinfo  {journal} {Journal of Physics A: Mathematical and
  Theoretical}\ }\textbf {\bibinfo {volume} {53}},\ \bibinfo {pages} {335301}
  (\bibinfo {year} {2020}{\natexlab{b}})}\BibitemShut {NoStop}%
\bibitem [{\citenamefont {Maity}\ \emph {et~al.}(2020)\citenamefont {Maity},
  \citenamefont {Bhattacharya},\ and\ \citenamefont {Majumdar}}]{sam6}%
  \BibitemOpen
  \bibfield  {author} {\bibinfo {author} {\bibfnamefont {Ananda~G}\
  \bibnamefont {Maity}}, \bibinfo {author} {\bibfnamefont {Samyadeb}\
  \bibnamefont {Bhattacharya}}, \ and\ \bibinfo {author} {\bibfnamefont {A~S}\
  \bibnamefont {Majumdar}},\ }\bibfield  {title} {\enquote {\bibinfo {title}
  {Detecting non-markovianity via uncertainty relations},}\ }\href {\doibase
  10.1088/1751-8121/ab7135} {\bibfield  {journal} {\bibinfo  {journal} {Journal
  of Physics A: Mathematical and Theoretical}\ }\textbf {\bibinfo {volume}
  {53}},\ \bibinfo {pages} {175301} (\bibinfo {year} {2020})}\BibitemShut
  {NoStop}%
\bibitem [{\citenamefont {Bhattacharya}\ and\ \citenamefont
  {Bhattacharya}(2021)}]{sam7}%
  \BibitemOpen
  \bibfield  {author} {\bibinfo {author} {\bibfnamefont {Bihalan}\ \bibnamefont
  {Bhattacharya}}\ and\ \bibinfo {author} {\bibfnamefont {Samyadeb}\
  \bibnamefont {Bhattacharya}},\ }\bibfield  {title} {\enquote {\bibinfo
  {title} {Convex geometry of markovian lindblad dynamics and witnessing
  non-markovianity},}\ }\href {\doibase
  https://doi.org/10.1007/s11128-021-03177-y} {\bibfield  {journal} {\bibinfo
  {journal} {Quantum Information Processing}\ }\textbf {\bibinfo {volume}
  {20}},\ \bibinfo {pages} {253} (\bibinfo {year} {2021})}\BibitemShut
  {NoStop}%
\bibitem [{\citenamefont {Bhattacharya}\ \emph {et~al.}(2018)\citenamefont
  {Bhattacharya}, \citenamefont {Banerjee},\ and\ \citenamefont {Pati}}]{sam8}%
  \BibitemOpen
  \bibfield  {author} {\bibinfo {author} {\bibfnamefont {Samyadeb}\
  \bibnamefont {Bhattacharya}}, \bibinfo {author} {\bibfnamefont {Subhashish}\
  \bibnamefont {Banerjee}}, \ and\ \bibinfo {author} {\bibfnamefont {A~K}\
  \bibnamefont {Pati}},\ }\bibfield  {title} {\enquote {\bibinfo {title}
  {Evolution of coherence and non-classicality under global environmental
  interaction},}\ }\href {\doibase https://doi.org/10.1007/s11128-018-1989-4}
  {\bibfield  {journal} {\bibinfo  {journal} {Quantum Information Processing}\
  }\textbf {\bibinfo {volume} {17}},\ \bibinfo {pages} {236} (\bibinfo {year}
  {2018})}\BibitemShut {NoStop}%
\bibitem [{\citenamefont {Das}\ \emph {et~al.}(2021)\citenamefont {Das},
  \citenamefont {Roy}, \citenamefont {Bhattacharya},\ and\ \citenamefont
  {Sen}}]{sam9}%
  \BibitemOpen
  \bibfield  {author} {\bibinfo {author} {\bibfnamefont {Sreetama}\
  \bibnamefont {Das}}, \bibinfo {author} {\bibfnamefont {Sudipto~Singha}\
  \bibnamefont {Roy}}, \bibinfo {author} {\bibfnamefont {Samyadeb}\
  \bibnamefont {Bhattacharya}}, \ and\ \bibinfo {author} {\bibfnamefont
  {Ujjwal}\ \bibnamefont {Sen}},\ }\bibfield  {title} {\enquote {\bibinfo
  {title} {Nearly markovian maps and entanglement-based bound on corresponding
  non-markovianity},}\ }\href {\doibase 10.1088/1751-8121/ac1d8b} {\bibfield
  {journal} {\bibinfo  {journal} {Journal of Physics A: Mathematical and
  Theoretical}\ }\textbf {\bibinfo {volume} {54}},\ \bibinfo {pages} {395301}
  (\bibinfo {year} {2021})}\BibitemShut {NoStop}%
\bibitem [{\citenamefont {Giarmatzi}\ and\ \citenamefont
  {Costa}(2021)}]{Giarmatzi2021witnessingquantum}%
  \BibitemOpen
  \bibfield  {author} {\bibinfo {author} {\bibfnamefont {Christina}\
  \bibnamefont {Giarmatzi}}\ and\ \bibinfo {author} {\bibfnamefont {Fabio}\
  \bibnamefont {Costa}},\ }\bibfield  {title} {\enquote {\bibinfo {title}
  {Witnessing quantum memory in non-{M}arkovian processes},}\ }\href {\doibase
  10.22331/q-2021-04-26-440} {\bibfield  {journal} {\bibinfo  {journal}
  {{Quantum}}\ }\textbf {\bibinfo {volume} {5}},\ \bibinfo {pages} {440}
  (\bibinfo {year} {2021})}\BibitemShut {NoStop}%
\bibitem [{\citenamefont {Mallick}\ \emph {et~al.}(2024)\citenamefont
  {Mallick}, \citenamefont {Mukherjee}, \citenamefont {Maity},\ and\
  \citenamefont {Majumdar}}]{mallick2023}%
  \BibitemOpen
  \bibfield  {author} {\bibinfo {author} {\bibfnamefont {Bivas}\ \bibnamefont
  {Mallick}}, \bibinfo {author} {\bibfnamefont {Saheli}\ \bibnamefont
  {Mukherjee}}, \bibinfo {author} {\bibfnamefont {Ananda~G.}\ \bibnamefont
  {Maity}}, \ and\ \bibinfo {author} {\bibfnamefont {A.~S.}\ \bibnamefont
  {Majumdar}},\ }\bibfield  {title} {\enquote {\bibinfo {title} {Assessing
  non-markovian dynamics through moments of the choi state},}\ }\href {\doibase
  10.1103/PhysRevA.109.022247} {\bibfield  {journal} {\bibinfo  {journal}
  {Phys. Rev. A}\ }\textbf {\bibinfo {volume} {109}},\ \bibinfo {pages}
  {022247} (\bibinfo {year} {2024})}\BibitemShut {NoStop}%
\bibitem [{\citenamefont {Alicki}\ and\ \citenamefont {Lendi}(2007)}]{alicki}%
  \BibitemOpen
  \bibfield  {author} {\bibinfo {author} {\bibfnamefont {R}~\bibnamefont
  {Alicki}}\ and\ \bibinfo {author} {\bibfnamefont {K}~\bibnamefont {Lendi}},\
  }\href@noop {} {\emph {\bibinfo {title} {Quantum Dynamical Semigroups and
  Applications}}},\ Lecture notes in Physics\ (\bibinfo  {publisher}
  {Springer-Verlag Berlin Heidelberg},\ \bibinfo {year} {2007})\BibitemShut
  {NoStop}%
\bibitem [{\citenamefont {Laine}\ \emph {et~al.}(2014)\citenamefont {Laine},
  \citenamefont {Breuer},\ and\ \citenamefont {Piilo}}]{task1}%
  \BibitemOpen
  \bibfield  {author} {\bibinfo {author} {\bibfnamefont {Elsi-Mari}\
  \bibnamefont {Laine}}, \bibinfo {author} {\bibfnamefont {Heinz-Peter}\
  \bibnamefont {Breuer}}, \ and\ \bibinfo {author} {\bibfnamefont {Jyrki}\
  \bibnamefont {Piilo}},\ }\bibfield  {title} {\enquote {\bibinfo {title}
  {Nonlocal memory effects allow perfect teleportation with mixed states},}\
  }\href {https://www.nature.com/articles/srep04620} {\bibfield  {journal}
  {\bibinfo  {journal} {Scientific Reports}\ }\textbf {\bibinfo {volume} {4}},\
  \bibinfo {pages} {4620} (\bibinfo {year} {2014})}\BibitemShut {NoStop}%
\bibitem [{\citenamefont {Xiang}\ \emph {et~al.}(2014)\citenamefont {Xiang},
  \citenamefont {Hou}, \citenamefont {Li}, \citenamefont {Guo}, \citenamefont
  {Breuer}, \citenamefont {Laine},\ and\ \citenamefont {Piilo}}]{task3}%
  \BibitemOpen
  \bibfield  {author} {\bibinfo {author} {\bibfnamefont {Guo-Yong}\
  \bibnamefont {Xiang}}, \bibinfo {author} {\bibfnamefont {Zhi-Bo}\
  \bibnamefont {Hou}}, \bibinfo {author} {\bibfnamefont {Chuan-Feng}\
  \bibnamefont {Li}}, \bibinfo {author} {\bibfnamefont {Guang-Can}\
  \bibnamefont {Guo}}, \bibinfo {author} {\bibfnamefont {Heinz-Peter}\
  \bibnamefont {Breuer}}, \bibinfo {author} {\bibfnamefont {Elsi-Mari}\
  \bibnamefont {Laine}}, \ and\ \bibinfo {author} {\bibfnamefont {Jyrki}\
  \bibnamefont {Piilo}},\ }\bibfield  {title} {\enquote {\bibinfo {title}
  {Entanglement distribution in optical fibers assisted by nonlocal memory
  effects},}\ }\href {http://stacks.iop.org/0295-5075/107/i=5/a=54006}
  {\bibfield  {journal} {\bibinfo  {journal} {EPL (Europhysics Letters)}\
  }\textbf {\bibinfo {volume} {107}},\ \bibinfo {pages} {54006} (\bibinfo
  {year} {2014})}\BibitemShut {NoStop}%
\bibitem [{\citenamefont {Thomas}\ \emph {et~al.}(2018)\citenamefont {Thomas},
  \citenamefont {Siddharth}, \citenamefont {Banerjee},\ and\ \citenamefont
  {Ghosh}}]{task4}%
  \BibitemOpen
  \bibfield  {author} {\bibinfo {author} {\bibfnamefont {George}\ \bibnamefont
  {Thomas}}, \bibinfo {author} {\bibfnamefont {Nana}\ \bibnamefont
  {Siddharth}}, \bibinfo {author} {\bibfnamefont {Subhashish}\ \bibnamefont
  {Banerjee}}, \ and\ \bibinfo {author} {\bibfnamefont {Sibasish}\ \bibnamefont
  {Ghosh}},\ }\bibfield  {title} {\enquote {\bibinfo {title} {Thermodynamics of
  non-markovian reservoirs and heat engines},}\ }\href {\doibase
  10.1103/PhysRevE.97.062108} {\bibfield  {journal} {\bibinfo  {journal} {Phys.
  Rev. E}\ }\textbf {\bibinfo {volume} {97}},\ \bibinfo {pages} {062108}
  (\bibinfo {year} {2018})}\BibitemShut {NoStop}%
\bibitem [{\citenamefont {Reich}\ \emph {et~al.}(2015)\citenamefont {Reich},
  \citenamefont {Katz},\ and\ \citenamefont {Koch}}]{task5}%
  \BibitemOpen
  \bibfield  {author} {\bibinfo {author} {\bibfnamefont {Daneil~M}\
  \bibnamefont {Reich}}, \bibinfo {author} {\bibfnamefont {Nadav}\ \bibnamefont
  {Katz}}, \ and\ \bibinfo {author} {\bibfnamefont {C~P}\ \bibnamefont
  {Koch}},\ }\bibfield  {title} {\enquote {\bibinfo {title} {Exploiting
  non-markovianity for quantum control},}\ }\href
  {https://www.nature.com/articles/srep12430} {\bibfield  {journal} {\bibinfo
  {journal} {Scientific Reports}\ }\textbf {\bibinfo {volume} {5}},\ \bibinfo
  {pages} {12430} (\bibinfo {year} {2015})}\BibitemShut {NoStop}%
\bibitem [{\citenamefont {Breuer}\ and\ \citenamefont
  {Petruccione}(2002)}]{breuer}%
  \BibitemOpen
  \bibfield  {author} {\bibinfo {author} {\bibfnamefont {H.~P.}\ \bibnamefont
  {Breuer}}\ and\ \bibinfo {author} {\bibfnamefont {F.}~\bibnamefont
  {Petruccione}},\ }\href@noop {} {\emph {\bibinfo {title} {The theory of open
  quantum systems}}}\ (\bibinfo  {publisher} {Oxford University Press},\
  \bibinfo {address} {Great Clarendon Street},\ \bibinfo {year}
  {2002})\BibitemShut {NoStop}%
\bibitem [{\citenamefont {de~Vega}\ and\ \citenamefont
  {Alonso}(2017)}]{Vegareview}%
  \BibitemOpen
  \bibfield  {author} {\bibinfo {author} {\bibfnamefont {In\'es}\ \bibnamefont
  {de~Vega}}\ and\ \bibinfo {author} {\bibfnamefont {Daniel}\ \bibnamefont
  {Alonso}},\ }\bibfield  {title} {\enquote {\bibinfo {title} {Dynamics of
  non-markovian open quantum systems},}\ }\href {\doibase
  10.1103/RevModPhys.89.015001} {\bibfield  {journal} {\bibinfo  {journal}
  {Rev. Mod. Phys.}\ }\textbf {\bibinfo {volume} {89}},\ \bibinfo {pages}
  {015001} (\bibinfo {year} {2017})}\BibitemShut {NoStop}%
\bibitem [{\citenamefont {Breuer}\ \emph {et~al.}(2016)\citenamefont {Breuer},
  \citenamefont {Laine}, \citenamefont {Piilo},\ and\ \citenamefont
  {Vacchini}}]{BLPreview}%
  \BibitemOpen
  \bibfield  {author} {\bibinfo {author} {\bibfnamefont {Heinz-Peter}\
  \bibnamefont {Breuer}}, \bibinfo {author} {\bibfnamefont {Elsi-Mari}\
  \bibnamefont {Laine}}, \bibinfo {author} {\bibfnamefont {Jyrki}\ \bibnamefont
  {Piilo}}, \ and\ \bibinfo {author} {\bibfnamefont {Bassano}\ \bibnamefont
  {Vacchini}},\ }\bibfield  {title} {\enquote {\bibinfo {title} {Colloquium},}\
  }\href {\doibase 10.1103/RevModPhys.88.021002} {\bibfield  {journal}
  {\bibinfo  {journal} {Rev. Mod. Phys.}\ }\textbf {\bibinfo {volume} {88}},\
  \bibinfo {pages} {021002} (\bibinfo {year} {2016})}\BibitemShut {NoStop}%
\bibitem [{\citenamefont {Chiribella}\ \emph
  {et~al.}(2008{\natexlab{a}})\citenamefont {Chiribella}, \citenamefont
  {D’Ariano},\ and\ \citenamefont {Perinotti}}]{Chiribella08(1)}%
  \BibitemOpen
  \bibfield  {author} {\bibinfo {author} {\bibfnamefont {G.}~\bibnamefont
  {Chiribella}}, \bibinfo {author} {\bibfnamefont {G.~M.}\ \bibnamefont
  {D’Ariano}}, \ and\ \bibinfo {author} {\bibfnamefont {P.}~\bibnamefont
  {Perinotti}},\ }\bibfield  {title} {\enquote {\bibinfo {title} {Quantum
  circuit architecture},}\ }\href
  {https://doi.org/10.1103/PhysRevLett.101.060401} {\bibfield  {journal}
  {\bibinfo  {journal} {Phys. Rev. Lett.}\ }\textbf {\bibinfo {volume} {101}},\
  \bibinfo {pages} {060401} (\bibinfo {year} {2008}{\natexlab{a}})}\BibitemShut
  {NoStop}%
\bibitem [{\citenamefont {Chiribella}\ \emph {et~al.}(2009)\citenamefont
  {Chiribella}, \citenamefont {D'Ariano},\ and\ \citenamefont
  {Perinotti}}]{Chiribella09}%
  \BibitemOpen
  \bibfield  {author} {\bibinfo {author} {\bibfnamefont {Giulio}\ \bibnamefont
  {Chiribella}}, \bibinfo {author} {\bibfnamefont {Giacomo~Mauro}\ \bibnamefont
  {D'Ariano}}, \ and\ \bibinfo {author} {\bibfnamefont {Paolo}\ \bibnamefont
  {Perinotti}},\ }\bibfield  {title} {\enquote {\bibinfo {title} {Theoretical
  framework for quantum networks},}\ }\href {\doibase
  10.1103/PhysRevA.80.022339} {\bibfield  {journal} {\bibinfo  {journal} {Phys.
  Rev. A}\ }\textbf {\bibinfo {volume} {80}},\ \bibinfo {pages} {022339}
  (\bibinfo {year} {2009})}\BibitemShut {NoStop}%
\bibitem [{\citenamefont {White}\ \emph {et~al.}(2020)\citenamefont {White},
  \citenamefont {Hill}, \citenamefont {Pollock}, \citenamefont {Hollenberg},\
  and\ \citenamefont {Modi}}]{White2020}%
  \BibitemOpen
  \bibfield  {author} {\bibinfo {author} {\bibfnamefont {G.~A.~L.}\
  \bibnamefont {White}}, \bibinfo {author} {\bibfnamefont {C.~D.}\ \bibnamefont
  {Hill}}, \bibinfo {author} {\bibfnamefont {F.~A.}\ \bibnamefont {Pollock}},
  \bibinfo {author} {\bibfnamefont {L.~C.~L.}\ \bibnamefont {Hollenberg}}, \
  and\ \bibinfo {author} {\bibfnamefont {K.}~\bibnamefont {Modi}},\ }\bibfield
  {title} {\enquote {\bibinfo {title} {Demonstration of non-markovian process
  characterisation and control on a quantum processor},}\ }\href {\doibase
  10.1038/s41467-020-20113-3} {\bibfield  {journal} {\bibinfo  {journal}
  {Nature Communications}\ }\textbf {\bibinfo {volume} {11}} (\bibinfo {year}
  {2020}),\ 10.1038/s41467-020-20113-3}\BibitemShut {NoStop}%
\bibitem [{\citenamefont {Barreiro}\ \emph {et~al.}(2010)\citenamefont
  {Barreiro}, \citenamefont {Schindler}, \citenamefont {G\"{u}hne},
  \citenamefont {Monz}, \citenamefont {Chwalla}, \citenamefont {Roos},
  \citenamefont {Hennrich},\ and\ \citenamefont {Blatt}}]{Barreiro2010}%
  \BibitemOpen
  \bibfield  {author} {\bibinfo {author} {\bibfnamefont {Julio~T.}\
  \bibnamefont {Barreiro}}, \bibinfo {author} {\bibfnamefont {Philipp}\
  \bibnamefont {Schindler}}, \bibinfo {author} {\bibfnamefont {Otfried}\
  \bibnamefont {G\"{u}hne}}, \bibinfo {author} {\bibfnamefont {Thomas}\
  \bibnamefont {Monz}}, \bibinfo {author} {\bibfnamefont {Michael}\
  \bibnamefont {Chwalla}}, \bibinfo {author} {\bibfnamefont {Christian~F.}\
  \bibnamefont {Roos}}, \bibinfo {author} {\bibfnamefont {Markus}\ \bibnamefont
  {Hennrich}}, \ and\ \bibinfo {author} {\bibfnamefont {Rainer}\ \bibnamefont
  {Blatt}},\ }\bibfield  {title} {\enquote {\bibinfo {title} {Experimental
  multiparticle entanglement dynamics induced by decoherence},}\ }\href
  {\doibase 10.1038/nphys1781} {\bibfield  {journal} {\bibinfo  {journal}
  {Nature Physics}\ }\textbf {\bibinfo {volume} {6}},\ \bibinfo {pages}
  {943--946} (\bibinfo {year} {2010})}\BibitemShut {NoStop}%
\bibitem [{\citenamefont {Ball}\ \emph {et~al.}(2016)\citenamefont {Ball},
  \citenamefont {Stace}, \citenamefont {Flammia},\ and\ \citenamefont
  {Biercuk}}]{Ball16}%
  \BibitemOpen
  \bibfield  {author} {\bibinfo {author} {\bibfnamefont {Harrison}\
  \bibnamefont {Ball}}, \bibinfo {author} {\bibfnamefont {Thomas~M.}\
  \bibnamefont {Stace}}, \bibinfo {author} {\bibfnamefont {Steven~T.}\
  \bibnamefont {Flammia}}, \ and\ \bibinfo {author} {\bibfnamefont
  {Michael~J.}\ \bibnamefont {Biercuk}},\ }\bibfield  {title} {\enquote
  {\bibinfo {title} {Effect of noise correlations on randomized
  benchmarking},}\ }\href {\doibase 10.1103/PhysRevA.93.022303} {\bibfield
  {journal} {\bibinfo  {journal} {Phys. Rev. A}\ }\textbf {\bibinfo {volume}
  {93}},\ \bibinfo {pages} {022303} (\bibinfo {year} {2016})}\BibitemShut
  {NoStop}%
\bibitem [{\citenamefont {Figueroa-Romero}\ \emph {et~al.}(2021)\citenamefont
  {Figueroa-Romero}, \citenamefont {Modi}, \citenamefont {Harris},
  \citenamefont {Stace},\ and\ \citenamefont {Hsieh}}]{Figueroa-Romero21}%
  \BibitemOpen
  \bibfield  {author} {\bibinfo {author} {\bibfnamefont {Pedro}\ \bibnamefont
  {Figueroa-Romero}}, \bibinfo {author} {\bibfnamefont {Kavan}\ \bibnamefont
  {Modi}}, \bibinfo {author} {\bibfnamefont {Robert~J.}\ \bibnamefont
  {Harris}}, \bibinfo {author} {\bibfnamefont {Thomas~M.}\ \bibnamefont
  {Stace}}, \ and\ \bibinfo {author} {\bibfnamefont {Min-Hsiu}\ \bibnamefont
  {Hsieh}},\ }\bibfield  {title} {\enquote {\bibinfo {title} {Randomized
  benchmarking for non-markovian noise},}\ }\href {\doibase
  10.1103/PRXQuantum.2.040351} {\bibfield  {journal} {\bibinfo  {journal} {PRX
  Quantum}\ }\textbf {\bibinfo {volume} {2}},\ \bibinfo {pages} {040351}
  (\bibinfo {year} {2021})}\BibitemShut {NoStop}%
\bibitem [{\citenamefont {Figueroa-Romero}\ \emph {et~al.}(2022)\citenamefont
  {Figueroa-Romero}, \citenamefont {Modi},\ and\ \citenamefont
  {Hsieh}}]{FigueroaRomero2022}%
  \BibitemOpen
  \bibfield  {author} {\bibinfo {author} {\bibfnamefont {Pedro}\ \bibnamefont
  {Figueroa-Romero}}, \bibinfo {author} {\bibfnamefont {Kavan}\ \bibnamefont
  {Modi}}, \ and\ \bibinfo {author} {\bibfnamefont {Min-Hsiu}\ \bibnamefont
  {Hsieh}},\ }\bibfield  {title} {\enquote {\bibinfo {title} {Towards a general
  framework of {R}andomized {B}enchmarking incorporating non-{M}arkovian
  {N}oise},}\ }\href {\doibase 10.22331/q-2022-12-01-868} {\bibfield  {journal}
  {\bibinfo  {journal} {{Quantum}}\ }\textbf {\bibinfo {volume} {6}},\ \bibinfo
  {pages} {868} (\bibinfo {year} {2022})}\BibitemShut {NoStop}%
\bibitem [{\citenamefont {Addis}\ \emph {et~al.}(2015)\citenamefont {Addis},
  \citenamefont {Ciccarello}, \citenamefont {Cascio}, \citenamefont {Palma},\
  and\ \citenamefont {Maniscalco}}]{Addis2015}%
  \BibitemOpen
  \bibfield  {author} {\bibinfo {author} {\bibfnamefont {Carole}\ \bibnamefont
  {Addis}}, \bibinfo {author} {\bibfnamefont {Francesco}\ \bibnamefont
  {Ciccarello}}, \bibinfo {author} {\bibfnamefont {Michele}\ \bibnamefont
  {Cascio}}, \bibinfo {author} {\bibfnamefont {G~Massimo}\ \bibnamefont
  {Palma}}, \ and\ \bibinfo {author} {\bibfnamefont {Sabrina}\ \bibnamefont
  {Maniscalco}},\ }\bibfield  {title} {\enquote {\bibinfo {title} {Dynamical
  decoupling efficiency versus quantum non-markovianity},}\ }\href {\doibase
  10.1088/1367-2630/17/12/123004} {\bibfield  {journal} {\bibinfo  {journal}
  {New Journal of Physics}\ }\textbf {\bibinfo {volume} {17}},\ \bibinfo
  {pages} {123004} (\bibinfo {year} {2015})}\BibitemShut {NoStop}%
\bibitem [{\citenamefont {Biercuk}\ \emph {et~al.}(2009)\citenamefont
  {Biercuk}, \citenamefont {Uys}, \citenamefont {VanDevender}, \citenamefont
  {Shiga}, \citenamefont {Itano},\ and\ \citenamefont
  {Bollinger}}]{Biercuk2009}%
  \BibitemOpen
  \bibfield  {author} {\bibinfo {author} {\bibfnamefont {Michael~J.}\
  \bibnamefont {Biercuk}}, \bibinfo {author} {\bibfnamefont {Hermann}\
  \bibnamefont {Uys}}, \bibinfo {author} {\bibfnamefont {Aaron~P.}\
  \bibnamefont {VanDevender}}, \bibinfo {author} {\bibfnamefont {Nobuyasu}\
  \bibnamefont {Shiga}}, \bibinfo {author} {\bibfnamefont {Wayne~M.}\
  \bibnamefont {Itano}}, \ and\ \bibinfo {author} {\bibfnamefont {John~J.}\
  \bibnamefont {Bollinger}},\ }\bibfield  {title} {\enquote {\bibinfo {title}
  {Optimized dynamical decoupling in a model quantum memory},}\ }\href
  {\doibase 10.1038/nature07951} {\bibfield  {journal} {\bibinfo  {journal}
  {Nature}\ }\textbf {\bibinfo {volume} {458}},\ \bibinfo {pages} {996--1000}
  (\bibinfo {year} {2009})}\BibitemShut {NoStop}%
\bibitem [{\citenamefont {Awasthi}\ \emph {et~al.}(2018)\citenamefont
  {Awasthi}, \citenamefont {Bhattacharya}, \citenamefont {Sen(De)},\ and\
  \citenamefont {Sen}}]{SamyaErg}%
  \BibitemOpen
  \bibfield  {author} {\bibinfo {author} {\bibfnamefont {Natasha}\ \bibnamefont
  {Awasthi}}, \bibinfo {author} {\bibfnamefont {Samyadeb}\ \bibnamefont
  {Bhattacharya}}, \bibinfo {author} {\bibfnamefont {Aditi}\ \bibnamefont
  {Sen(De)}}, \ and\ \bibinfo {author} {\bibfnamefont {Ujjwal}\ \bibnamefont
  {Sen}},\ }\bibfield  {title} {\enquote {\bibinfo {title} {Universal quantum
  uncertainty relations between nonergodicity and loss of information},}\
  }\href {\doibase 10.1103/PhysRevA.97.032103} {\bibfield  {journal} {\bibinfo
  {journal} {Phys. Rev. A}\ }\textbf {\bibinfo {volume} {97}},\ \bibinfo
  {pages} {032103} (\bibinfo {year} {2018})}\BibitemShut {NoStop}%
\bibitem [{\citenamefont {Breuer}\ \emph {et~al.}(2009)\citenamefont {Breuer},
  \citenamefont {Laine},\ and\ \citenamefont {Piilo}}]{BLP2}%
  \BibitemOpen
  \bibfield  {author} {\bibinfo {author} {\bibfnamefont {Heinz-Peter}\
  \bibnamefont {Breuer}}, \bibinfo {author} {\bibfnamefont {Elsi-Mari}\
  \bibnamefont {Laine}}, \ and\ \bibinfo {author} {\bibfnamefont {Jyrki}\
  \bibnamefont {Piilo}},\ }\bibfield  {title} {\enquote {\bibinfo {title}
  {Measure for the degree of non-markovian behavior of quantum processes in
  open systems},}\ }\href {\doibase 10.1103/PhysRevLett.103.210401} {\bibfield
  {journal} {\bibinfo  {journal} {Phys. Rev. Lett.}\ }\textbf {\bibinfo
  {volume} {103}},\ \bibinfo {pages} {210401} (\bibinfo {year}
  {2009})}\BibitemShut {NoStop}%
\bibitem [{\citenamefont {Chiribella}\ \emph
  {et~al.}(2008{\natexlab{b}})\citenamefont {Chiribella}, \citenamefont
  {D'Ariano},\ and\ \citenamefont {Perinotti}}]{Chiribella2008}%
  \BibitemOpen
  \bibfield  {author} {\bibinfo {author} {\bibfnamefont {G.}~\bibnamefont
  {Chiribella}}, \bibinfo {author} {\bibfnamefont {G.~M.}\ \bibnamefont
  {D'Ariano}}, \ and\ \bibinfo {author} {\bibfnamefont {P.}~\bibnamefont
  {Perinotti}},\ }\bibfield  {title} {\enquote {\bibinfo {title} {Transforming
  quantum operations: Quantum supermaps},}\ }\href {\doibase
  10.1209/0295-5075/83/30004} {\bibfield  {journal} {\bibinfo  {journal}
  {Europhysics Letters}\ }\textbf {\bibinfo {volume} {83}},\ \bibinfo {pages}
  {30004} (\bibinfo {year} {2008}{\natexlab{b}})}\BibitemShut {NoStop}%
\bibitem [{\citenamefont {Gour}(2019)}]{Gour2019}%
  \BibitemOpen
  \bibfield  {author} {\bibinfo {author} {\bibfnamefont {Gilad}\ \bibnamefont
  {Gour}},\ }\bibfield  {title} {\enquote {\bibinfo {title} {Comparison of
  quantum channels by superchannels},}\ }\href {\doibase
  10.1109/TIT.2019.2907989} {\bibfield  {journal} {\bibinfo  {journal} {IEEE
  Transactions on Information Theory}\ }\textbf {\bibinfo {volume} {65}},\
  \bibinfo {pages} {5880--5904} (\bibinfo {year} {2019})}\BibitemShut {NoStop}%
\bibitem [{\citenamefont {Quintino}\ \emph {et~al.}(2019)\citenamefont
  {Quintino}, \citenamefont {Dong}, \citenamefont {Shimbo}, \citenamefont
  {Soeda},\ and\ \citenamefont {Murao}}]{Quintino2019}%
  \BibitemOpen
  \bibfield  {author} {\bibinfo {author} {\bibfnamefont {Marco~T\'ulio}\
  \bibnamefont {Quintino}}, \bibinfo {author} {\bibfnamefont {Qingxiuxiong}\
  \bibnamefont {Dong}}, \bibinfo {author} {\bibfnamefont {Atsushi}\
  \bibnamefont {Shimbo}}, \bibinfo {author} {\bibfnamefont {Akihito}\
  \bibnamefont {Soeda}}, \ and\ \bibinfo {author} {\bibfnamefont {Mio}\
  \bibnamefont {Murao}},\ }\bibfield  {title} {\enquote {\bibinfo {title}
  {Probabilistic exact universal quantum circuits for transforming unitary
  operations},}\ }\href {\doibase 10.1103/PhysRevA.100.062339} {\bibfield
  {journal} {\bibinfo  {journal} {Phys. Rev. A}\ }\textbf {\bibinfo {volume}
  {100}},\ \bibinfo {pages} {062339} (\bibinfo {year} {2019})}\BibitemShut
  {NoStop}%
\bibitem [{\citenamefont {Ara{\'{u}}jo}\ \emph {et~al.}(2017)\citenamefont
  {Ara{\'{u}}jo}, \citenamefont {Feix}, \citenamefont {Navascu{\'{e}}s},\ and\
  \citenamefont {Brukner}}]{Araujo2017}%
  \BibitemOpen
  \bibfield  {author} {\bibinfo {author} {\bibfnamefont {Mateus}\ \bibnamefont
  {Ara{\'{u}}jo}}, \bibinfo {author} {\bibfnamefont {Adrien}\ \bibnamefont
  {Feix}}, \bibinfo {author} {\bibfnamefont {Miguel}\ \bibnamefont
  {Navascu{\'{e}}s}}, \ and\ \bibinfo {author} {\bibfnamefont {{\v{C}}aslav}\
  \bibnamefont {Brukner}},\ }\bibfield  {title} {\enquote {\bibinfo {title} {A
  purification postulate for quantum mechanics with indefinite causal order},}\
  }\href {\doibase 10.22331/q-2017-04-26-10} {\bibfield  {journal} {\bibinfo
  {journal} {{Quantum}}\ }\textbf {\bibinfo {volume} {1}},\ \bibinfo {pages}
  {10} (\bibinfo {year} {2017})}\BibitemShut {NoStop}%
\bibitem [{\citenamefont {Guerin}\ \emph {et~al.}(2019)\citenamefont {Guerin},
  \citenamefont {Rubino},\ and\ \citenamefont {\v{C} Brukner}}]{Guerin18}%
  \BibitemOpen
  \bibfield  {author} {\bibinfo {author} {\bibfnamefont {P.~A.}\ \bibnamefont
  {Guerin}}, \bibinfo {author} {\bibfnamefont {G.}~\bibnamefont {Rubino}}, \
  and\ \bibinfo {author} {\bibnamefont {\v{C} Brukner}},\ }\bibfield  {title}
  {\enquote {\bibinfo {title} {Communication through quantum-controlled
  noise},}\ }\href {https://doi.org/10.1103/PhysRevA.99.062317} {\bibfield
  {journal} {\bibinfo  {journal} {Phys. Rev. A}\ }\textbf {\bibinfo {volume}
  {99}},\ \bibinfo {pages} {062317} (\bibinfo {year} {2019})}\BibitemShut
  {NoStop}%
\bibitem [{\citenamefont {Dong}\ \emph {et~al.}(2023)\citenamefont {Dong},
  \citenamefont {Quintino}, \citenamefont {Soeda},\ and\ \citenamefont
  {Murao}}]{Dong2023}%
  \BibitemOpen
  \bibfield  {author} {\bibinfo {author} {\bibfnamefont {Qingxiuxiong}\
  \bibnamefont {Dong}}, \bibinfo {author} {\bibfnamefont {Marco~T{\'{u}}lio}\
  \bibnamefont {Quintino}}, \bibinfo {author} {\bibfnamefont {Akihito}\
  \bibnamefont {Soeda}}, \ and\ \bibinfo {author} {\bibfnamefont {Mio}\
  \bibnamefont {Murao}},\ }\bibfield  {title} {\enquote {\bibinfo {title} {The
  quantum switch is uniquely defined by its action on unitary operations},}\
  }\href {\doibase 10.22331/q-2023-11-07-1169} {\bibfield  {journal} {\bibinfo
  {journal} {{Quantum}}\ }\textbf {\bibinfo {volume} {7}},\ \bibinfo {pages}
  {1169} (\bibinfo {year} {2023})}\BibitemShut {NoStop}%
\bibitem [{\citenamefont {Araújo}\ \emph {et~al.}(2015)\citenamefont
  {Araújo}, \citenamefont {Branciard}, \citenamefont {Costa}, \citenamefont
  {Feix}, \citenamefont {Giarmatzi},\ and\ \citenamefont {Časlav
  Brukner}}]{Araujo2015}%
  \BibitemOpen
  \bibfield  {author} {\bibinfo {author} {\bibfnamefont {Mateus}\ \bibnamefont
  {Araújo}}, \bibinfo {author} {\bibfnamefont {Cyril}\ \bibnamefont
  {Branciard}}, \bibinfo {author} {\bibfnamefont {Fabio}\ \bibnamefont
  {Costa}}, \bibinfo {author} {\bibfnamefont {Adrien}\ \bibnamefont {Feix}},
  \bibinfo {author} {\bibfnamefont {Christina}\ \bibnamefont {Giarmatzi}}, \
  and\ \bibinfo {author} {\bibnamefont {Časlav Brukner}},\ }\bibfield  {title}
  {\enquote {\bibinfo {title} {Witnessing causal nonseparability},}\ }\href
  {\doibase 10.1088/1367-2630/17/10/102001} {\bibfield  {journal} {\bibinfo
  {journal} {New Journal of Physics}\ }\textbf {\bibinfo {volume} {17}},\
  \bibinfo {pages} {102001} (\bibinfo {year} {2015})}\BibitemShut {NoStop}%
\bibitem [{\citenamefont {Branciard}(2016)}]{Branciard2016}%
  \BibitemOpen
  \bibfield  {author} {\bibinfo {author} {\bibfnamefont {Cyril}\ \bibnamefont
  {Branciard}},\ }\bibfield  {title} {\enquote {\bibinfo {title} {Witnesses of
  causal nonseparability: an introduction and a few case studies},}\ }\href
  {\doibase 10.1038/srep26018} {\bibfield  {journal} {\bibinfo  {journal}
  {Scientific Reports}\ }\textbf {\bibinfo {volume} {6}} (\bibinfo {year}
  {2016}),\ 10.1038/srep26018}\BibitemShut {NoStop}%
\bibitem [{\citenamefont {Bavaresco}\ \emph {et~al.}(2019)\citenamefont
  {Bavaresco}, \citenamefont {Ara{\'{u}}jo}, \citenamefont {Brukner},\ and\
  \citenamefont {Quintino}}]{Bavaresco2019}%
  \BibitemOpen
  \bibfield  {author} {\bibinfo {author} {\bibfnamefont {Jessica}\ \bibnamefont
  {Bavaresco}}, \bibinfo {author} {\bibfnamefont {Mateus}\ \bibnamefont
  {Ara{\'{u}}jo}}, \bibinfo {author} {\bibfnamefont {{\v{C}}aslav}\
  \bibnamefont {Brukner}}, \ and\ \bibinfo {author} {\bibfnamefont
  {Marco~T{\'{u}}lio}\ \bibnamefont {Quintino}},\ }\bibfield  {title} {\enquote
  {\bibinfo {title} {Semi-device-independent certification of indefinite causal
  order},}\ }\href {\doibase 10.22331/q-2019-08-19-176} {\bibfield  {journal}
  {\bibinfo  {journal} {{Quantum}}\ }\textbf {\bibinfo {volume} {3}},\ \bibinfo
  {pages} {176} (\bibinfo {year} {2019})}\BibitemShut {NoStop}%
\bibitem [{\citenamefont {Dourdent}\ \emph {et~al.}(2022)\citenamefont
  {Dourdent}, \citenamefont {Abbott}, \citenamefont {Brunner}, \citenamefont
  {\ifmmode \check{S}\else \v{S}\fi{}upi\ifmmode~\acute{c}\else \'{c}\fi{}},\
  and\ \citenamefont {Branciard}}]{Dourdent22}%
  \BibitemOpen
  \bibfield  {author} {\bibinfo {author} {\bibfnamefont {Hippolyte}\
  \bibnamefont {Dourdent}}, \bibinfo {author} {\bibfnamefont {Alastair~A.}\
  \bibnamefont {Abbott}}, \bibinfo {author} {\bibfnamefont {Nicolas}\
  \bibnamefont {Brunner}}, \bibinfo {author} {\bibfnamefont {Ivan}\
  \bibnamefont {\ifmmode \check{S}\else \v{S}\fi{}upi\ifmmode~\acute{c}\else
  \'{c}\fi{}}}, \ and\ \bibinfo {author} {\bibfnamefont {Cyril}\ \bibnamefont
  {Branciard}},\ }\bibfield  {title} {\enquote {\bibinfo {title}
  {Semi-device-independent certification of causal nonseparability with trusted
  quantum inputs},}\ }\href {\doibase 10.1103/PhysRevLett.129.090402}
  {\bibfield  {journal} {\bibinfo  {journal} {Phys. Rev. Lett.}\ }\textbf
  {\bibinfo {volume} {129}},\ \bibinfo {pages} {090402} (\bibinfo {year}
  {2022})}\BibitemShut {NoStop}%
\bibitem [{\citenamefont {van~der Lugt}\ \emph {et~al.}(2023)\citenamefont
  {van~der Lugt}, \citenamefont {Barrett},\ and\ \citenamefont
  {Chiribella}}]{vanderLugt2023}%
  \BibitemOpen
  \bibfield  {author} {\bibinfo {author} {\bibfnamefont {Tein}\ \bibnamefont
  {van~der Lugt}}, \bibinfo {author} {\bibfnamefont {Jonathan}\ \bibnamefont
  {Barrett}}, \ and\ \bibinfo {author} {\bibfnamefont {Giulio}\ \bibnamefont
  {Chiribella}},\ }\bibfield  {title} {\enquote {\bibinfo {title}
  {Device-independent certification of indefinite causal order in the quantum
  switch},}\ }\href {\doibase 10.1038/s41467-023-40162-8} {\bibfield  {journal}
  {\bibinfo  {journal} {Nature Communications}\ }\textbf {\bibinfo {volume}
  {14}} (\bibinfo {year} {2023}),\ 10.1038/s41467-023-40162-8}\BibitemShut
  {NoStop}%
\bibitem [{\citenamefont {Burgarth}\ \emph {et~al.}(2013)\citenamefont
  {Burgarth}, \citenamefont {Chiribella}, \citenamefont {Giovannetti},
  \citenamefont {Perinotti},\ and\ \citenamefont {Yuasa}}]{Burgarth_2013}%
  \BibitemOpen
  \bibfield  {author} {\bibinfo {author} {\bibfnamefont {D}~\bibnamefont
  {Burgarth}}, \bibinfo {author} {\bibfnamefont {G}~\bibnamefont {Chiribella}},
  \bibinfo {author} {\bibfnamefont {V}~\bibnamefont {Giovannetti}}, \bibinfo
  {author} {\bibfnamefont {P}~\bibnamefont {Perinotti}}, \ and\ \bibinfo
  {author} {\bibfnamefont {K}~\bibnamefont {Yuasa}},\ }\bibfield  {title}
  {\enquote {\bibinfo {title} {Ergodic and mixing quantum channels in finite
  dimensions},}\ }\href {\doibase 10.1088/1367-2630/15/7/073045} {\bibfield
  {journal} {\bibinfo  {journal} {New Journal of Physics}\ }\textbf {\bibinfo
  {volume} {15}},\ \bibinfo {pages} {073045} (\bibinfo {year}
  {2013})}\BibitemShut {NoStop}%
\bibitem [{\citenamefont {Chru\ifmmode \acute{s}\else
  \'{s}\fi{}ci\ifmmode~\acute{n}\else \'{n}\fi{}ski}\ and\ \citenamefont
  {Siudzi\ifmmode~\acute{n}\else \'{n}\fi{}ska}(2016)}]{genPauli}%
  \BibitemOpen
  \bibfield  {author} {\bibinfo {author} {\bibfnamefont {Dariusz}\ \bibnamefont
  {Chru\ifmmode \acute{s}\else \'{s}\fi{}ci\ifmmode~\acute{n}\else
  \'{n}\fi{}ski}}\ and\ \bibinfo {author} {\bibfnamefont {Katarzyna}\
  \bibnamefont {Siudzi\ifmmode~\acute{n}\else \'{n}\fi{}ska}},\ }\bibfield
  {title} {\enquote {\bibinfo {title} {Generalized pauli channels and a class
  of non-markovian quantum evolution},}\ }\href {\doibase
  10.1103/PhysRevA.94.022118} {\bibfield  {journal} {\bibinfo  {journal} {Phys.
  Rev. A}\ }\textbf {\bibinfo {volume} {94}},\ \bibinfo {pages} {022118}
  (\bibinfo {year} {2016})}\BibitemShut {NoStop}%
\bibitem [{\citenamefont {{Lindblad}}(1976)}]{lindblad}%
  \BibitemOpen
  \bibfield  {author} {\bibinfo {author} {\bibfnamefont {G.}~\bibnamefont
  {{Lindblad}}},\ }\bibfield  {title} {\enquote {\bibinfo {title} {{On the
  generators of quantum dynamical semigroups}},}\ }\href {\doibase
  10.1007/BF01608499} {\bibfield  {journal} {\bibinfo  {journal}
  {Communications in Mathematical Physics}\ }\textbf {\bibinfo {volume} {48}},\
  \bibinfo {pages} {119--130} (\bibinfo {year} {1976})}\BibitemShut {NoStop}%
\bibitem [{\citenamefont {{Gorini}}\ \emph {et~al.}(1976)\citenamefont
  {{Gorini}}, \citenamefont {{Kossakowski}},\ and\ \citenamefont
  {{Sudarshan}}}]{gorini}%
  \BibitemOpen
  \bibfield  {author} {\bibinfo {author} {\bibfnamefont {V.}~\bibnamefont
  {{Gorini}}}, \bibinfo {author} {\bibfnamefont {A.}~\bibnamefont
  {{Kossakowski}}}, \ and\ \bibinfo {author} {\bibfnamefont {E.~C.~G.}\
  \bibnamefont {{Sudarshan}}},\ }\bibfield  {title} {\enquote {\bibinfo {title}
  {{Completely positive dynamical semigroups of N-level systems}},}\ }\href
  {\doibase 10.1063/1.522979} {\bibfield  {journal} {\bibinfo  {journal}
  {Journal of Mathematical Physics}\ }\textbf {\bibinfo {volume} {17}},\
  \bibinfo {pages} {821--825} (\bibinfo {year} {1976})}\BibitemShut {NoStop}%
\bibitem [{\citenamefont {Wechs}\ \emph {et~al.}(2021)\citenamefont {Wechs},
  \citenamefont {Dourdent}, \citenamefont {Abbott},\ and\ \citenamefont
  {Branciard}}]{Wechs21}%
  \BibitemOpen
  \bibfield  {author} {\bibinfo {author} {\bibfnamefont {Julian}\ \bibnamefont
  {Wechs}}, \bibinfo {author} {\bibfnamefont {Hippolyte}\ \bibnamefont
  {Dourdent}}, \bibinfo {author} {\bibfnamefont {Alastair~A.}\ \bibnamefont
  {Abbott}}, \ and\ \bibinfo {author} {\bibfnamefont {Cyril}\ \bibnamefont
  {Branciard}},\ }\bibfield  {title} {\enquote {\bibinfo {title} {Quantum
  circuits with classical versus quantum control of causal order},}\ }\href
  {\doibase 10.1103/PRXQuantum.2.030335} {\bibfield  {journal} {\bibinfo
  {journal} {PRX Quantum}\ }\textbf {\bibinfo {volume} {2}},\ \bibinfo {pages}
  {030335} (\bibinfo {year} {2021})}\BibitemShut {NoStop}%
\bibitem [{\citenamefont {Loizeau}\ and\ \citenamefont
  {Grinbaum}(2020)}]{sup-switch1}%
  \BibitemOpen
  \bibfield  {author} {\bibinfo {author} {\bibfnamefont {Nicolas}\ \bibnamefont
  {Loizeau}}\ and\ \bibinfo {author} {\bibfnamefont {Alexei}\ \bibnamefont
  {Grinbaum}},\ }\bibfield  {title} {\enquote {\bibinfo {title} {Channel
  capacity enhancement with indefinite causal order},}\ }\href {\doibase
  10.1103/PhysRevA.101.012340} {\bibfield  {journal} {\bibinfo  {journal}
  {Phys. Rev. A}\ }\textbf {\bibinfo {volume} {101}},\ \bibinfo {pages}
  {012340} (\bibinfo {year} {2020})}\BibitemShut {NoStop}%
\bibitem [{\citenamefont {Abbott}\ \emph {et~al.}(2020)\citenamefont {Abbott},
  \citenamefont {Wechs}, \citenamefont {Horsman}, \citenamefont {Mhalla},\ and\
  \citenamefont {Branciard}}]{Abbott20}%
  \BibitemOpen
  \bibfield  {author} {\bibinfo {author} {\bibfnamefont {Alastair~A.}\
  \bibnamefont {Abbott}}, \bibinfo {author} {\bibfnamefont {Julian}\
  \bibnamefont {Wechs}}, \bibinfo {author} {\bibfnamefont {Dominic}\
  \bibnamefont {Horsman}}, \bibinfo {author} {\bibfnamefont {Mehdi}\
  \bibnamefont {Mhalla}}, \ and\ \bibinfo {author} {\bibfnamefont {Cyril}\
  \bibnamefont {Branciard}},\ }\bibfield  {title} {\enquote {\bibinfo {title}
  {Communication through coherent control of quantum channels},}\ }\href
  {\doibase 10.22331/q-2020-09-24-333} {\bibfield  {journal} {\bibinfo
  {journal} {Quantum}\ }\textbf {\bibinfo {volume} {4}},\ \bibinfo {pages}
  {333} (\bibinfo {year} {2020})}\BibitemShut {NoStop}%
\bibitem [{\citenamefont {Fellous-Asiani}\ \emph {et~al.}(2023)\citenamefont
  {Fellous-Asiani}, \citenamefont {Mothe}, \citenamefont {Bresque},
  \citenamefont {Dourdent}, \citenamefont {Camati}, \citenamefont {Abbott},
  \citenamefont {Auff\`eves},\ and\ \citenamefont {Branciard}}]{sup-switch2}%
  \BibitemOpen
  \bibfield  {author} {\bibinfo {author} {\bibfnamefont {Marco}\ \bibnamefont
  {Fellous-Asiani}}, \bibinfo {author} {\bibfnamefont {Rapha\"el}\ \bibnamefont
  {Mothe}}, \bibinfo {author} {\bibfnamefont {L\'ea}\ \bibnamefont {Bresque}},
  \bibinfo {author} {\bibfnamefont {Hippolyte}\ \bibnamefont {Dourdent}},
  \bibinfo {author} {\bibfnamefont {Patrice~A.}\ \bibnamefont {Camati}},
  \bibinfo {author} {\bibfnamefont {Alastair~A.}\ \bibnamefont {Abbott}},
  \bibinfo {author} {\bibfnamefont {Alexia}\ \bibnamefont {Auff\`eves}}, \ and\
  \bibinfo {author} {\bibfnamefont {Cyril}\ \bibnamefont {Branciard}},\
  }\bibfield  {title} {\enquote {\bibinfo {title} {Comparing the quantum switch
  and its simulations with energetically constrained operations},}\ }\href
  {\doibase 10.1103/PhysRevResearch.5.023111} {\bibfield  {journal} {\bibinfo
  {journal} {Phys. Rev. Res.}\ }\textbf {\bibinfo {volume} {5}},\ \bibinfo
  {pages} {023111} (\bibinfo {year} {2023})}\BibitemShut {NoStop}%
\bibitem [{\citenamefont {Hall}\ \emph {et~al.}(2014)\citenamefont {Hall},
  \citenamefont {Cresser}, \citenamefont {Li},\ and\ \citenamefont
  {Andersson}}]{Hall14}%
  \BibitemOpen
  \bibfield  {author} {\bibinfo {author} {\bibfnamefont {Michael J.~W.}\
  \bibnamefont {Hall}}, \bibinfo {author} {\bibfnamefont {James~D.}\
  \bibnamefont {Cresser}}, \bibinfo {author} {\bibfnamefont {Li}~\bibnamefont
  {Li}}, \ and\ \bibinfo {author} {\bibfnamefont {Erika}\ \bibnamefont
  {Andersson}},\ }\bibfield  {title} {\enquote {\bibinfo {title} {Canonical
  form of master equations and characterization of non-markovianity},}\ }\href
  {\doibase 10.1103/PhysRevA.89.042120} {\bibfield  {journal} {\bibinfo
  {journal} {Phys. Rev. A}\ }\textbf {\bibinfo {volume} {89}},\ \bibinfo
  {pages} {042120} (\bibinfo {year} {2014})}\BibitemShut {NoStop}%
\bibitem [{\citenamefont {Bhattacharya}\ \emph {et~al.}(2017)\citenamefont
  {Bhattacharya}, \citenamefont {Misra}, \citenamefont {Mukhopadhyay},\ and\
  \citenamefont {Pati}}]{Bhattacharya17}%
  \BibitemOpen
  \bibfield  {author} {\bibinfo {author} {\bibfnamefont {Samyadeb}\
  \bibnamefont {Bhattacharya}}, \bibinfo {author} {\bibfnamefont {Avijit}\
  \bibnamefont {Misra}}, \bibinfo {author} {\bibfnamefont {Chiranjib}\
  \bibnamefont {Mukhopadhyay}}, \ and\ \bibinfo {author} {\bibfnamefont
  {Arun~Kumar}\ \bibnamefont {Pati}},\ }\bibfield  {title} {\enquote {\bibinfo
  {title} {Exact master equation for a spin interacting with a spin bath:
  Non-markovianity and negative entropy production rate},}\ }\href {\doibase
  10.1103/PhysRevA.95.012122} {\bibfield  {journal} {\bibinfo  {journal} {Phys.
  Rev. A}\ }\textbf {\bibinfo {volume} {95}},\ \bibinfo {pages} {012122}
  (\bibinfo {year} {2017})}\BibitemShut {NoStop}%
\end{thebibliography}%

\end{document}